\documentclass[10pt,a4paper]{article}

\usepackage[round]{natbib}
\usepackage{amsfonts}
\usepackage{dsfont}
\usepackage{physics}
\usepackage{xcolor}
\usepackage{algorithm}
\usepackage{indentfirst}
\usepackage{booktabs}
\usepackage{graphicx}
\usepackage{subcaption}

\newtheorem{assumption}{Assumption}
\usepackage{authblk}
\usepackage{parskip}
\usepackage{xurl}

\usepackage[bottom=2.5cm,top=2.5cm,left=2.5cm,right=2.5cm]{geometry}

\usepackage[utf8]{inputenc}  
\usepackage[T1]{fontenc}

\begin{document}

\providecommand{\keywords}[1]{\textbf{\textit{Keywords --}} #1}

\title{Learning the Spoofability of Limit Order Books With Interpretable Probabilistic Neural Networks}
\author[1,2]{Timothée Fabre}
\author[1]{Damien Challet}
\affil[1]{Laboratoire MICS, CentraleSupélec, Université Paris-Saclay, France.}
\affil[2]{SUN ZU Lab, Paris, France.}
\date{\today}

\maketitle
%\newpage
\begin{abstract}
\noindent 

This paper investigates real-time detection of spoofing activity in limit order books, focusing on cryptocurrency centralized exchanges. We first introduce novel order flow variables based on multi-scale Hawkes processes that account both for the size and placement distance from current best prices of new limit orders. Using a Level-3 data set, we train a neural network model to predict the conditional probability distribution of mid price movements based on these features. Our empirical analysis highlights the critical role of the posting distance of limit orders in the price formation process, showing that spoofing detection models that do not take the posting distance into account are inadequate to describe the data. Next, we propose a spoofing detection framework based on the probabilistic market manipulation gain of a spoofing agent and use the previously trained neural network to compute the expected gain. Running this algorithm on all submitted limit orders in the period 2024-12-04 to 2024-12-07, we find that 31\% of large orders could spoof the market. Because of its simple neuronal architecture, our model can be run in real time. This work contributes to enhancing market integrity by providing a robust tool for monitoring and mitigating spoofing in both cryptocurrency exchanges and traditional financial markets.

\vspace{0.5cm}

\noindent \keywords{High-Frequency, Market Manipulation, Spoofing, Neural Network, Limit Order Book, Cryptocurrency}
\end{abstract}

\setcounter{tocdepth}{2}
\tableofcontents

\section{Introduction}

Modern electronic financial markets use  limit order books (LOBs) that gather the buy and sell intentions of market participants. Because traders react to LOB updates in real time, some agents try to manipulate microstructural features of LOBs to tweak the dynamics of LOBs with  sophisticated high-frequency techniques. One such manipulative practice that has attracted significant attention is spoofing. In its simplest form, spoofing is an attempt to deceive market participants about the true supply and demand by placing large orders on one side of the LOB to trigger executions on the opposite side. Spoofing can involve various tactics, including layering and vacuuming. Layering consists in placing multiple orders at different price levels to create an impression of increase in market liquidity, and then canceling or modifying these orders once the desired market reaction is observed. Vacuuming occurs when a spoofer places substantial orders to absorb liquidity at specific price levels, subsequently canceling them to form large liquidity gaps and profit from a price movement.

LOBs in cryptocurrency centralized exchanges (CEXs) work similarly to those of traditional financial markets. However, the absence of market surveillance and strong regulations on CEXs opens the door to manipulative behaviors. Such unregulated markets are thereby a fascinating playground for researchers working on market manipulation detection. Intuitively, it is reasonable to postulate that spoofing activity is substantially larger in CEXs than in regulated equity markets.
Detecting spoofing in a live environment is exceptionally challenging due to the multidimensional aspects of the problem, hence the relative scarcity of scientific literature on this topic. Indeed, the development of a realistic and practical spoofing model should specify three key components:
\begin{itemize}
    \item \textbf{pricing} which involves the distance of placement of spoof orders and the market depth they aim to manipulate;
    \item \textbf{sizing} which refers to the amount of fake liquidity injected into the market;
    \item \textbf{horizon} which relates to the time frame over which the spoofer operates, which can range from milliseconds to seconds and depends on many variables such as volatility, trading activity, targeted price impact, etc.
\end{itemize}

The complexity of spoofing detection is further exacerbated for small-tick assets as the distance of placement cannot be represented by a discrete variable taking only a few values, \textit{e.g.} 2 ticks from the best price, but is best represented as a continuous variable, complicating the detection process. Traditional detection methods, which are often developed for large-tick assets, \textit{i.e.}, on discrete price level frameworks, are thus inadequate for small-tick assets.

Past works have focused on the analysis of spoofing and the implication of the microstructure of the electronic markets on potential manipulative behaviour of trading engines. For example,  \cite{lee2013microstructure} analyze the behaviour of anonymous market participants and identify spoofing orders with classification rules based on pre-defined thresholds. Their labelling methodology is based on the characteristics of the orders and a sequence of actions made by agents. They find that spoofing orders are placed deep in the book.  \cite{wang2015strategic} show that spoofing is profitable, destabilizes the market, and that it is more likely to happen in periods of high activity. Prediction of manipulative behaviours often relies on a data set of prosecuted cases, as in \cite{do2023detecting}. While Machine Learning models can be successfully applied to the detection of some suspicious trading patterns \citep{martinez2019order}, more recent works have developed sophisticated tools to complete this task, based on anomaly detection, see \cite{kang2023conspiracy,poutre2024deep} for instance.

An interesting perspective is introduced by \cite{tao2022detecting} who focus on a price formation model built upon a multi-level imbalance measure. The authors take the point of view of the spoofer and see the problem as an expected cost optimization. The simplicity of this framework enables them to find the optimal strategy a spoofer must have to profit from a spoofing order, which is then applied to the detection of suspicious behaviours in a trading environment. The main novelty of this detection tool is its unsupervised nature, as it does not rely on a labelled data set. Nevertheless, the lack of dependency of the price dynamic over the past order flow and the hypothesis that spoofers  place their order optimally leaves room for improvement. %In addition, except for the latter work, the unsupervised detection of suspicious orders remains mostly unexplored and opens the way to the development of new methodologies.

\citet{cartea2020spoofing, cartea2023spoofing} use the classic liquidity imbalance to characterize spoofing,  defined by the relative liquidity of the best queues: denote by $V^b$ the volume at the best bid and $V^a$ the volume at the best ask, then the liquidity imbalance is 
\begin{equation}
    \mathcal{I}:=\frac{V^b-V^a}{V^b+V^a}.
\end{equation}

An abundant literature has been written about the use of $\mathcal{I}$ as a high-frequency predictor for the next trade side or the direction of the next price move \citep{gould2016queue, lehalle2017limit, pulido2023understanding}, which has been known and applied for a while now. The problem with this measure for spoofing is that spoofers hardly place non-\textit{bona fide} orders at the best prices. In fact, even though placing a large order at one best queue generally reverses the imbalance, thereby bumping the probability of observing a trade on the other side of the book, it is still an extremely risky practice for the manipulative agent. For example, her manipulative order could be filled by any execution algorithm looking for the right opportunity to get a price discount on a large trade. In practice, spoofers place orders  deeper in the book, adjusting the distance in order to maximize the price impact while minimizing the probability of execution of the non-\textit{bona fide} order. Therefore, the imbalance measure $\mathcal{I}$ cannot be used, and a more sophisticated predictive variable ---computed over more than one price limit--- needs to be defined. In this spirit, \cite{tao2022detecting} construct a large tick model with a price dynamic that depends on a new liquidity imbalance that is built upon several price limits up to a fixed book depth. Let $K$ be a fixed depth, \textit{e.g.} 5 price limits, for $1\leq k\leq K$, $V_k^b$ the volume at the $k$th bid price limit ($k=1$ corresponds to the best price), $V_k^a$ the volume at the $k$th ask price limit, and define
\begin{equation}
    \mathcal{I}_K:=\frac{\sum_{k=1}^K\omega_kV_k^b}{\sum_{k=1}^K\omega_k(V_k^b+V_k^a)},
\end{equation}
where $(\omega_k)_{1\leq k\leq K}$ is a sequence of weights applied to the limits. The authors specify a model for the price dynamic and estimate the liquidity weights via likelihood maximization. Interestingly, they find that the weights are not decreasing with the price limit, indicating that price impact could be greater with limit orders posted far away from the best limit. But this framework is only suited for large tick assets and the generalized liquidity imbalance $\mathcal{I}_K$ alone is certainly not enough to capture the intricacies of the price dynamic at the high-frequency scale as it is a snapshot of the market state at the time of observation.

In this work, we define new order flow measures and leverage them as a set of features in order to predict the mid price movement distribution after the insertion of spoofing orders, allowing us to compute the expected cost function of the spoofer. Our specification is inspired by the literature on  Hawkes processes applied to limit order books. Let us recall its main concepts: a univariate linear Hawkes process $(N_t)_{t\geq0}$ is a self-exciting point process whose conditional intensity $\lambda_t$ at time $t\geq0$ can be written
\begin{equation}
    \lambda_t=\mu+\int_{(-\infty,t]}h(t-s)\mathrm{d}N_s,
\end{equation}
where $\mu$ is the baseline, and $h$ a positive function. Theses point processes and their application to high-frequency data have been extensively studied in the past decade, and their predictive power has been demonstrated when being applied to events that tend to cluster over time, see \cite{muni2020analyzing, muni2022marked, sfendourakis2023lob} for instance. Once a parametric Hawkes process is estimated on data, \textit{e.g.} in the case of a parametric kernel, the computation of the probability of observing a new event in a small interval $[t,t+\mathrm{d}t)$ is made simple using the approximation
\begin{equation}
    \mathbb{P}(N_{t+\mathrm{d}t}-N_t=1\,| \mathcal{F}_t)\simeq\lambda_t\,\mathrm{d}t\simeq\mu\mathrm{d}t+\left(\int_{(-\infty,t]}h(t-s)\mathrm{d}N_s\right)\mathrm{d}t.
\end{equation}

The order flow measures that we use in this work are directly inspired from the linear self-exciting part of this intensity, which encapsulates the predictive power of the endogenous part of the system. Once the measures are defined, the next question is the choice of the parametric form of the kernel functions. In fact, the true kernel of the intensity of order book events ---which relates to the order flow measures in our framework--- was found to be slowly decreasing in time \citep{bacry2012non, bacry2016first, fosset2020endogenous, fosset2022non}. Some studies have shown the dependence of the kernel in the size of the events \citep{rambaldi2017role}, recent elements pointing towards a non-decreasing and concave volume kernel \citep{fabre2024neural}. The literature about the influence of the distance of limit orders on the arrival rate of events is rather sparse and to the best of our knowledge, no non-parametric study has revealed any shape for the distance kernel. Hence, our work is the first to study the structure of dependence of the price formation with respect to the distance of placement of limit orders, and its implication in the detection of spoofing.

The outline of the paper is as follows. Section 2 focuses on a new family of order flow variables that are used as predictors. Then, a probabilistic neural network is trained on a data set of limit orders, and a sensitivity analysis is conducted on the resulting price move distribution with respect to the variables and the characteristics of limit orders. Section 3 introduces a spoofing framework in which an agent artificially inflates the liquidity on one side of the book. Introducing the cost function of the spoofer provides a natural real-time spoofing detection method whose effectiveness is assessed using several examples of suspicious orders. Section 4 finally provides elements of discussion and concludes.

\section{An order-driven price formation model}

In this Section, we define Hawkes-inspired order flow variables, some of which are directly sensitive to the action of the spoofer. We then build a model of the price movement distribution induced by the insertion of a new order  and study how  spoofers can distort  the price formation process to their advantage.

\subsection{A marked Hawkes-inspired set of features}

We place ourselves in a filtered probability space $\left(\Omega, \mathcal{F}, (\mathcal{F}_t)_{t\geq 0}, \mathbb{P}\right)$. We set a trading horizon denoted by $T$. We further denote by $(p_t)_{0\leq t\leq T}$ the fair price process of the asset over the execution time frame, and $\Delta p:=p_T-p_0$ the fair price variation over the trading period.  In the rest of the paper, the fair price is the mid price, \textit{i.e.} the average between the best bid price and the best ask price. The framework can be adapted to any other definition of the fair price such as the micro price of \cite{stoikov2018micro}.

Let $(N_t^{\text{L},b})_{t\geq0}$ and $(N_t^{\text{L},a})_{t\geq0}$ be the point process counting the flow of buy, and respectively sell, limit orders, and let $(N_t^{\text{M},b})_{t\geq0}$ and $(N_t^{\text{M},a})_{t\geq0}$ analogously be the point processes of the flow of sell, resp. buy, marketable orders, \textit{i.e.} orders that cause transactions on the bid, resp. ask, side of the LOB. Denote by $(v_t^{\text{L},b})_{t\geq0}$ and $(v_t^{\text{L}, a})_{t\geq0}$ the mark process of the volume of buy, resp. sell, limit orders, $(\delta_t^{\text{L},b})_{t\geq0}$ and $(\delta_t^{\text{L},a})_{t\geq0}$ the mark processes of the distance of placement of buy, resp. sell, limit orders, and $(v_t^{\text{M}, b})_{t\geq0}$ and $(v_t^{\text{M},a})_{t\geq0}$ the mark process of the volume of sell, resp. buy, marketable orders. The marks are positive \textit{a.s.} for all $t\geq 0$, and $v_t^{\text{L},b}=0$ for all $t\geq 0$ such that $N_t^{\text{L},b}-N_{t^-}^{\text{L}, b}=0$ (the same holds for sell limit orders), and $v_t^{\text{M}, b}=0$ for all $t\geq 0$ such that $N_t^{\text{M}, b}-N_{t^-}^{\text{M}, b}=0$ (the same holds for buy marketable orders). Note that all processes are $(\mathcal{F}_t)_{t\geq 0}$-adapted.

We define the following order flow variables, for all $t\geq 0$,

\begin{equation}\label{eq:limit_order_flow_bid}
    L_t^b(\phi):=\int_{(-\infty,t]}\phi(t-s,v_s^{\text{L}, b},\delta_s^{\text{L},b})\mathrm{d}N_s^{\text{L}, b},
\end{equation}

\begin{equation}
    L_t^a(\phi):=\int_{(-\infty,t]}\phi(t-s,v_s^{\text{L}, a},\delta_s^{\text{L},a})\mathrm{d}N_s^{\text{L}, a},
\end{equation}

\begin{equation}
    M_t^b(\psi):=\int_{(-\infty,t]}\psi(t-s,v_s^{\text{M}, b})\mathrm{d}N_s^{\text{M}, b},
\end{equation}

\begin{equation}
    M_t^a(\psi):=\int_{(-\infty,t]}\psi(t-s,v_s^{\text{M}, a})\mathrm{d}N_s^{\text{M}, a},
\end{equation}

where $\phi$ and $\psi$ are positive functions. The kernel $\phi$ characterizes the impact of past arrivals of limit orders over the limit order flow variable and depends on the elapsed time, the size of the order and its distance of placement in the LOB. The kernel $\psi$ characterizes the impact of past arrivals of marketable orders over the marketable order flow variable and depends on the elapsed time and the size of the order. 

We aim at using these order flow measures to predict the distribution of the fair price movement $\Delta p$, by using a family of kernels with different parameter configurations. Each kernel function applies a weight to each past event with respect to its characteristics --- size and distance of placement for a limit order and size for a marketable order --- and encodes a specific time scale, such that a combination of kernels will reproduce the memory of order flow.

Let $f$ be a positive non-decreasing function such that $f(0)=0$, $\boldsymbol{\beta}$ and $\boldsymbol{\eta}$ two finite sets of positive elements, and define the following families of kernels,
\begin{equation}
    \boldsymbol{\phi}:=\{\phi_{\beta,\eta}, \beta\in\boldsymbol{\beta}, \eta\in\boldsymbol{\eta}\},
\end{equation}
and
\begin{equation}
    \boldsymbol{\psi}:=\{\psi_{\beta}, \beta\in\boldsymbol{\beta}\},
\end{equation}
where for all $t\geq 0$, $v\geq0$, $\delta\geq0$, 

\begin{equation}\label{eq:kernel_phi_exp}
    \phi_{\beta,\eta}(t,v,\delta)=e^{-\beta t}f(v)e^{-\eta \delta},
\end{equation}

and for all $t\geq 0$, $v\geq0$,

\begin{equation}\label{eq:kernel_psi_exp}
    \psi_{\beta}(t,v)=e^{-\beta t}f(v).
\end{equation}

We now define, for all $t\geq 0$,

\begin{equation}\label{eq:limit_order_flow_family}
    \mathcal{L}_t:=\{(L_t^b(\phi), L_t^a(\phi)), \phi\in\boldsymbol{\phi}\},
\end{equation}

and

\begin{equation}\label{eq:market_order_flow_family}
    \mathcal{M}_t:=\{(M_t^b(\psi), M_t^a(\psi)), \psi\in\boldsymbol{\psi}\}.
\end{equation}

The families $\mathcal{L}_t$ and $\mathcal{M}_t$ together form a set of order flow variables at time $t$ that characterize different time scales (parameter $\beta$) and distance scales (parameter $\eta$). The idea is to use these families as a set of explanatory variables for the prediction of the distribution of $\Delta p$. Our choice of exponential kernels in Equations \eqref{eq:kernel_phi_exp} and \eqref{eq:kernel_psi_exp} is justified by two points. Firstly, the Markov property of the exponential kernel leads to fast computations in a live trading environment, which is mandatory at the high-frequency scale. Secondly, using multiple scale parameters $\beta$ and $\eta$ for the prediction makes it possible to approximate well  any empirical speed of decay of the order flow memory over several decades \citep{bochud2007optimal}.
%, the authors show that for a given power exponent $\alpha>0$ and a given set of $N$ different time scales $\beta^i$, there exists coefficients $c^i$ such that $\sum_{i=1}^Nc^ie^{-\beta^it}$ approximates $t^{-\alpha}$ fairly well over $N-1$ decades. Our setting is thus sufficiently flexible to approximate slow decays, provided that the sets $\boldsymbol{\beta}$ and $\boldsymbol{\eta}$ are large enough.

\subsection{A probabilistic neural network}

\subsubsection{Presentation of the model}

Let $x\in\mathbb{R}^d$ be a vector of $d$ features observed by the manipulative agent. In order to compute her expected cost and decide whether it is better to spoof or not, she needs to have a model for the distribution of $\Delta p$, conditionally on $x$. As such, the price movement variable $\Delta p$ is assumed to follow a probability distribution $\mathcal{D}\left(\Theta(x)\right)$, denoted by $\Delta p \sim\mathcal{D}\left(\Theta(x)\right)$, where $\mathcal{D}$ is a parametric distribution and $\Theta(x)$ is a vector of parameters that depends on the state vector $x$. When the structure of $\Theta$ with respect to $x$ is known, it is possible to analyze the impact of an action over the distribution of $\Delta p$.

When $d$ is large and if no simple relationship can be postulated between the distribution of $\Delta p$ and $x$, a natural idea is to train a neural network to learn the structure of the parameters of the distribution over the feature space. The estimation procedure of these probabilistic neural networks and their performance have been carefully studied in the literature \citep{williams1996using, pourkamali2024probabilistic}.

\subsubsection{Data}

This work uses market-by-event data, \textit{i.e.} Level-3 data, provided by SUN ZU Lab's feed handlers. We observe the insertion of each limit order, with its timestamp, size and price. We focus on the two most traded spot pairs on Coinbase, BTC-USD and ETH-USD, and we select an arbitrary week, from December 1st to December 7th, 2022. In order to remove noisy information from the data set, we keep orders that are posted not too far in the book. We set the distance threshold at $\pm$20\% of the mid price. Moreover, we remove orders whose size is smaller than 50 USD.

\subsubsection{Training}

The probabilistic neural network maps an input vector $x$ to an output vector of parameters $\Theta(x)$ of a chosen probability distribution $\mathcal{D}$, adapting the classical neural network architectures to the quantification of uncertainty. Its training can be tackled as a statistical estimation problem involving maximum likelihood estimation (MLE). Mathematically, assume that the conditional probability distribution function of $\Delta p$ evaluated at $y\in\mathbb{R}$ is $f(y;\Theta(x))$ for a given input vector $x\in\mathbb{R}^d$. For a given sample of $N$ observations $(\boldsymbol{x},\boldsymbol{y})$, where $\boldsymbol{x}:=(x_n)_{1\leq n\leq N}$ are the input vectors, and $\boldsymbol{y}:=(y_n)_{1\leq n\leq N}$ are the targets, the loss of the model, denoted by $\ell(\boldsymbol{x}, \boldsymbol{y})$, is the negative log-likelihood (NLL)
\begin{equation}
    \ell(\boldsymbol{x}, \boldsymbol{y})=-\sum_{n=1}^N\log{f\left(y_n;\Theta(x_n)\right)}.
\end{equation}

It is noteworthy that the framework still holds for the distribution of a vector of price moves $\boldsymbol{\Delta p}$. In such case, the predicted probability distribution becomes multivariate but the estimation procedure remains the same, see \textit{e.g.}  \cite{williams1996using} for the application of such models to the bivariate case.

\paragraph{The Gaussian case:} Let $\mathcal{D}$ be the Gaussian distribution and $\Theta(x)=\left(\mu(x),\sigma(x)\right)'$, where $\mu(x)$ is the location parameter and $\sigma(x)>0$ is the scale parameter. In this case, the loss of the model is
\begin{equation}
    \ell(\boldsymbol{x}, \boldsymbol{y})=\sum_{n=1}^N\left\{\log{\left(\sqrt{2\pi}\sigma(x_n)\right)}+\frac{1}{2}\left(\frac{y_n-\mu(x_n)}{\sigma(x_n)}\right)^2\right\}.
\end{equation}

\paragraph{Input normalization} We first extract the vectors of features $\boldsymbol{x}$ and the associated price moves $\boldsymbol{y}:=(y_n)_{1\leq n\leq N}$. Each observation $x_n$ is a vector of $d$ features that we denote by $x_n=(x_n^i)_{1\leq i\leq d}$, where $x_n^i$ is the $i$th feature of the $n$th observation. We proceed to a normalization of the data. Some features exhibit fat-tail distributions, which can lead to unstable weight updates during the training process. To counter this issue, we apply a Box-Cox transformation to the features, see \citet{box1964analysis}. For a review on the application of the Box-Cox transformation to real-world problems, we refer the reader to \citet{hossain2011use}. For any feature $i$ the associated Box-Cox transformation with normalization strength parameter $\lambda^i>0$ is denoted by $T_{\lambda^i}(x^i)$ and is defined as
\begin{equation}
    T_{\lambda^i}(x^i):=\frac{(x^i)^{\lambda^i}-1}{\lambda^i}.
\end{equation}

For each feature $i$, the hyperparameter $\lambda^i$ is chosen such that the Gaussian log-likelihood of the transformed feature $T_{\lambda^i}(x^i)$ is maximized. Note that the lambdas are determined in the training set and applied to transform the features of the test set. Once the optimal hyperparameters $\{\lambda_i\}$ are obtained, we compute the average and the standard deviation of each transformed feature over the training set, and use it for standardization by applying a $z$-score transformation. Denote by $\overline{T}_{\lambda^i}$ the empirical average of the Box-Cox transformation of the feature, and $\overline{S}_{\lambda^i}$ an estimator of its variance, then its $z$-score is defined as
\begin{equation}
    z_\lambda(x^i):=\frac{T_{\lambda^i}(x^i)-\overline{T}_{\lambda^i}}{\sqrt{\overline{S}_{\lambda^i}}}.
\end{equation}

We apply this normalization-standardization procedure to the input variables as a pre-processing layer of the neural network model.

\paragraph{Experiment setting} We conduct our numerical experiment with the following configuration:
\begin{itemize}
    \item \textbf{time horizon:} $T=1$ s;
    \item \textbf{volume kernel:} $f(v)=v$, for all $v> 0$;
    \item \textbf{(inverse) time scales:} $\boldsymbol{\beta}=\{10, 100, 1000\}$, expressed in units of $\text{s}^{-1}$;
    \item \textbf{(inverse) distance scales:} $\boldsymbol{\eta}=\{0.001, 0.1, 1, 10\}$, expressed in units of $\text{bp}^{-1}$ (basis points from the mid price).
\end{itemize}

We also use the bid-ask spread process as a feature, that we denote by $(\Psi_t)_{t\geq 0}$. Combining 3 time scales, 4 distance scales, the 2 sides of the book, we end up with 24 limit order flow features and 6 market order flow features, totalling 31 variables including the bid-ask spread. Each observation in the data set corresponds to the insertion of a limit order and the features are updated accordingly. At the $n$th data point, the set of features is
\begin{equation}
    x_n=\{\Psi_n\}\cup\mathcal{L}_n\cup\mathcal{M}_n,
\end{equation}
where $\mathcal{L}_n$ and $\mathcal{M}_n$ are the families of order flow variables that are defined in Equations \eqref{eq:limit_order_flow_family} and \eqref{eq:market_order_flow_family}. Note that in the multi-asset case, this set of features is computed for each asset. In the cross-asset experiment described below, we thus have 62 variables.

We  train the two following models separately:
\begin{itemize}
    \item \textbf{Univariate distribution:} We first focus on the univariate case where the goal is to predict the distribution of price moves of a single asset. We train the model to predict the parameters $(\mu, \sigma, \alpha)$ of a skewed Gaussian with probability density function $f$ defined, for all $x\in\mathbb{R}$, as
    \begin{equation}
        f(x)=\frac{2}{\sigma}\phi\left(\frac{x-\mu}{\sigma}\right)\Phi\left(\alpha\frac{x-\mu}{\sigma}\right),
    \end{equation}
    where $\phi(x):=\frac{1}{\sqrt{2\pi}}e^{-\frac{x^2}{2}}$ is the standard Gaussian density function and $\Phi(x):=\frac{1}{\sqrt{2\pi}}\int_{-\infty}^xe^{-\frac{t^2}{2}}\mathrm{d}t$ its cumulative distribution function. The parameter $\alpha$ adjusts the skew of the distribution.
    
    \item \textbf{Bivariate distribution:} We then train a bivariate model, in which the target is a vector of price moves of two assets. We use the features of both assets, thus totalling 62 variables. This model can be the foundation of a cross-asset spoofing detection tool as we will discuss later. We use a bivariate Gaussian distribution for this purpose.
\end{itemize}

In order to adapt the model to a high-frequency environment, we set a simple architecture for the neural network. The model is a feed-forward neural network with 1 hidden layer and 64 neurons, with a ReLU activation function for the hidden layer, and we apply an adapted transformation layer that maps the output of the neural network to the probability distribution parameters domain, \textit{i.e.} ensuring that $\mu\in\mathbb{R}$, $\sigma>0$ and $\alpha\in\mathbb{R}$ for the skewed Gaussian. We train the model with a batch size of $2^{12}$, Adam as gradient descent algorithm  \citep{kingma2014adam} with a learning rate of $10^{-3}$, over 1,000 epochs. The training is stopped if the validation loss has not improved within the last 100 epochs.

The training set is built using limit orders that were posted within three days, from 12/01/2022 to 12/03/2022. We randomly pick a group of 2,000,000 orders in this time frame, and split this group into two sets of 1,000,000 orders each. The first set is used for the training of the model, the second one for the validation. The orders are sorted with respect to time, such that the training set is composed of orders that were posted before the insertion of the first order of the validation set.

\subsection{Microstructure properties of the price formation model}

In this section, we study the dependence structure of the predicted moments of the distribution of $\Delta p$ with respect to some interpretable variables such as the bid-ask spread $\Psi$. We denote by $\mathcal{S}$ the standardized average of the distribution of $\Delta p$, defined as
\begin{equation}
    \mathcal{S}(\Delta p):=\frac{\mathbb{E}\left(\Delta p\right)}{\sqrt{\mathbb{V}\left(\Delta p\right)}}.
\end{equation}

\subsubsection{Partial dependence analysis}

\paragraph{Recovering the spread-volatility coupling of small tick assets} In the microstructure literature, the relationship between the volatility of price returns and the bid-ask spread has been extensively studied. In the market making game introduced by \citet{madhavan1997security}, the volatility is shown to be a linear function of the bid-ask spread. The validity of this result was also demonstrated on empirical data, see, \textit{e.g.}, \citet{wyart2008relation}. An adaptation of the result for large tick assets is proposed in \cite{robert2011new} using a measure of the implicit spread.

\begin{figure}[!ht]
    \centering
    \subfloat[BTC-USD]{%
        \includegraphics[width=0.4\linewidth]{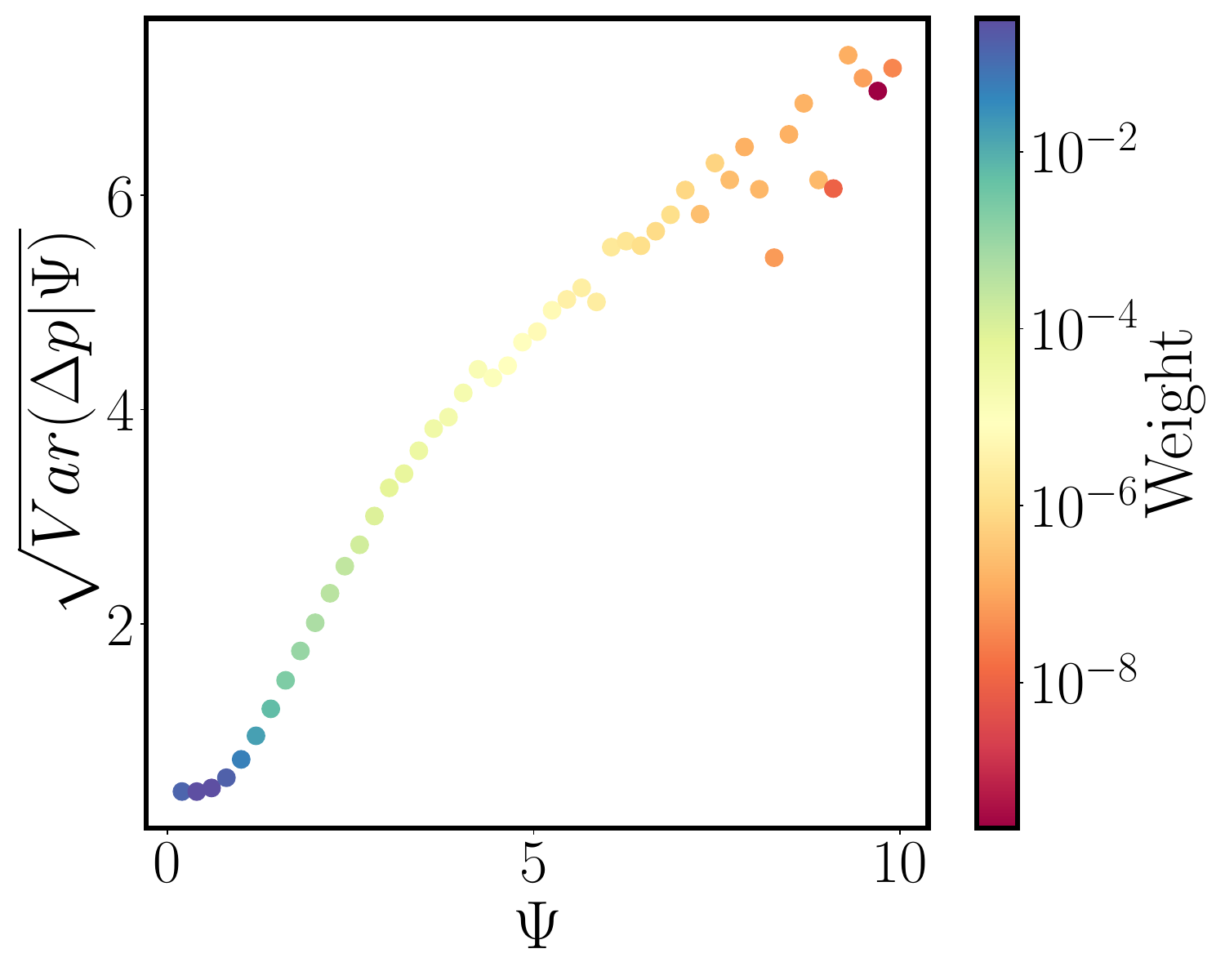}%
    }~~~~
    \subfloat[ETH-USD]{%
        \includegraphics[width=0.4\linewidth]{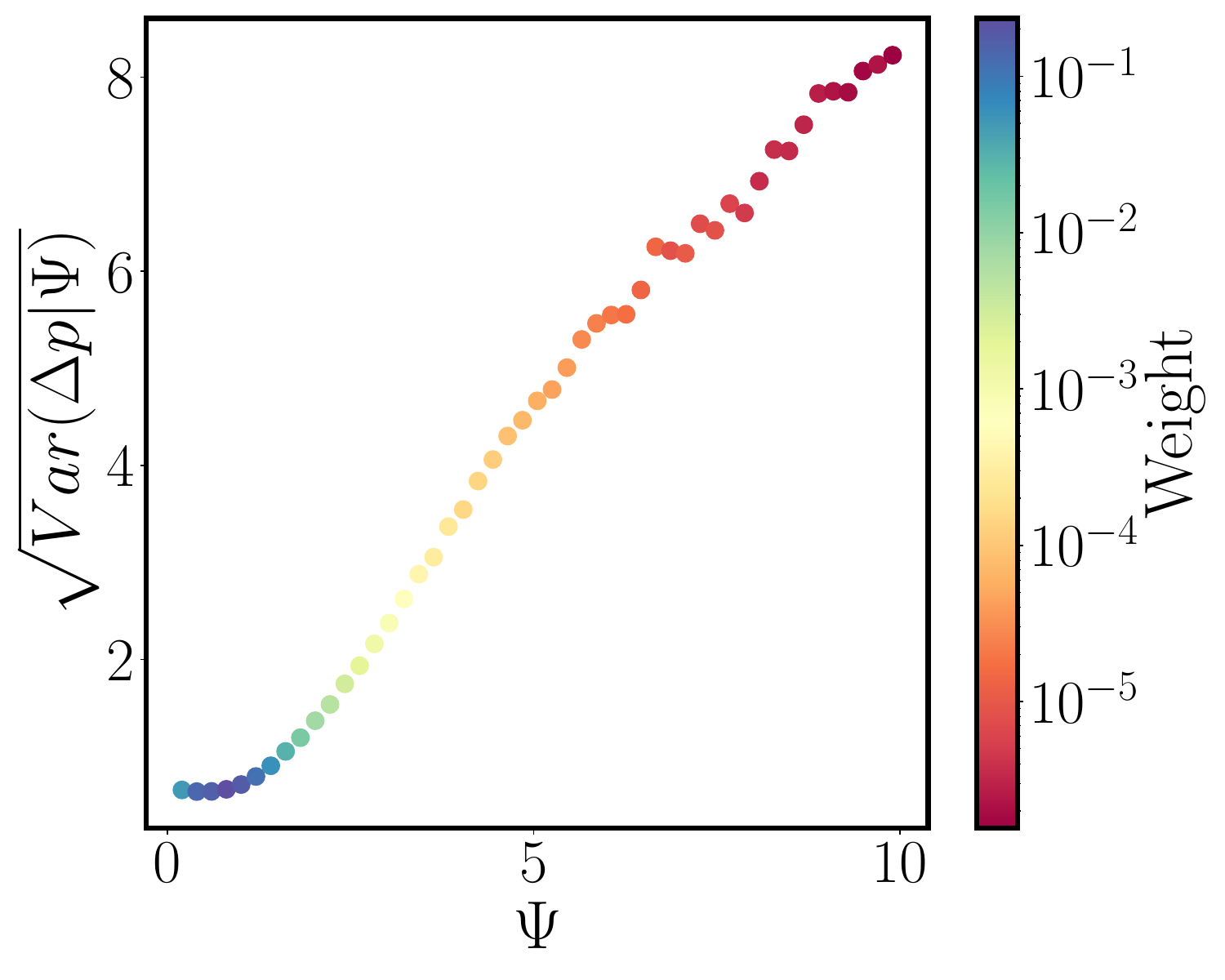}%
    }
    \caption{\textit{Partial dependence} --- Predicted standard deviation (bps) of the price move $\Delta p$ as a function of the bid-ask spread $\Psi$, expressed in basis points. Predictions are averaged per spread bin over the test set. Each point's weight reflects the proportion of time the corresponding bid-ask spread is observed.}
    \label{fig:pdplot_spread}
\end{figure}

In order to support the effectiveness of the model in reproducing the positive relationship between volatility and the bid-ask spread, we compute the partial dependence plot of the volatility with respect to the bid-ask spread. Here, the volatility is the predicted standard deviation of $\Delta p$, expressed in basis point of the mid price per second. We compute on the test set the average predicted volatility over bid-ask spread bins spanning from 0.1 bp to 10 bps. The results are displayed in Figure \ref{fig:pdplot_spread}. We observe a clean relationship, and, as expected, that the predicted standard deviation is monotonously increasing with the bid-ask spread. Although there is a boundary effect with a flat standard deviation for very small spreads, \textit{i.e.} $\Psi<1$ bp, the function is most of the time close to linear, with a slight concavity for large bid-ask spreads (larger than 4bps). Hence, the model is able to reproduce the spread-volatility relationship that has been previously observed in empirical studies.

\begin{figure}[!ht]
    \centering
    \subfloat[Average (bps)]{%
        \includegraphics[width=0.33\linewidth]{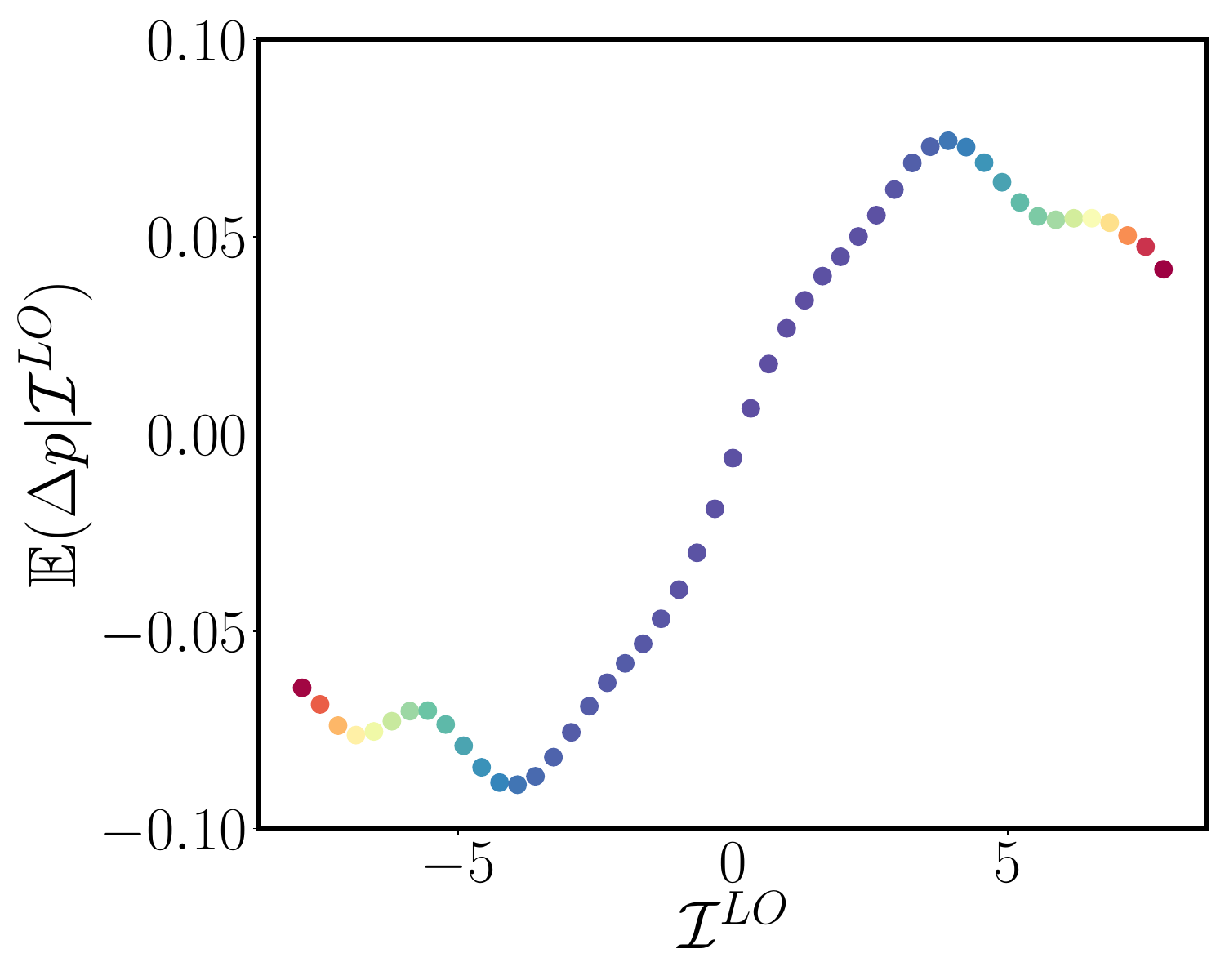}%
    }~~~~
    \subfloat[Standard deviation (bps)]{%
        \includegraphics[width=0.33\linewidth]{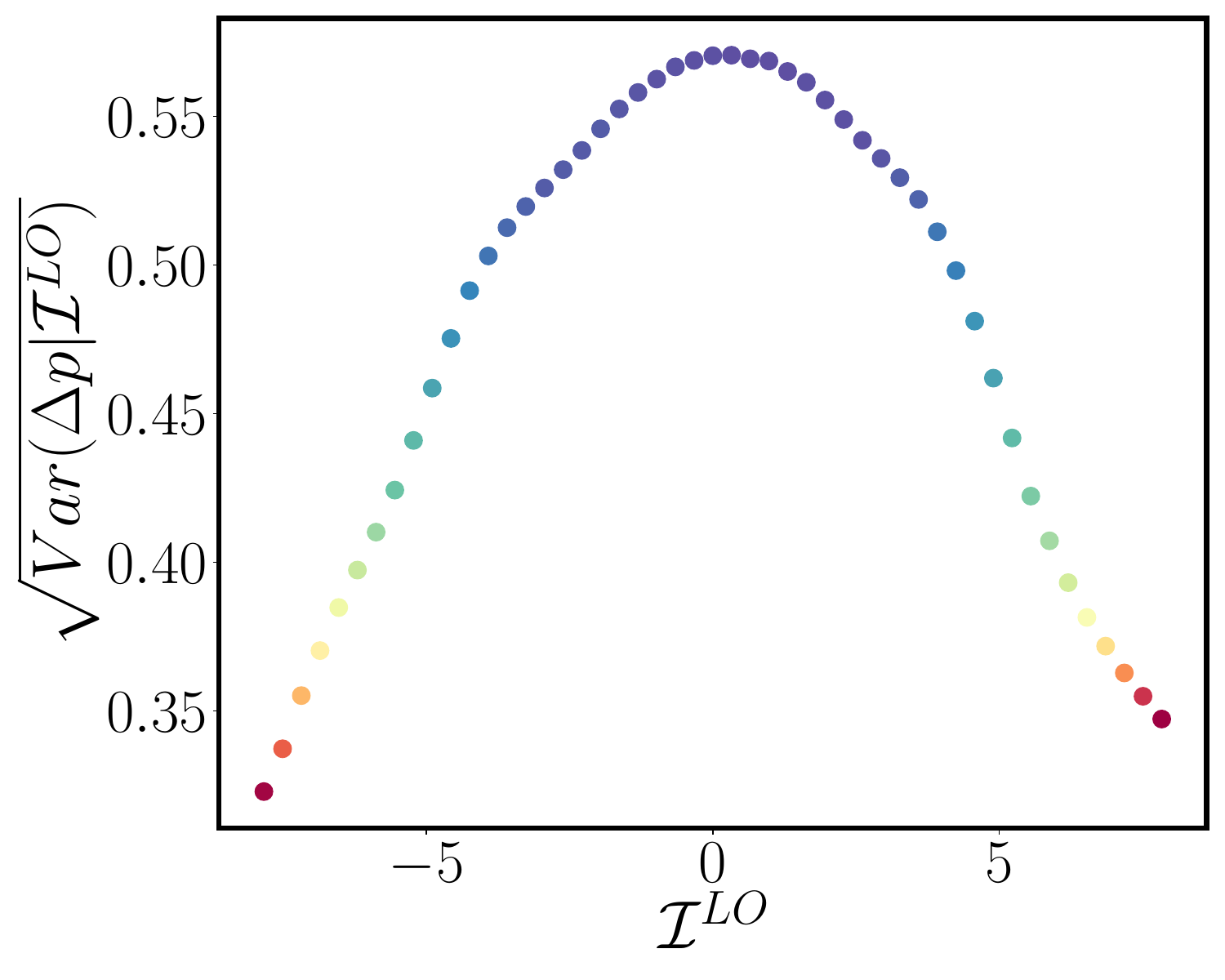}%
    }~~~~
    \subfloat[Sharpe ratio]{%
        \includegraphics[width=0.33\linewidth]{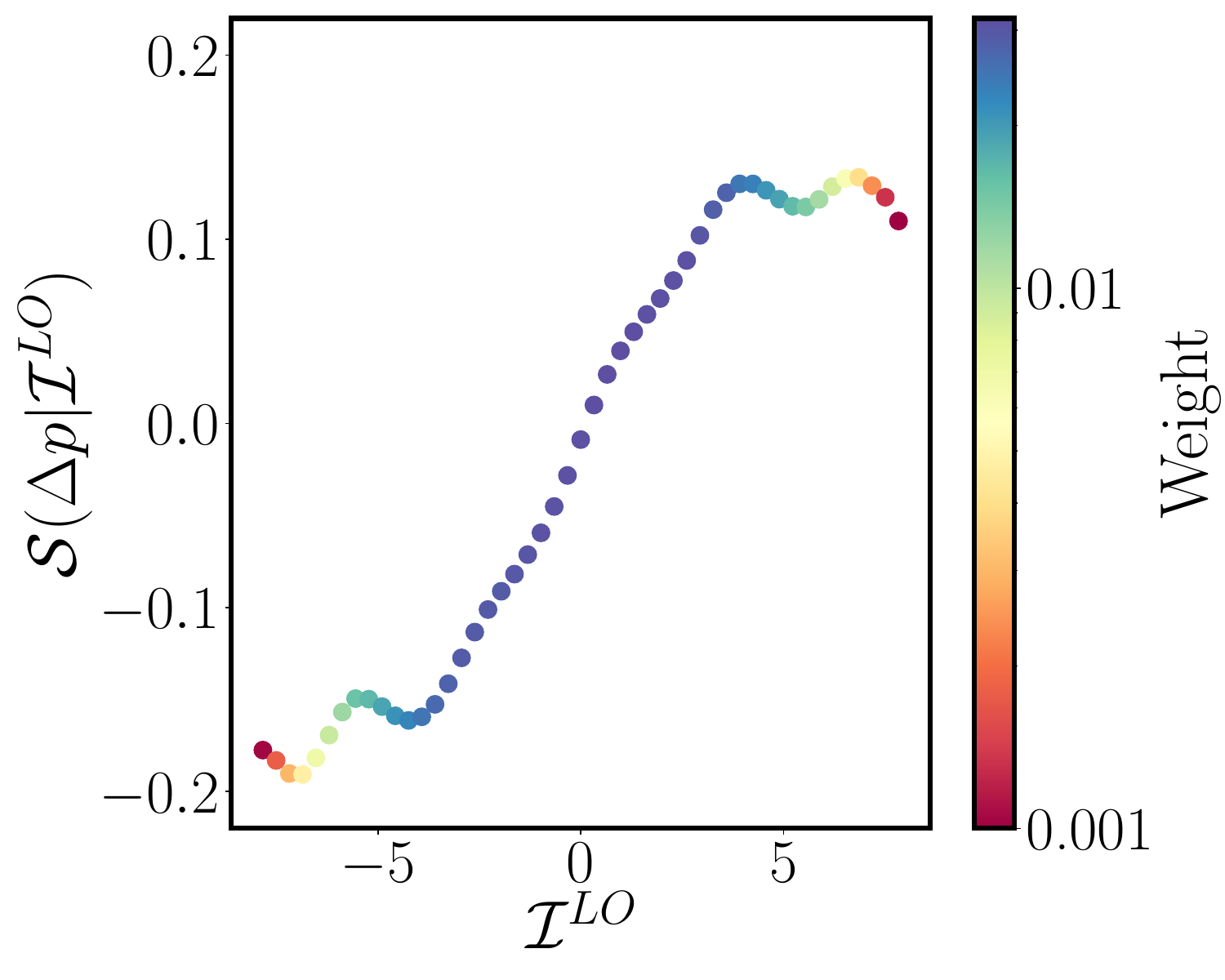}%
    }
    \caption{\textit{Partial dependence} --- Predicted moments of the price move $\Delta p$ as functions of the limit order flow imbalance $\mathcal{I}^{\text{LO}}$, BTC-USD. Predictions are averaged per imbalance bin over the test set. Each point's weight reflects the proportion of time the corresponding imbalance is observed.}
    \label{fig:btcusd_pdplot_limit_imbalance}
\end{figure}

\begin{figure}[!ht]
    \centering
    \subfloat[Average (bps)]{%
        \includegraphics[width=0.33\linewidth]{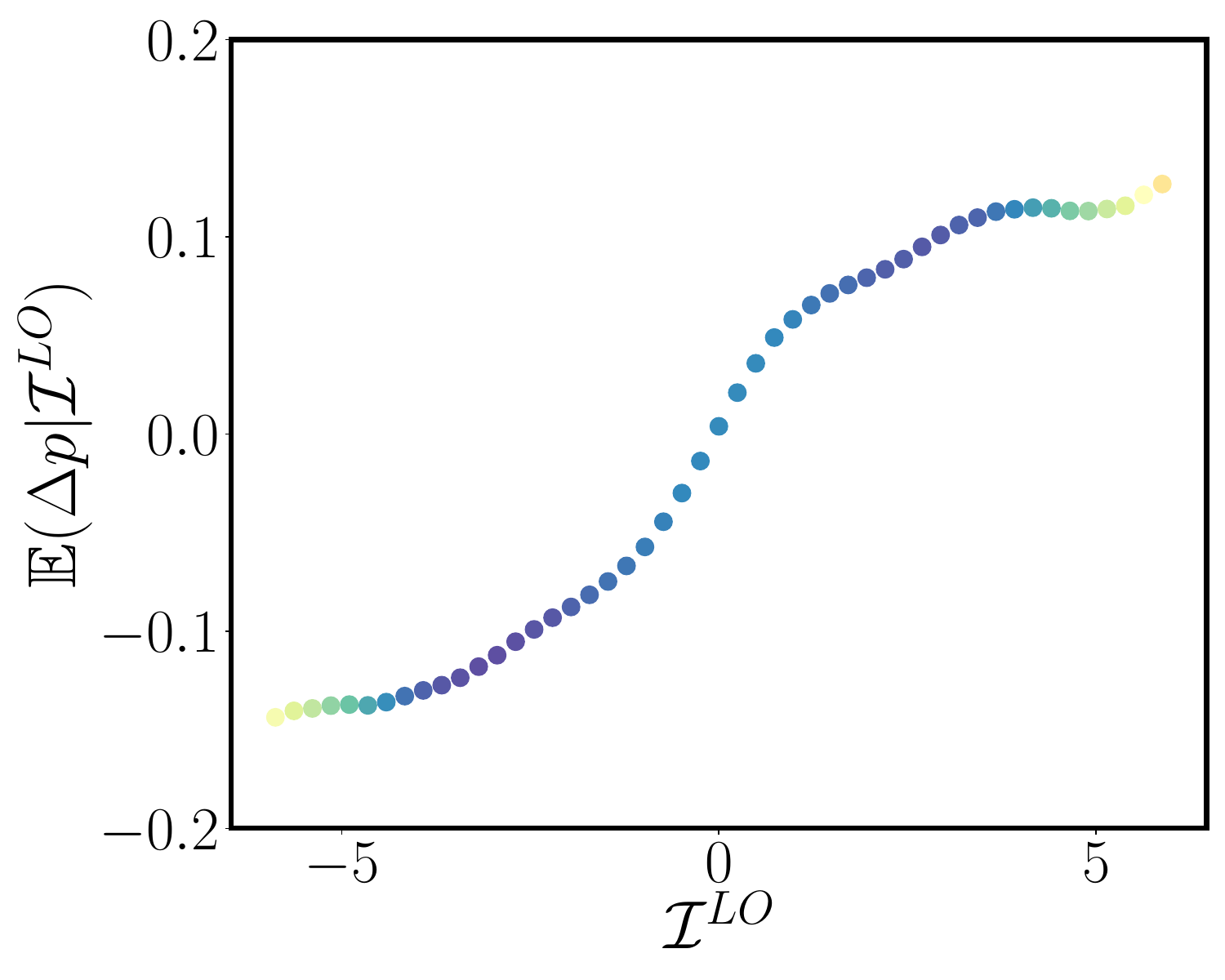}%
    }~~~~
    \subfloat[Standard deviation (bps)]{%
        \includegraphics[width=0.33\linewidth]{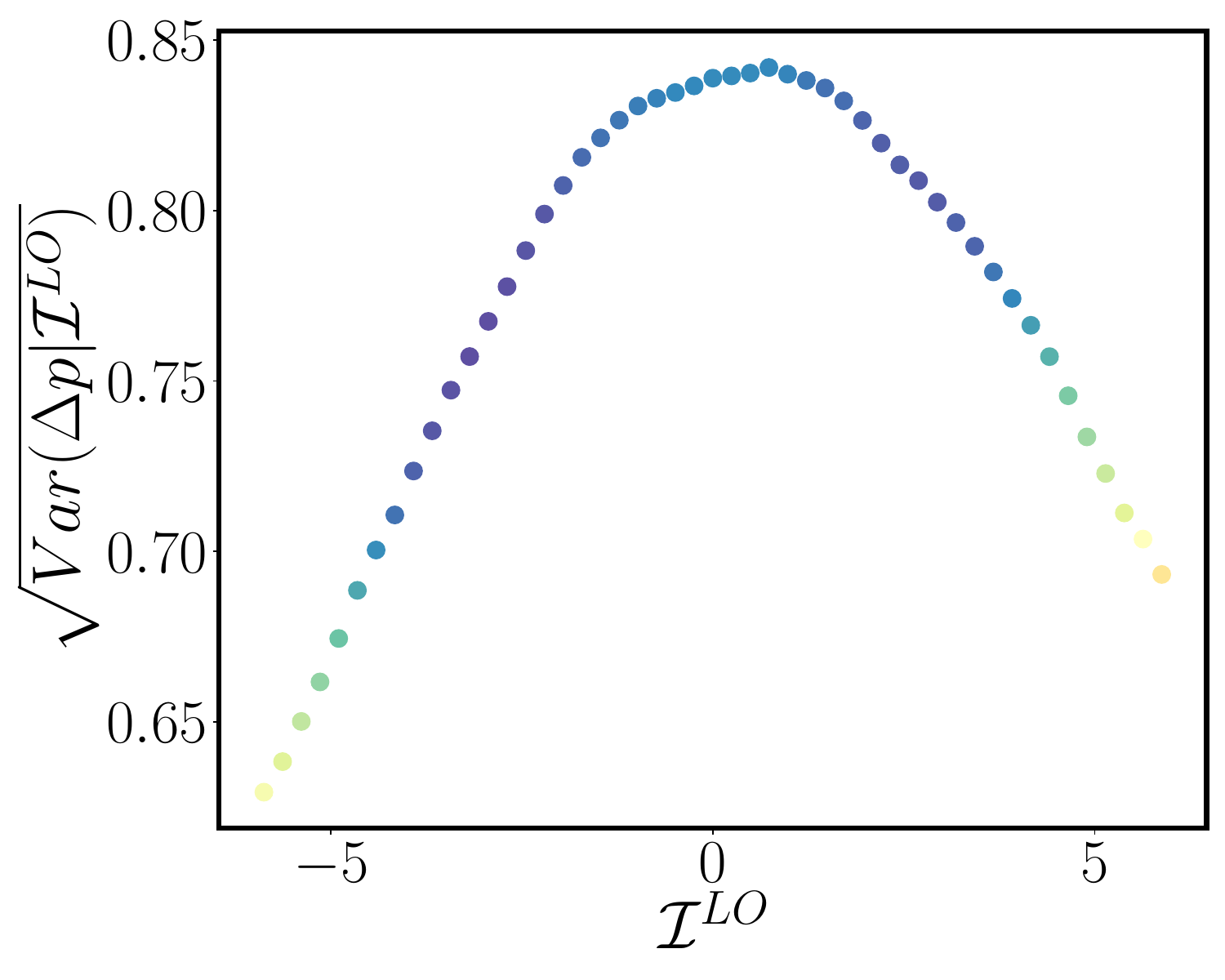}%
    }~~~~
    \subfloat[Sharpe ratio]{%
        \includegraphics[width=0.33\linewidth]{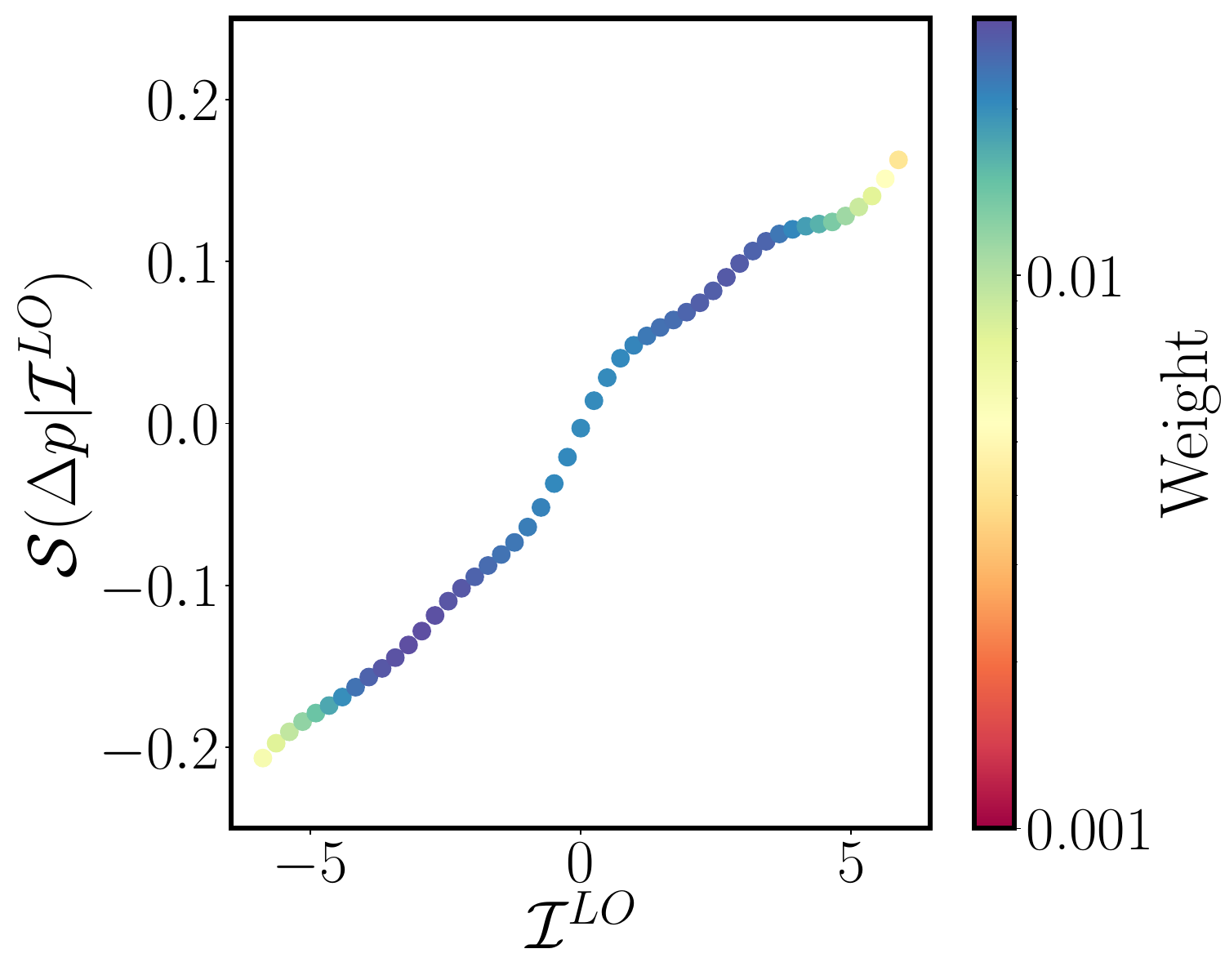}%
    }
    \caption{\textit{Partial dependence} --- Predicted moments of the price move $\Delta p$ as functions of the limit order flow imbalance $\mathcal{I}^{\text{LO}}$, ETH-USD. Predictions are averaged per imbalance bin over the test set. Each point's weight reflects the proportion of time the corresponding imbalance is observed.}
    \label{fig:ethusd_pdplot_limit_imbalance}
\end{figure}

\begin{figure}[!ht]
    \centering
    \subfloat[Average (bps)]{%
        \includegraphics[width=0.33\linewidth]{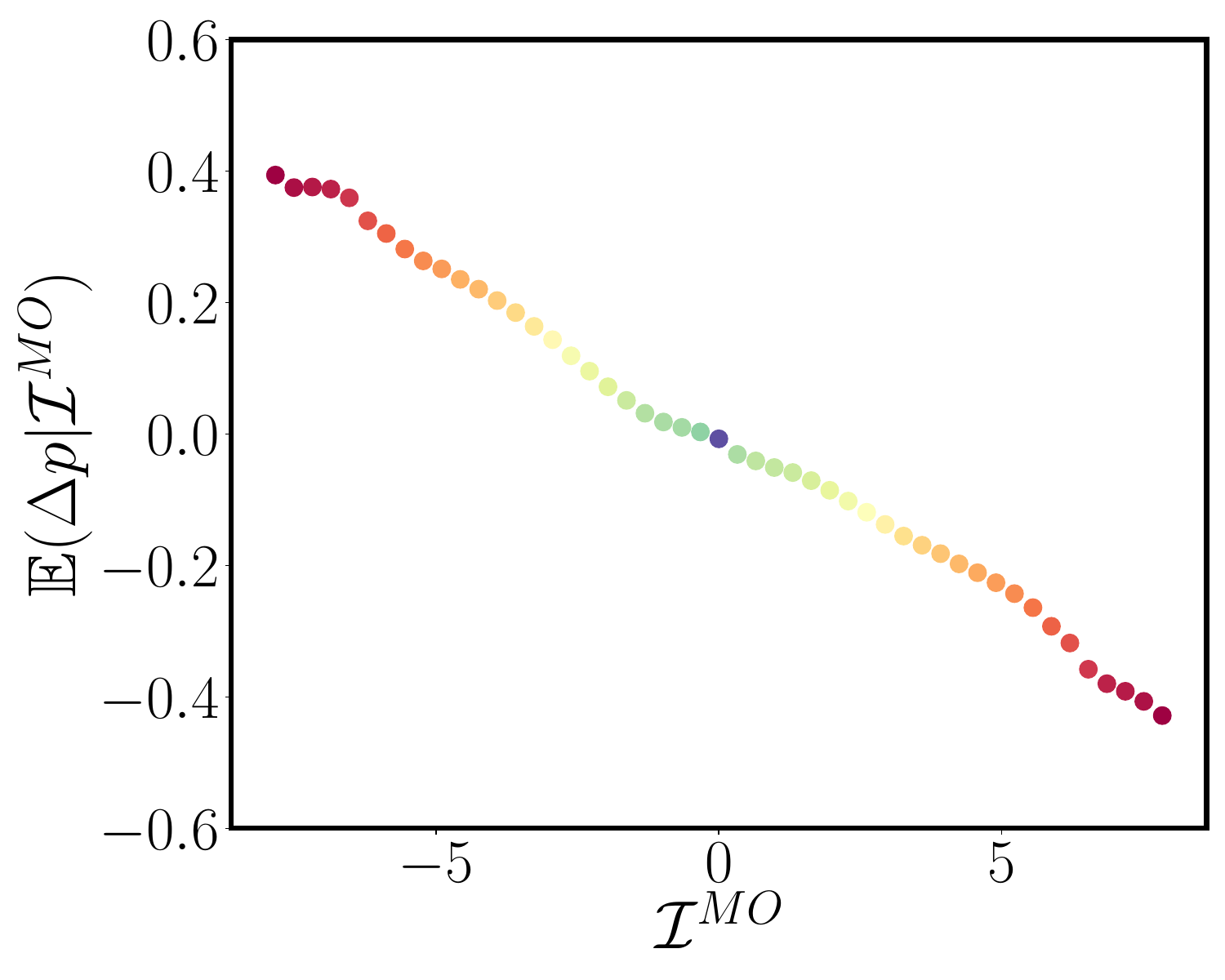}%
    }~~~~
    \subfloat[Standard deviation (bps)]{%
        \includegraphics[width=0.33\linewidth]{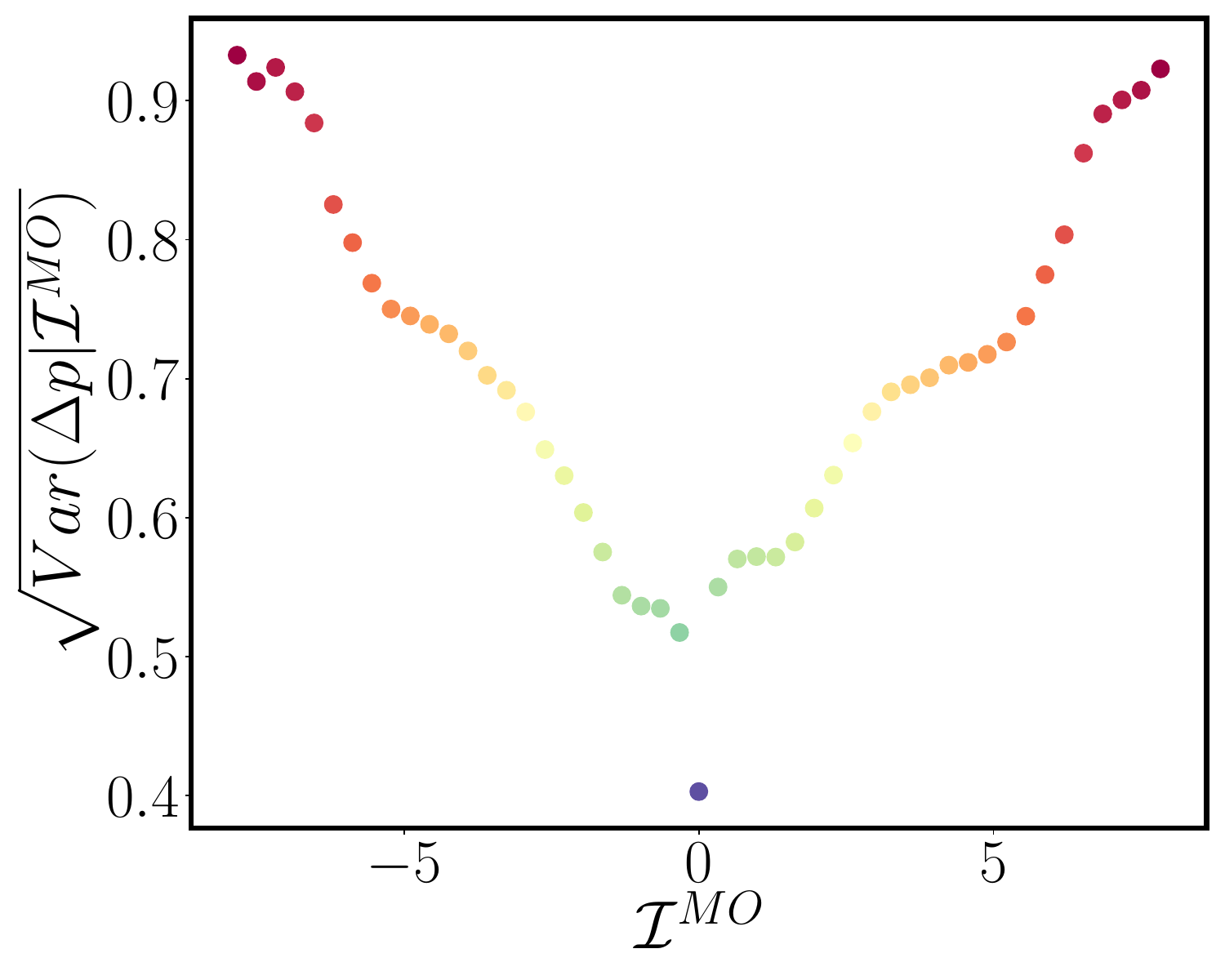}%
    }~~~~
    \subfloat[Sharpe ratio]{%
        \includegraphics[width=0.33\linewidth]{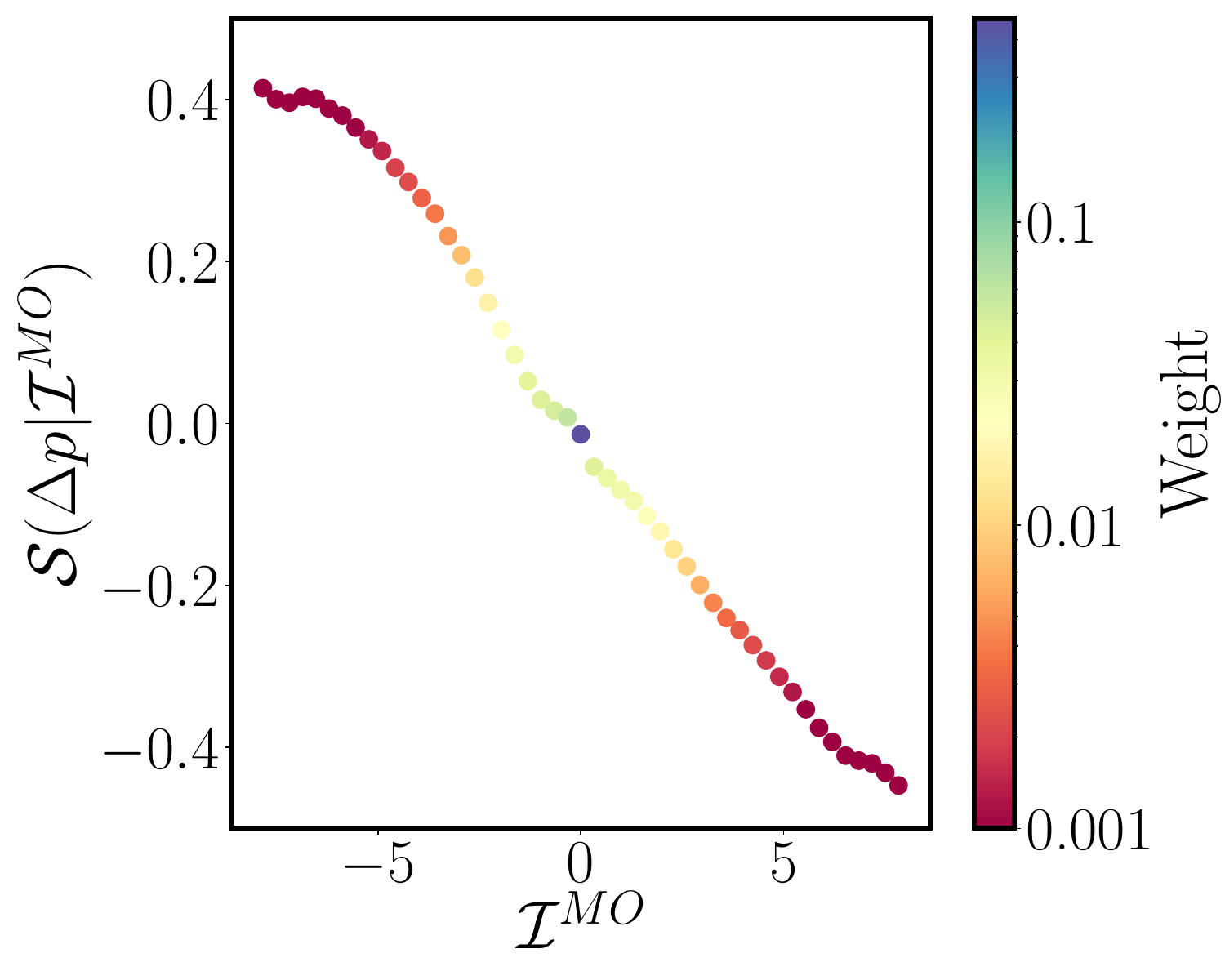}%
    }
    \caption{\textit{Partial dependence} --- Predicted moments of the price move $\Delta p$ as functions of the market order flow imbalance $\mathcal{I}^{\text{MO}}$, BTC-USD. Predictions are averaged per imbalance bin over the test set. Each point's weight reflects the proportion of time the corresponding imbalance is observed.}
    \label{fig:btcusd_pdplot_market_imbalance}
\end{figure}

\begin{figure}[!ht]
    \centering
    \subfloat[Average (bps)]{%
        \includegraphics[width=0.33\linewidth]{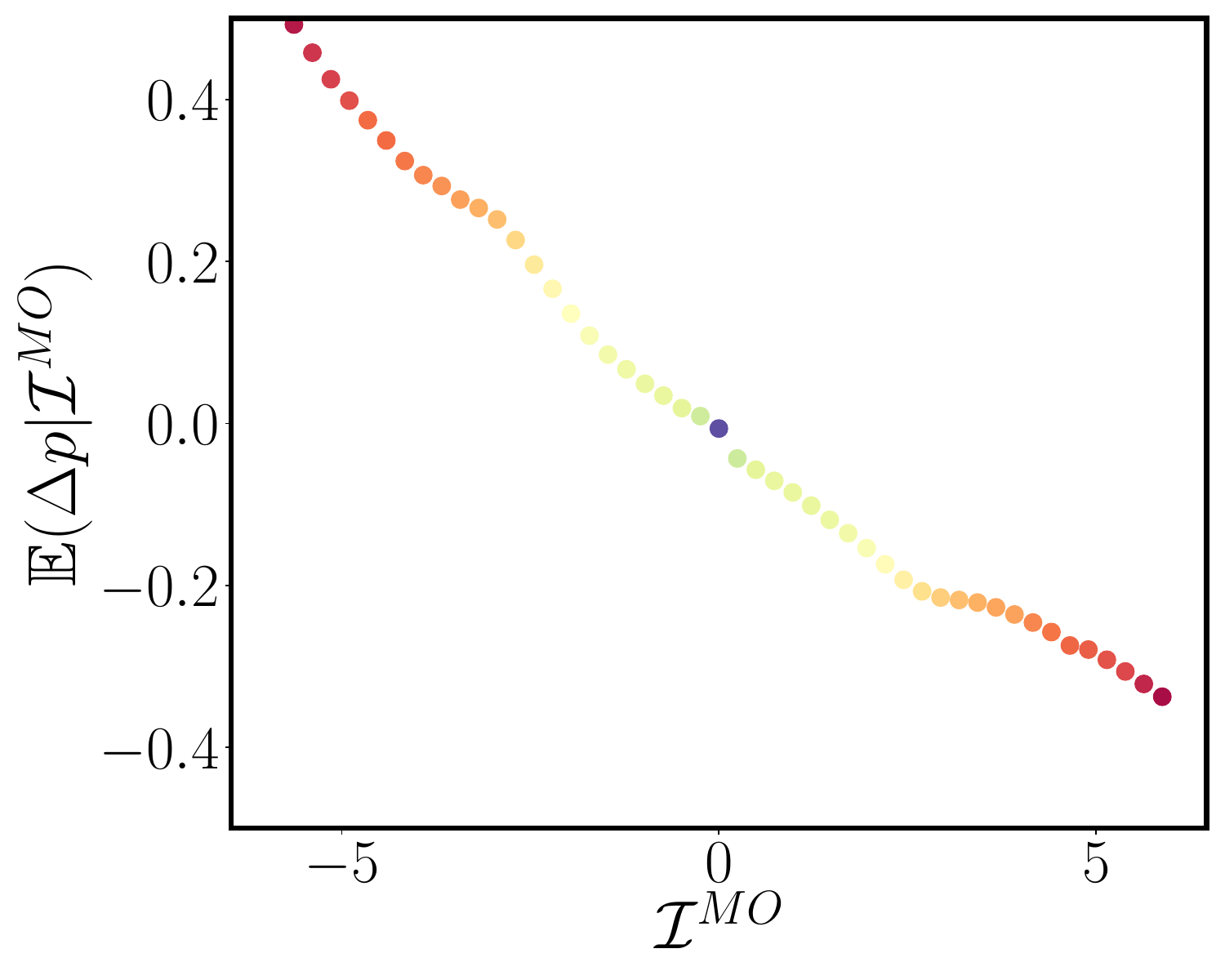}%
    }~~~~
    \subfloat[Standard deviation (bps)]{%
        \includegraphics[width=0.33\linewidth]{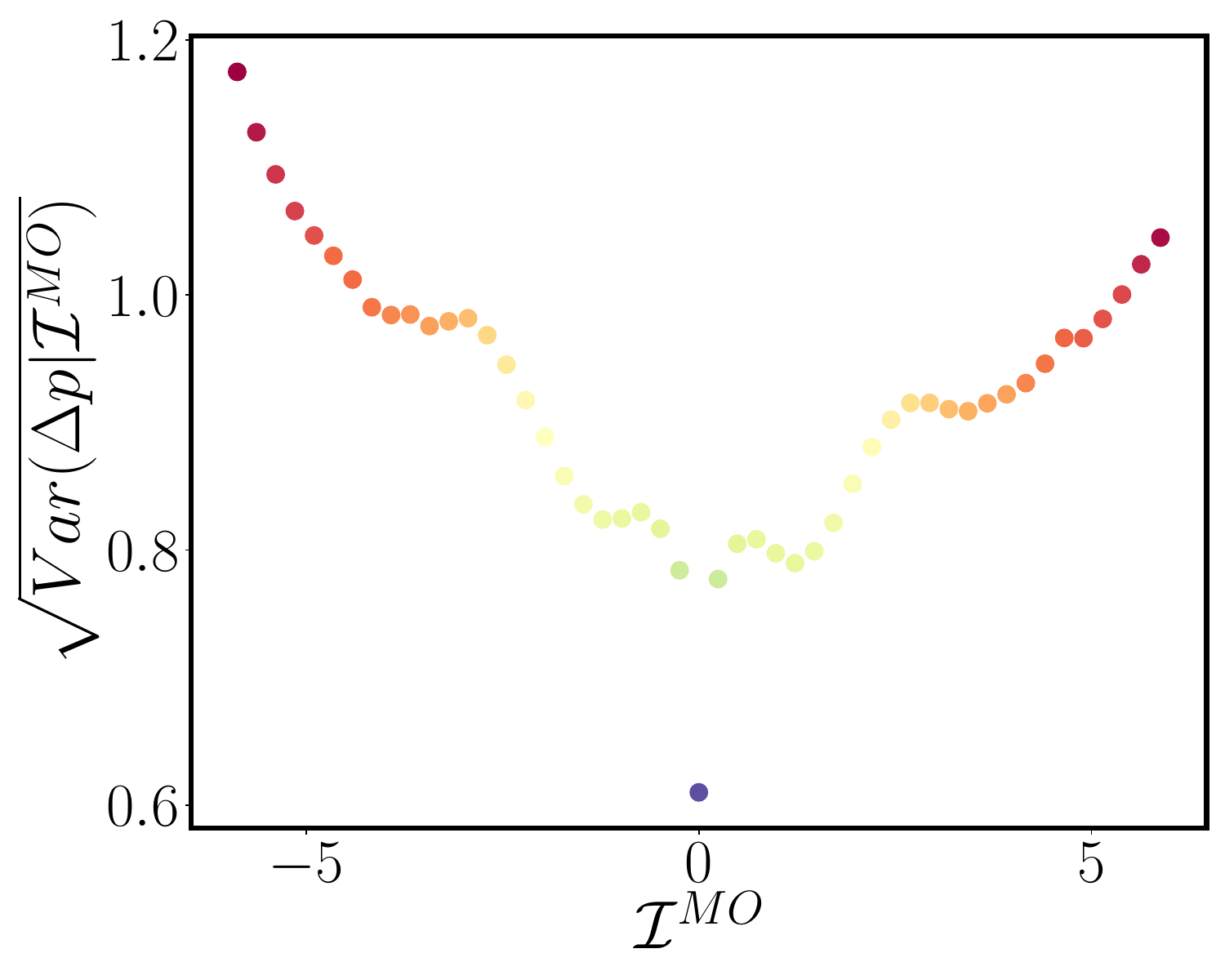}%
    }~~~~
    \subfloat[Sharpe ratio]{%
        \includegraphics[width=0.33\linewidth]{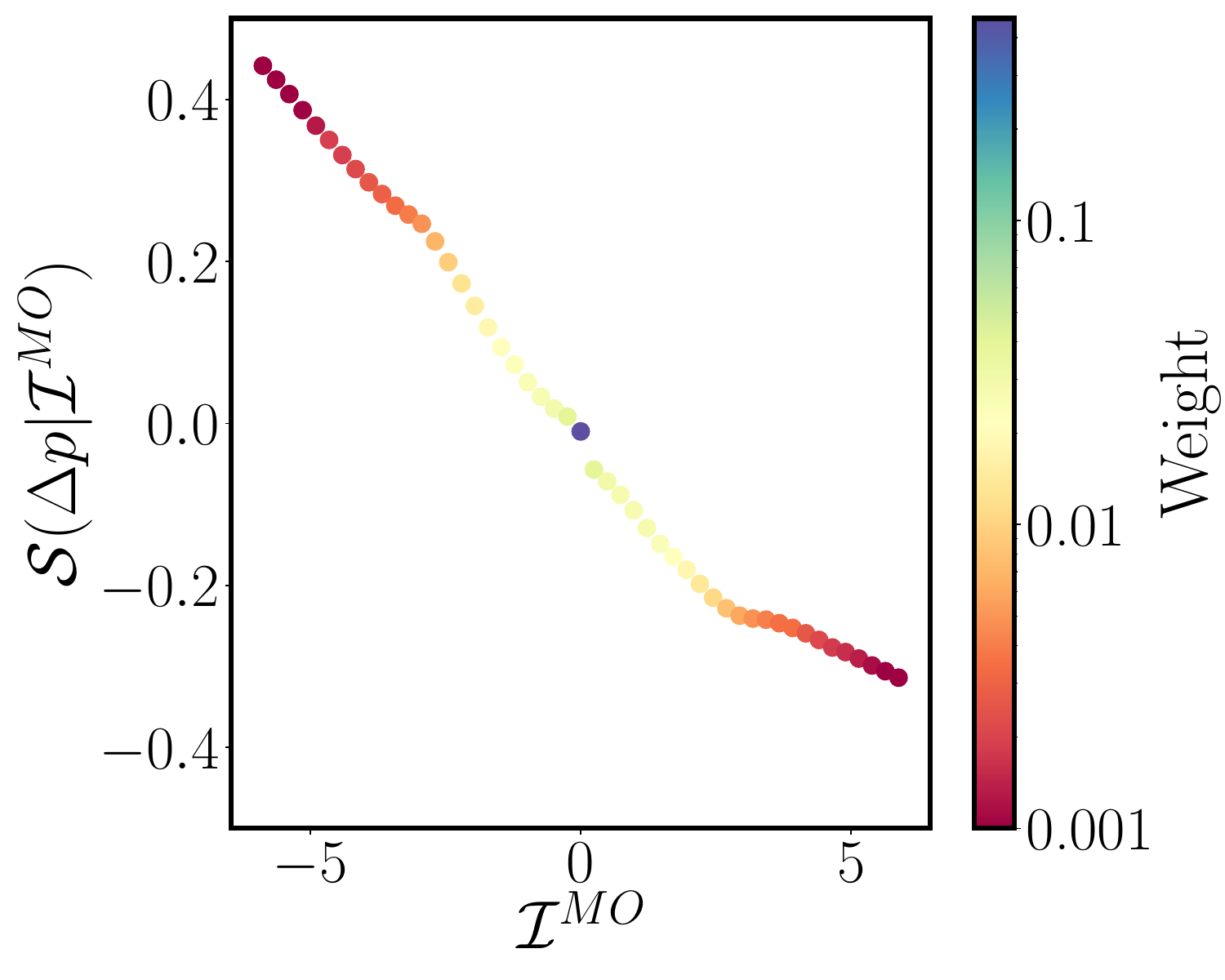}%
    }
    \caption{\textit{Partial dependence} --- Predicted moments of the price move $\Delta p$ as functions of the market order flow imbalance $\mathcal{I}^{\text{MO}}$, ETH-USD. Predictions are averaged per imbalance bin over the test set. Each point's weight reflects the proportion of time the corresponding imbalance is observed.}
    \label{fig:ethusd_pdplot_market_imbalance}
\end{figure}

\paragraph{An order flow imbalance variable} It is common practice to transform the flow of orders into interpretable and stationary variables, which has shown to improve explainability and performance of Machine Learning models in prediction tasks (see \citet{kolm2023deep, cont2023cross}). The order flow imbalance is a popular variable and indicates buy or sell pressure based on the flow of orders. It is usually computed using the observed volume variation at the price limits of the LOB, up to a specific depth. Using the order flow variables we defined in Equations \eqref{eq:limit_order_flow_family} and \eqref{eq:market_order_flow_family}, we are able to construct new order flow imbalance features that we define as follows 

\begin{equation}
    \mathcal{I}_t^{\text{LO}}:=\sum_{k=1}^K\omega_k^{\text{LO}}\log\left(\frac{1+L_t^b(\phi_k)}{1+L_t^a(\phi_k)}\right),
\end{equation}

\begin{equation}
    \mathcal{I}_t^{\text{MO}}:=\sum_{k=1}^K\omega_k^{\text{MO}}\log\left(\frac{1+M_t^b(\psi_k)}{1+M_t^a(\psi_k)}\right),
\end{equation}

where $(\omega_k^{\text{LO}})_{1\leq k\leq K}$, $(\omega_k^{\text{MO}})_{1\leq k\leq K}$ are two vectors of weights that are applied to each time scale and distance scale of the order flow features. Even if methodologies could be developed to tune the weights and maximize the predictive power of these measures, we focus on a uniform weighting for the sake of illustration and show the neural network is able to capture a structure of dependence between the parameters of the distribution and these features. Hence, in the following partial dependence experiment, we set $\omega_k^{\text{LO}}=\frac{1}{K}$ and $\omega_k^{\text{MO}}=\frac{1}{K}$, for all $1\leq k\leq K$. The results are displayed in Figures \ref{fig:btcusd_pdplot_limit_imbalance}, \ref{fig:ethusd_pdplot_limit_imbalance}, \ref{fig:btcusd_pdplot_market_imbalance}, \ref{fig:ethusd_pdplot_market_imbalance}. We notice a consistent sensitivity of the predicted moments of these variables. Although there are some boundaries effects for rare extreme imbalances, the mean of the distribution and its standardized version exhibit monotonic behaviour. Specifically, we observe that an increase in limit order flow imbalance corresponds to a larger expected price movement, while a higher market order flow imbalance leads to a smaller expected price movement. This highlights the model's ability to capture the impact of liquidity pressure: upward price movements are driven by pressure on the buy side (and conversely for the ask side), while downward price movements are likely to occur following trade executions on the bid side (and conversely for the ask side).

\subsubsection{Price response}

Being able to predict the impact of the insertion of a limit order on the predicted price move distribution is the key to computing the spoofability of the the market. Let us investigate the sensitivity of neural network to the modification an order size and distance.

Mathematically speaking, let $x_t$ be the input vector sampled at event time $t$ from real data. It contains both predictors independent from the new order and features directly caused by the order itself.  Equations \eqref{eq:limit_order_flow_bid} makes it easy to modify the inputs to account for different size and placement position of an order.
%We first neutralized its influence by   Let denote the neutralized input by $x_t^0$. By neutralized input, we mean that if a buy order is inserted with size $q_t$ and distance $\delta_t$, then for each of the columns of the vector $x_t$ that correspond to the limit order flow variables $L_t^b$ defined in Equation \eqref{eq:limit_order_flow_bid}, we remove the contribution $f(q_t)e^{-\eta\delta_t}$. Thus, the resulting corrected input $x_t^0$ is what $x_t$ would be if no order was inserted at time $t$.
Our neural network can therefore estimate the effect of the insertion of a hypothetical limit order. We use a grid of sizes $\boldsymbol{Q}$ and distances $\boldsymbol{\delta}$ and compute the associated limit order flow variables $L_t^b$ defined in Equation \eqref{eq:limit_order_flow_bid}. Let us denote by $x_t^+(Q,\delta)$ $x_t^0$  the input associated to a new limit order of size $Q$ placed at distance $\delta$  at time $t$, with the shorthand $x_t^0$ when $\delta$ and $Q=0$; their respective 
%Hence, the contribution to add in each configuration $(Q,\delta)$ is $f(Q)e^{-\eta\delta}$ for a distance scale parameter $\eta$. We finally obtain a corrected input vector $x_t^0$ and a hypothetical bumped input vector $x_t^+(Q,\delta)$, 
normalized-standardized versions are denoted by $z_t^+(Q,\delta)$ and $z_t^0$. The neural network predicts the parameters of the distribution of $\Delta p$ for both $z_t^0$ and $z_t^+(Q,\delta)$, and thus we obtain $\Theta(z_t^0)$ and $\Theta(z_t^+(Q,\delta))$.

Our goal is to study the influence of the characteristics $Q$ and $\delta$ of limit orders over the standardized average $\mathcal{S}$: we select $N$ random times $(t_n)_{1\leq n \leq N}$ and introduce the which quantifies the average impact of a limit order with size $Q$ and distance $\delta$ over the risk-adjusted expected price move
\begin{equation}
    \overline{\Delta\mathcal{S}_N}(Q,\delta):=\frac{1}{N}\sum_{n=1}^N\left(\mathcal{S}(\Theta(z_{t_n}^+(Q,\delta)))-\mathcal{S}(\Theta(z_{t_n}^0))\right)
\end{equation}
for $Q\in\boldsymbol{Q}$, $\delta\in\boldsymbol{\delta}$. We assess the statistical significance of the estimator $\overline{\Delta\mathcal{S}_N}(Q,\delta)$ using the $t$-statistic
\begin{equation}
    T_N(Q,\delta):=\sqrt{N}\frac{\overline{\Delta\mathcal{S}_N}(Q,\delta)}{\sqrt{\overline{V_N}(Q,\delta)}},
\end{equation}
where we define
\begin{equation}
    \overline{V_N}(Q,\delta):=\frac{1}{N-1}\sum_{n=1}^N\left(\mathcal{S}(\Theta(z_{t_n}^+(Q,\delta)))-\mathcal{S}(\Theta(z_{t_n}^0))-\overline{\Delta\mathcal{S}_N}(Q,\delta)\right)^2.
\end{equation}

\begin{figure}[!ht]
    \centering
    \subfloat[Impact, Bid]{%
        \includegraphics[width=0.25\linewidth]{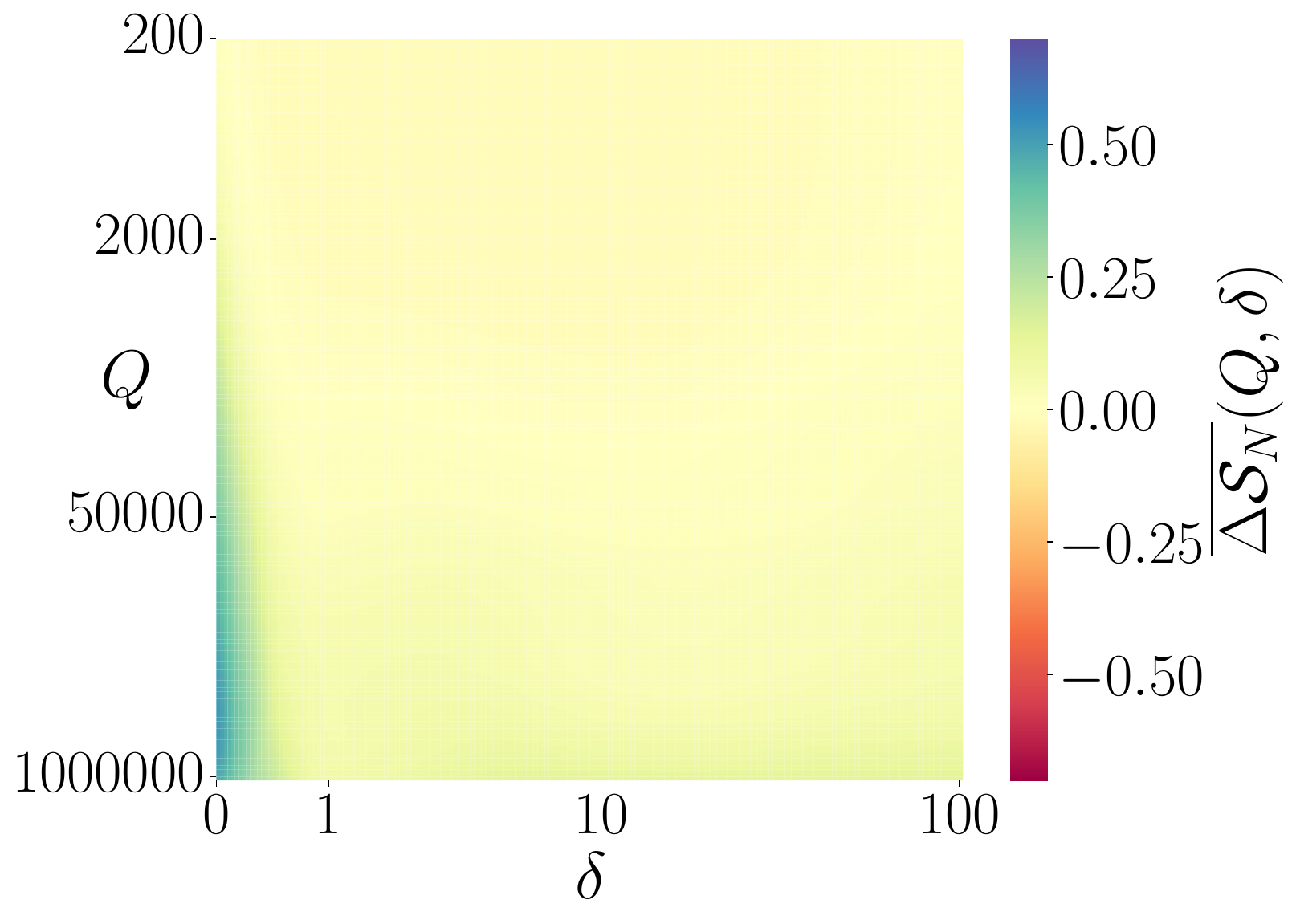}%
    }~~~
    \subfloat[Impact, Ask]{%
        \includegraphics[width=0.25\linewidth]{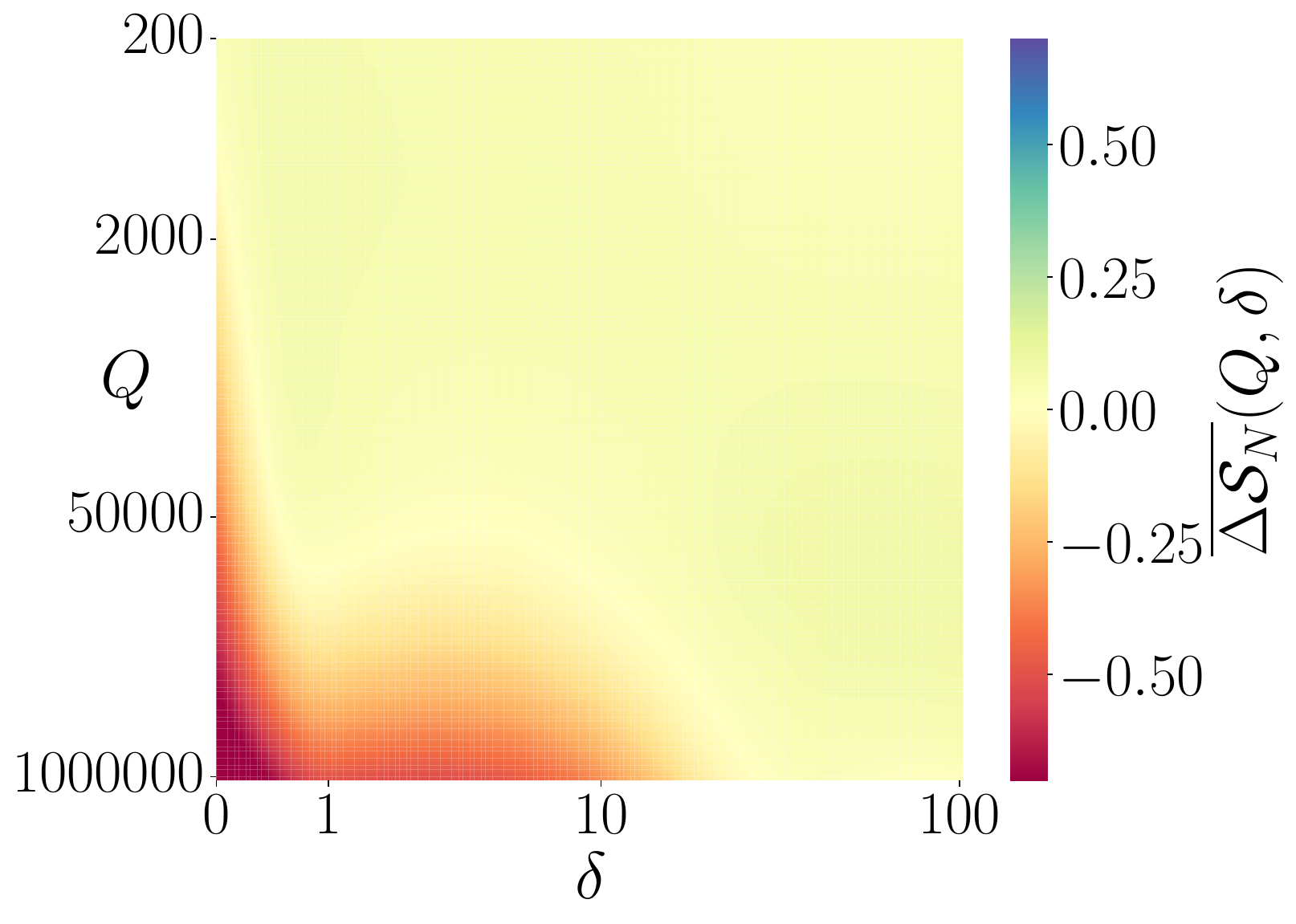}%
    }~~~
    \subfloat[$p$-value, Bid]{%
        \includegraphics[width=0.25\linewidth]{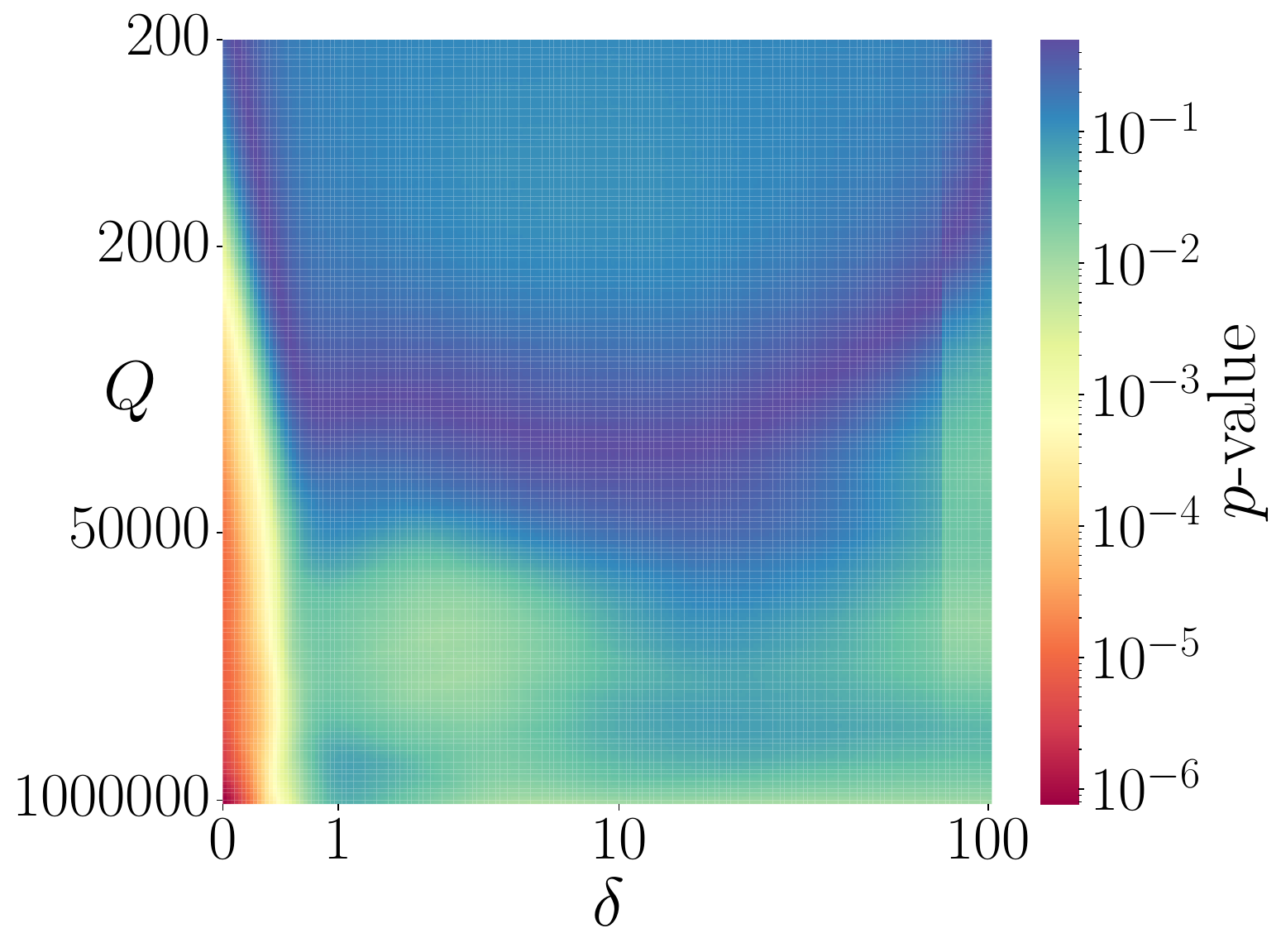}%
    }~~~
    \subfloat[$p$-value, Ask]{%
        \includegraphics[width=0.25\linewidth]{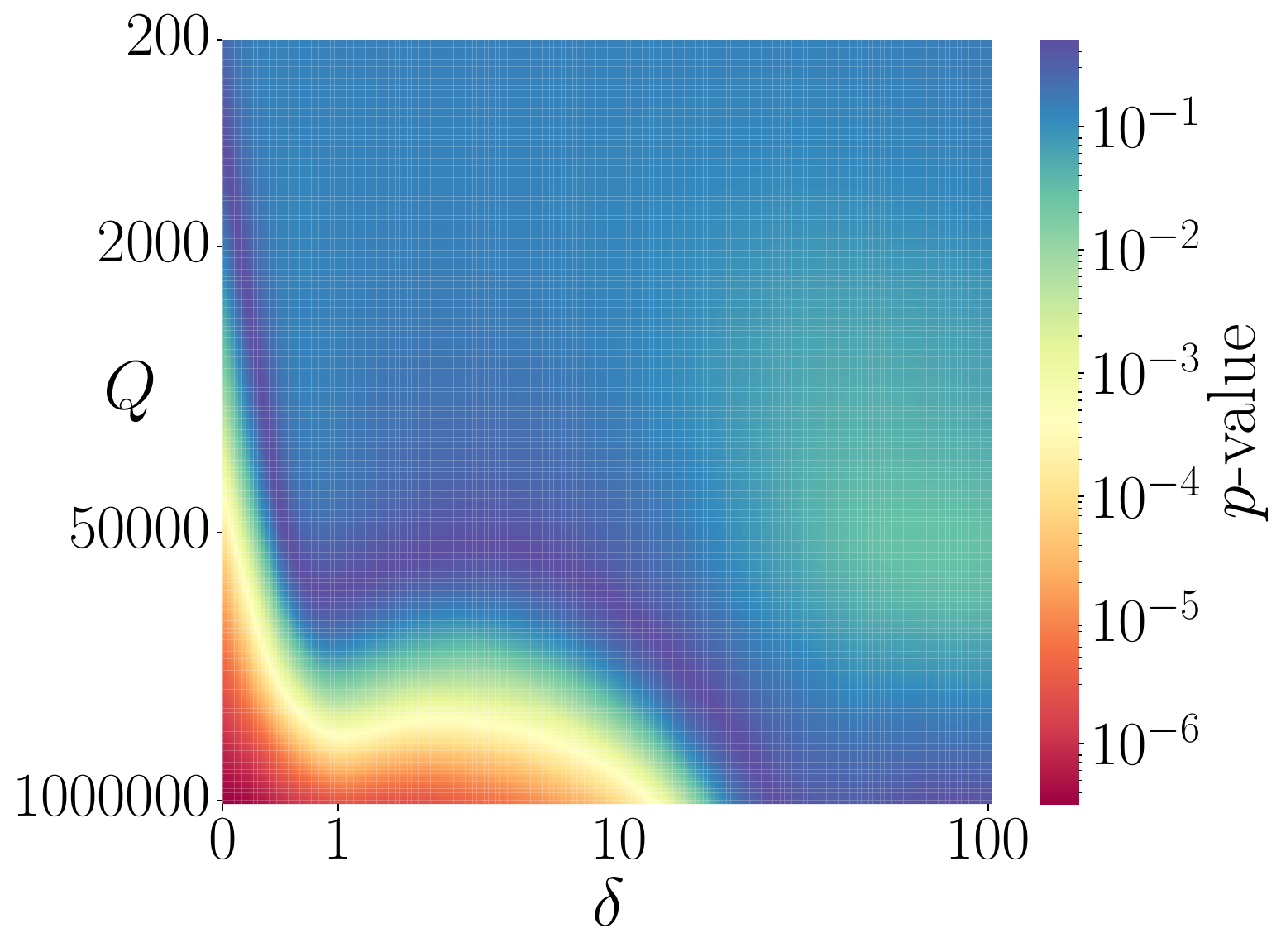}%
    }
    \caption{\textit{Single asset price response} --- Average variation of BTC-USD Sharpe ratio $\overline{\Delta\mathcal{S}_N}(Q,\delta)$ when an order of size $Q$ is posted in the BTC-USD LOB at a distance of $\delta$ basis points from the best queue, and $p$-value of the corresponding Student test.}
    \label{fig:btcusd_price_response}
\end{figure}

\begin{figure}[!ht]
    \centering
    \subfloat[Impact, Bid]{%
        \includegraphics[width=0.25\linewidth]{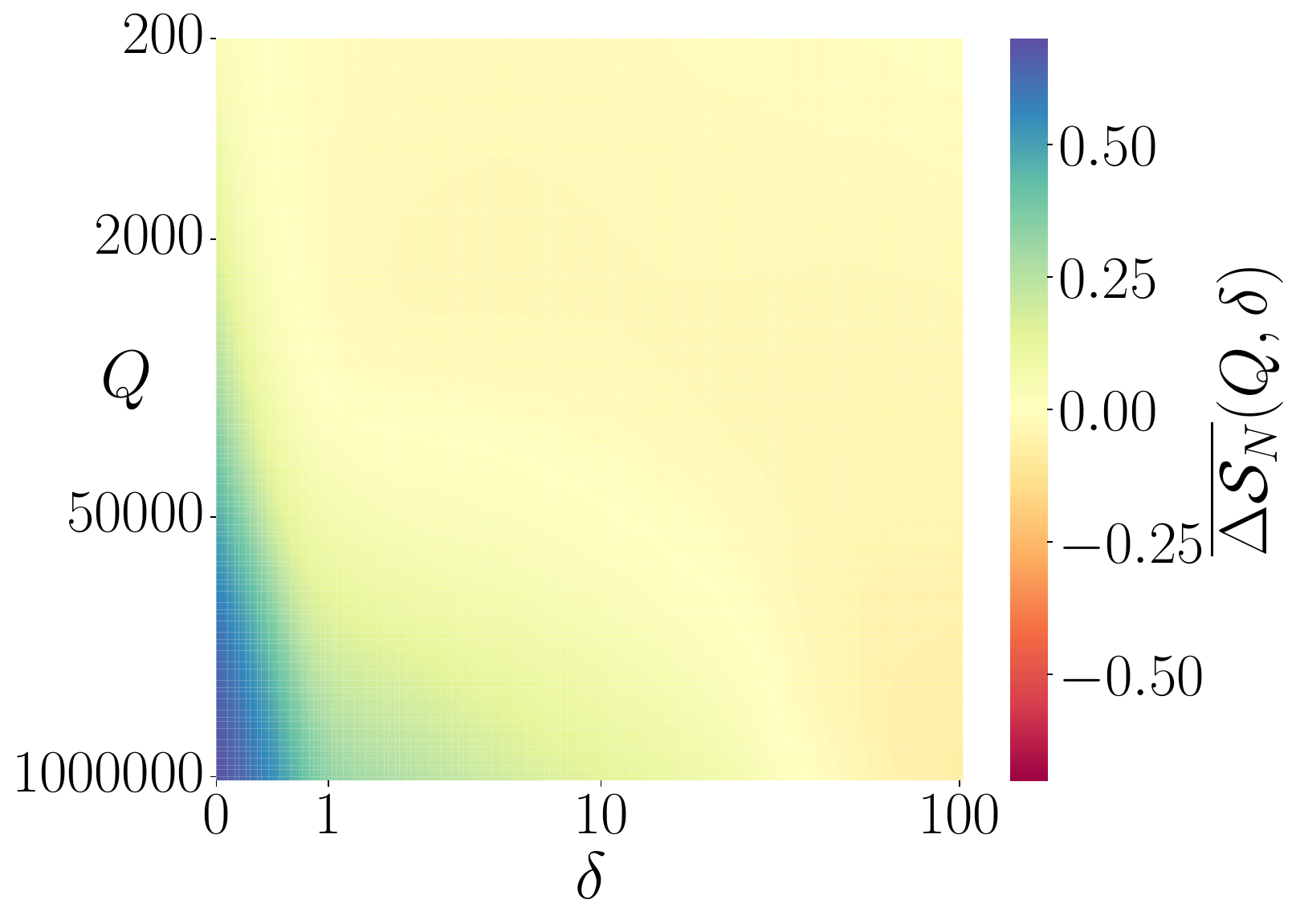}%
    }~~~
    \subfloat[Impact, Ask]{%
        \includegraphics[width=0.25\linewidth]{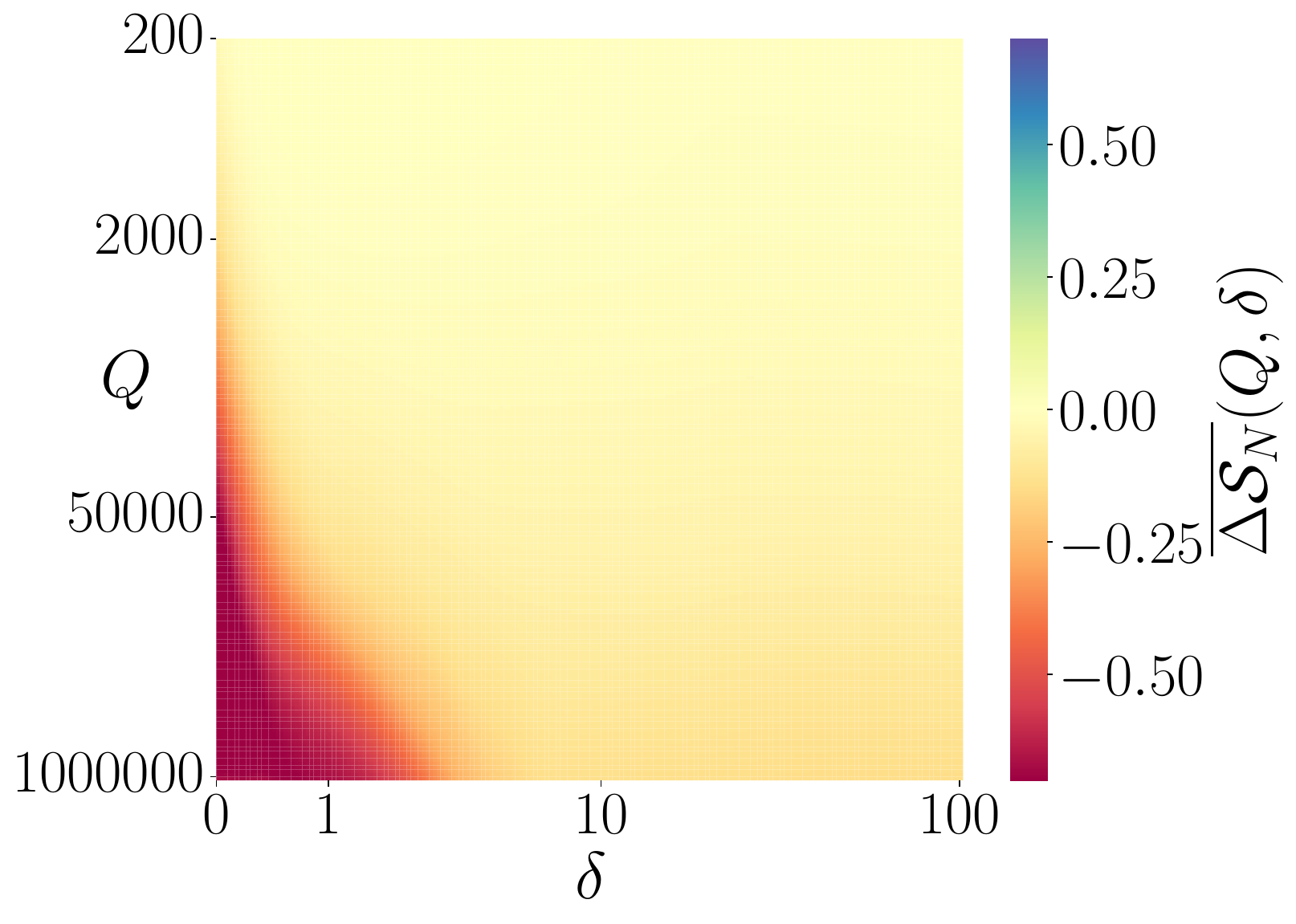}%
    }~~~
    \subfloat[$p$-value, Bid]{%
        \includegraphics[width=0.25\linewidth]{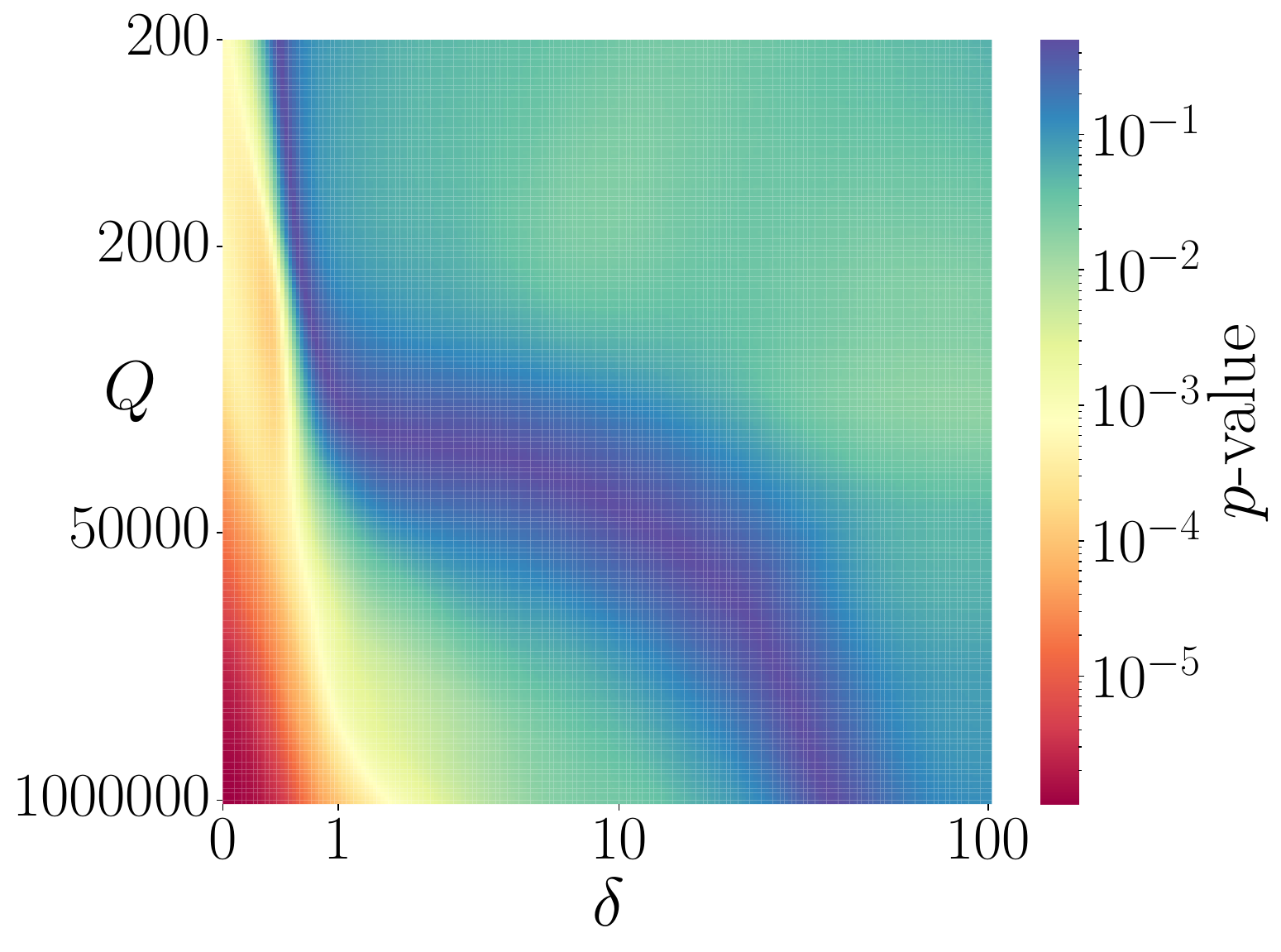}%
    }~~~
    \subfloat[$p$-value, Ask]{%
        \includegraphics[width=0.25\linewidth]{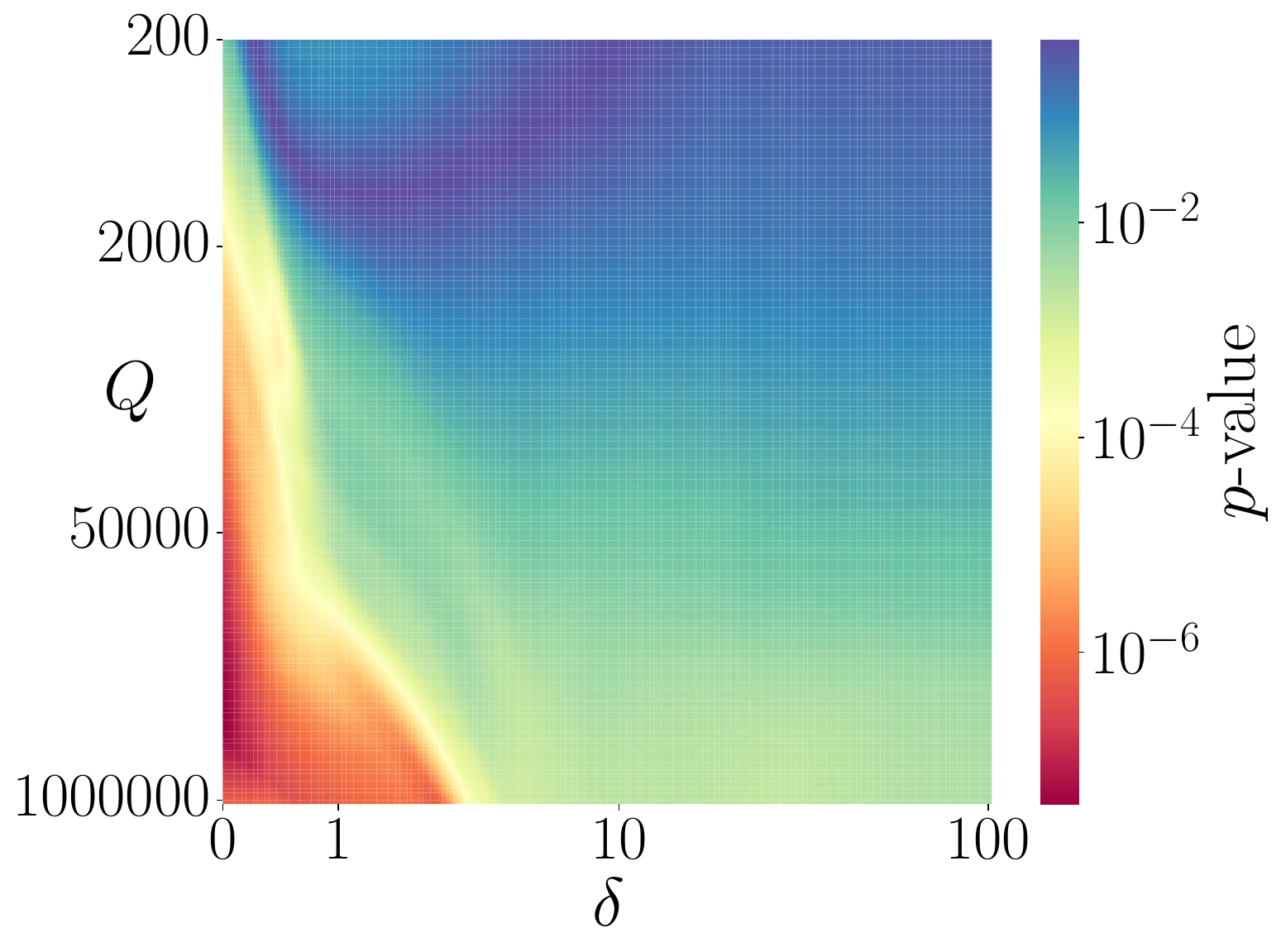}%
    }
    \caption{\textit{Single asset price response} --- Average variation of ETH-USD Sharpe ratio $\overline{\Delta\mathcal{S}_N}(Q,\delta)$ when an order of size $Q$ is posted in the ETH-USD LOB at a distance of $\delta$ basis points from the best queue, and $p$-value of the corresponding Student test.}
    \label{fig:ethusd_price_response}
\end{figure}

We sample $N=10$ observations over the test set for each couple $(Q,\delta)$. We observe in Figs\ \ref{fig:btcusd_price_response} and \ref{fig:ethusd_price_response} that the larger the posted size, the greater the impact over the Sharpe ratio. The distance of placement seems to have an inverse role over the Sharpe ratio, as expected: the smaller the distance, the larger the impact. Globally, there is a bid-ask symmetry: the impact is positive for bid orders, and negative for ask orders, meaning that large bid orders posted near the best queue will tend to push the price upward while large ask orders posted near the best queue will tend to push the price downward. Although the symmetry is not perfect, {\em e.g.} the ask side of BTC-USD exhibits statistically significant price impact for large orders posted at a distance up to 10 bps, we observe that in general, orders of size $Q\geq 50,000$ USD and posting distance $\delta$ smaller than several basis points generate maximum impact over the predicted risk-adjusted price move. Overall, this experiment demonstrates the spoofability of the limit order book, and shows that it is not necessary to post orders at the best queue if one seeks to generate short-term price impact. It also emphasizes the key role of the size and the distance of placement of limit orders in price dynamics at the high-frequency scale, indicating that spoofing detection models that do not take the posting distance into account are not suitable to describe the data.

\begin{figure}[!ht]
    \centering
    \subfloat[Impact, Bid]{%
        \includegraphics[width=0.25\linewidth]{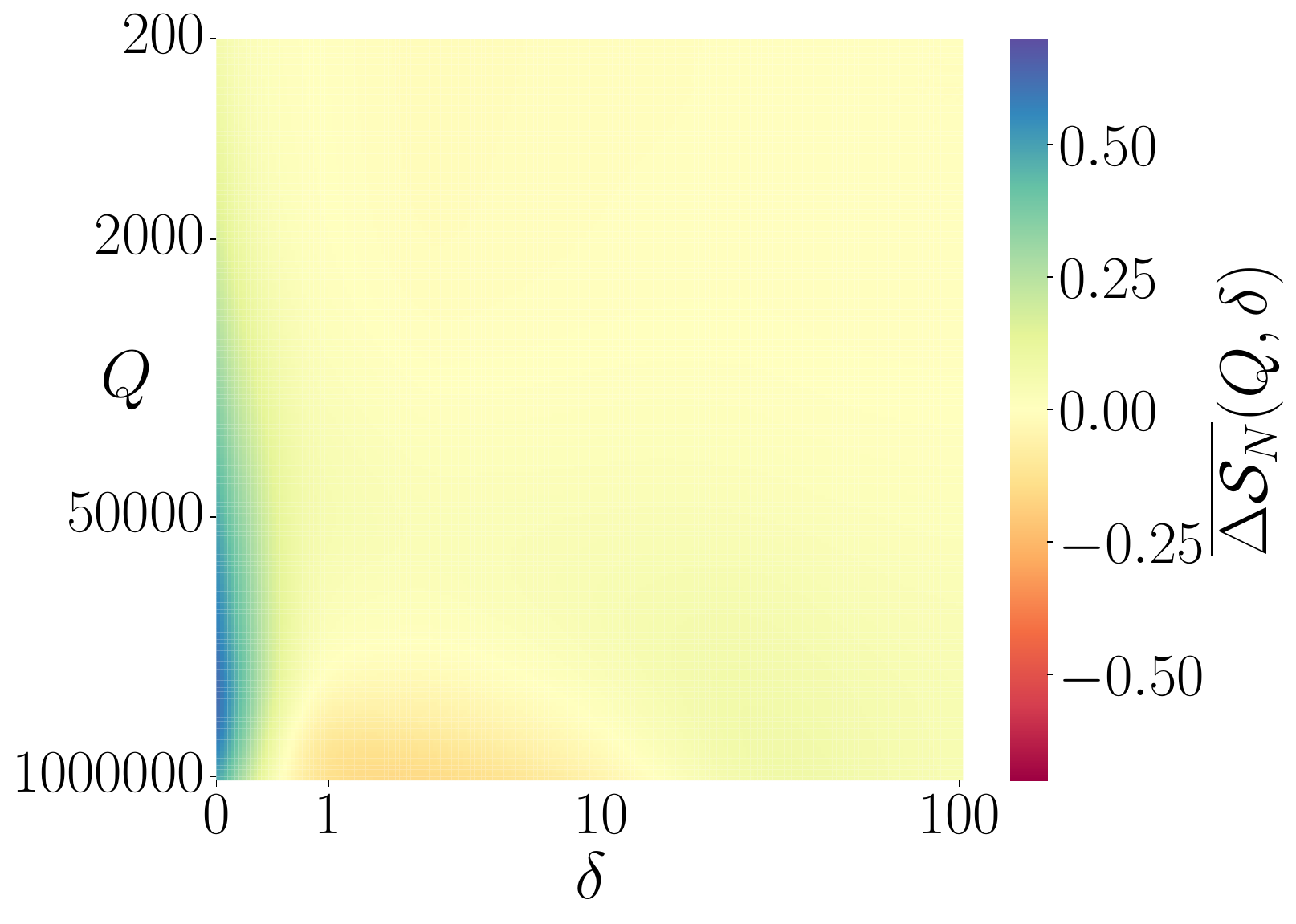}%
    }~~~
    \subfloat[Impact, Ask]{%
        \includegraphics[width=0.25\linewidth]{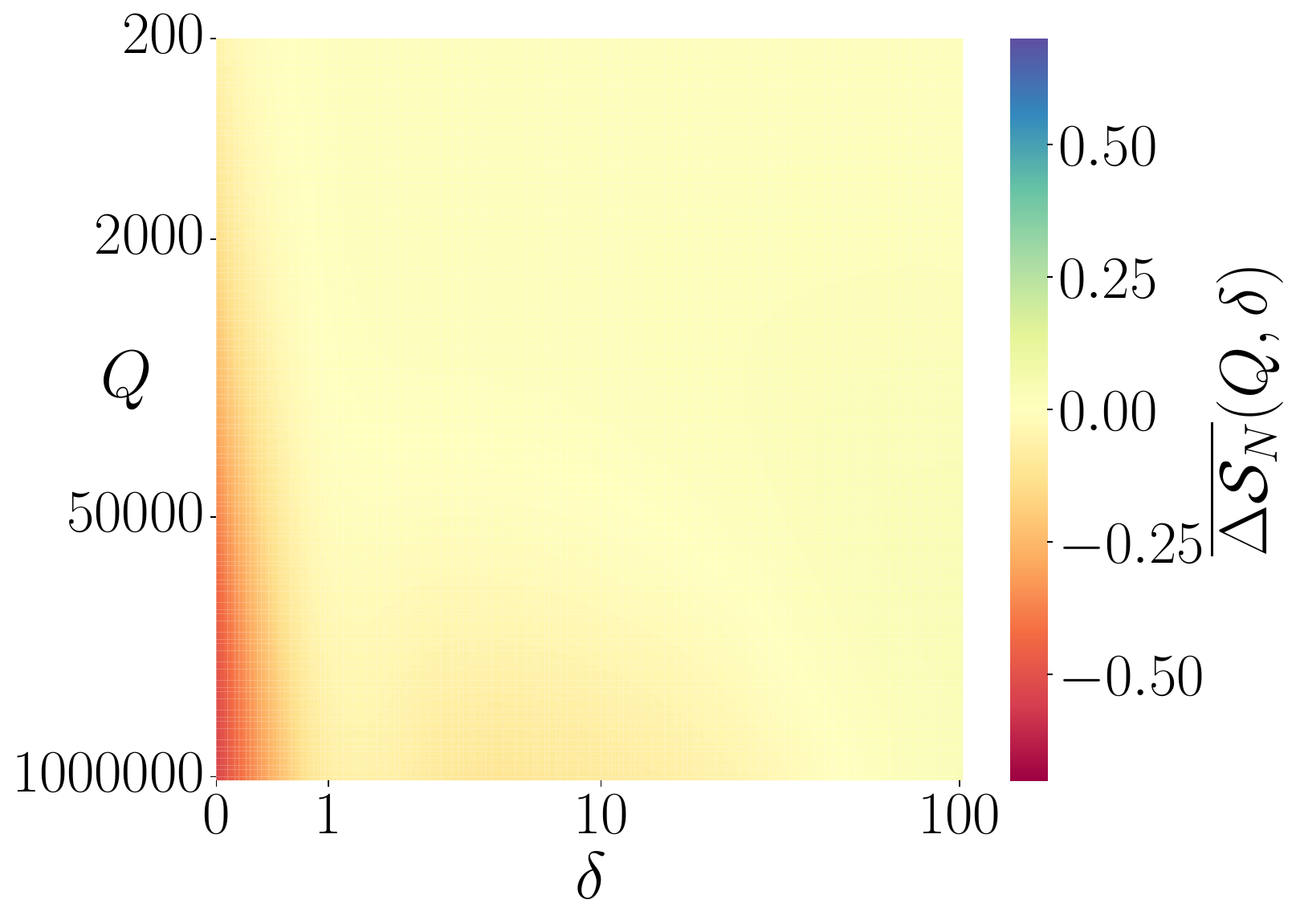}%
    }~~~
    \subfloat[$p$-value, Bid]{%
        \includegraphics[width=0.25\linewidth]{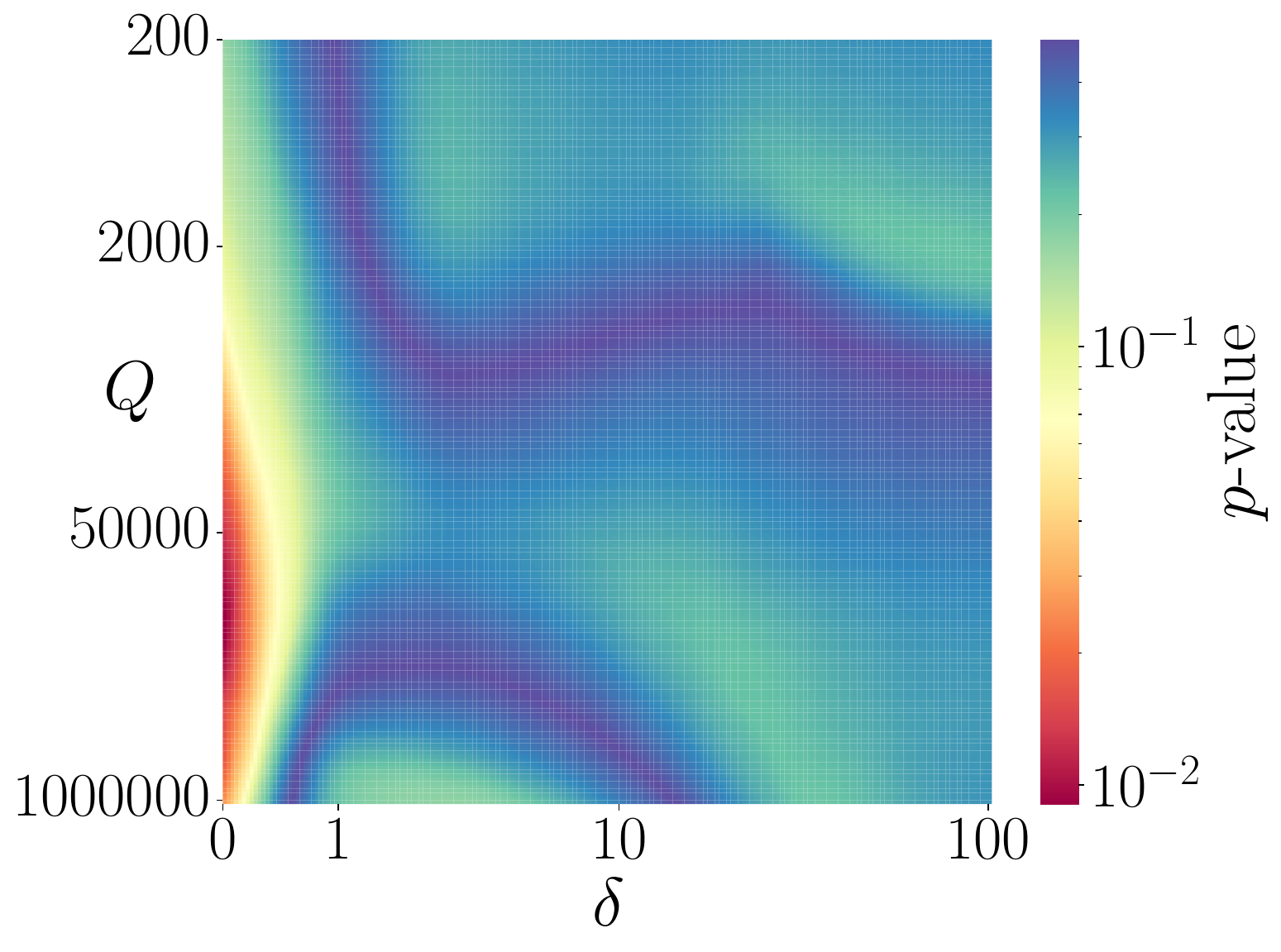}%
    }~~~
    \subfloat[$p$-value, Ask]{%
        \includegraphics[width=0.25\linewidth]{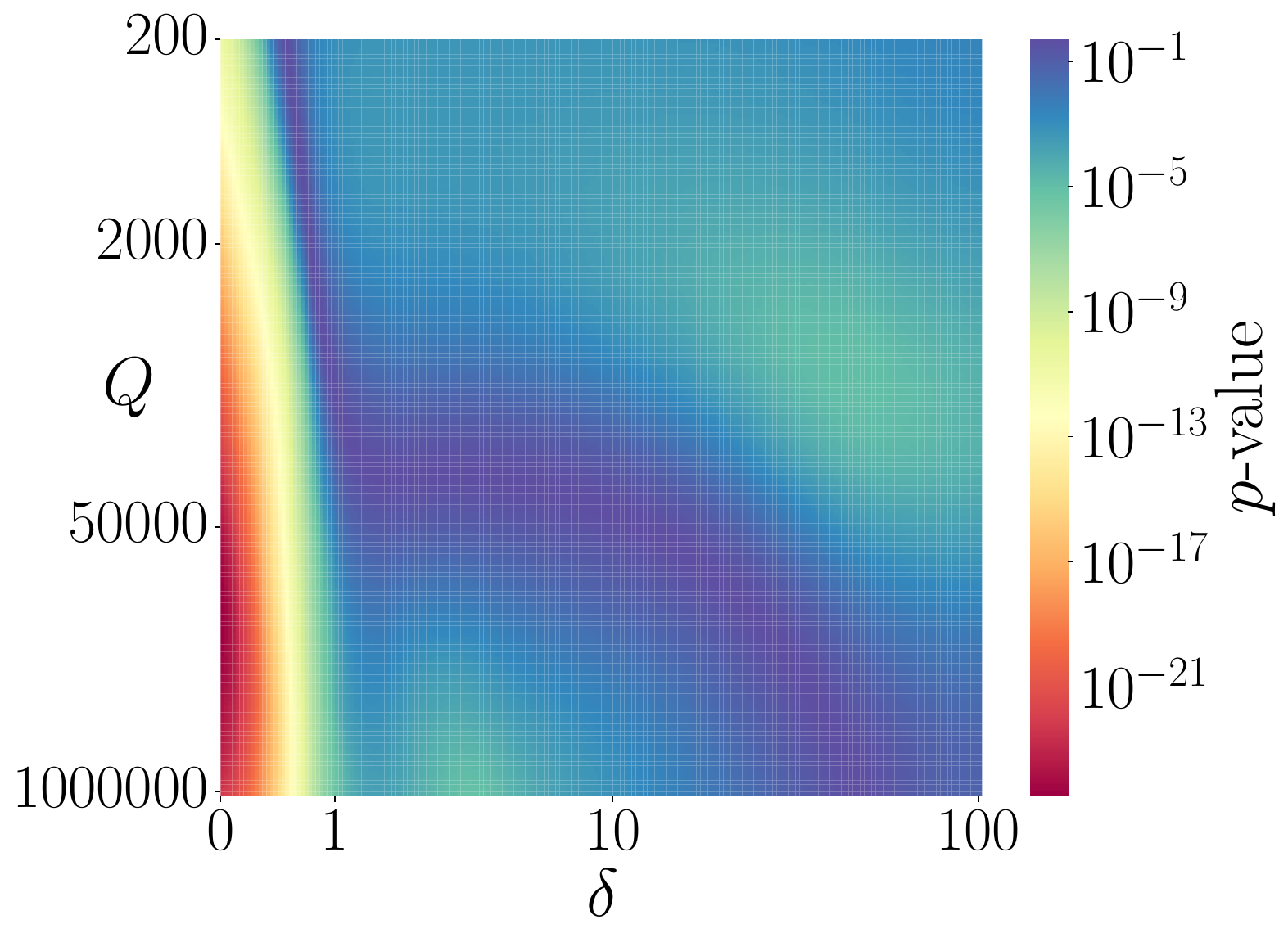}%
    }
    \caption{\textit{Cross-asset price response} --- Average variation of ETH-USD Sharpe ratio $\overline{\Delta\mathcal{S}_N}(Q,\delta)$ when an order of size $Q$ is posted in the BTC-USD LOB at a distance of $\delta$ basis points from the best queue, and $p$-value of the corresponding Student test.}
    \label{fig:btcusd_ethusd_price_response}
\end{figure}

\begin{figure}[!ht]
    \centering
    \subfloat[Impact, Bid]{%
        \includegraphics[width=0.25\linewidth]{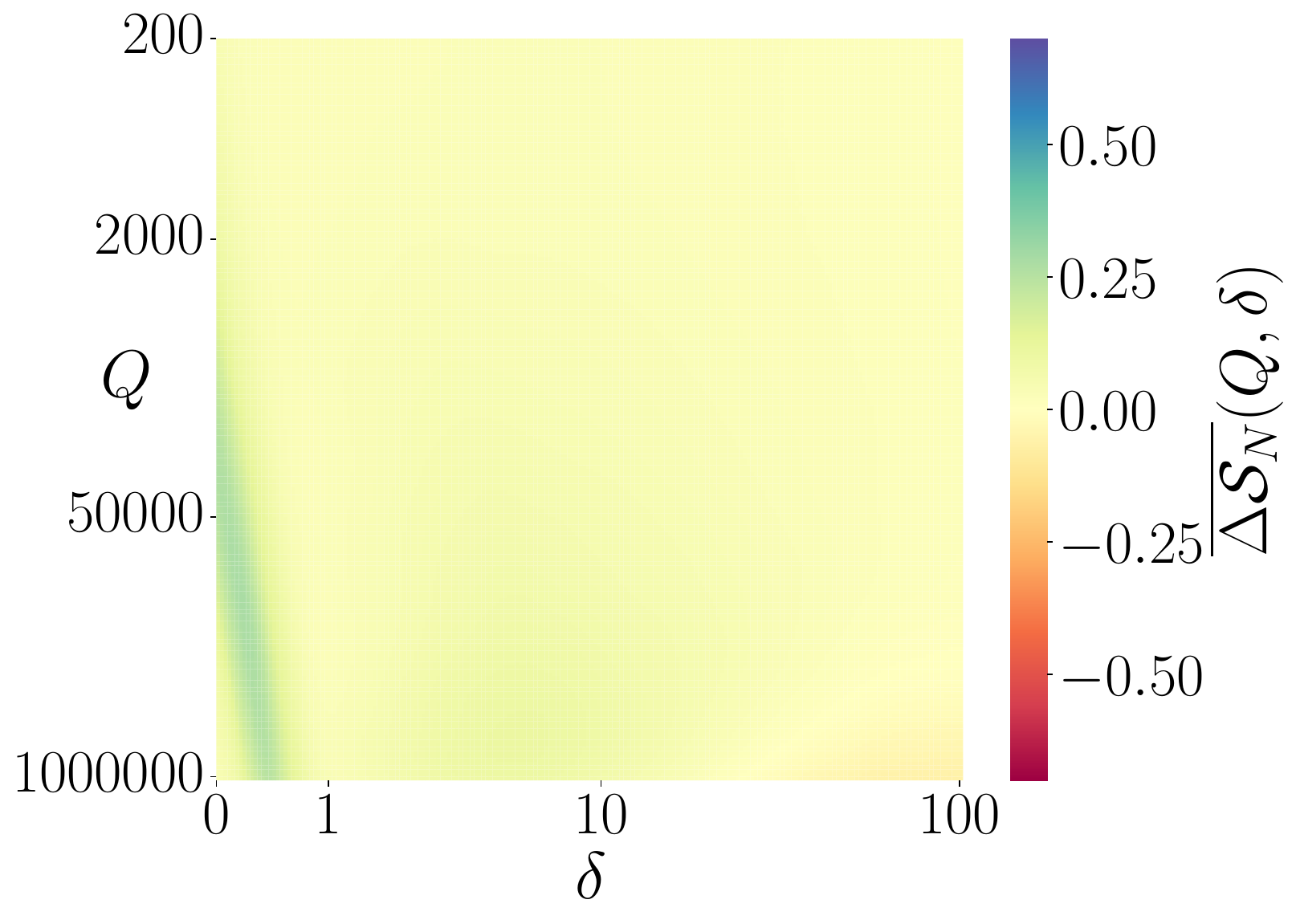}%
    }~~~
    \subfloat[Impact, Ask]{%
        \includegraphics[width=0.25\linewidth]{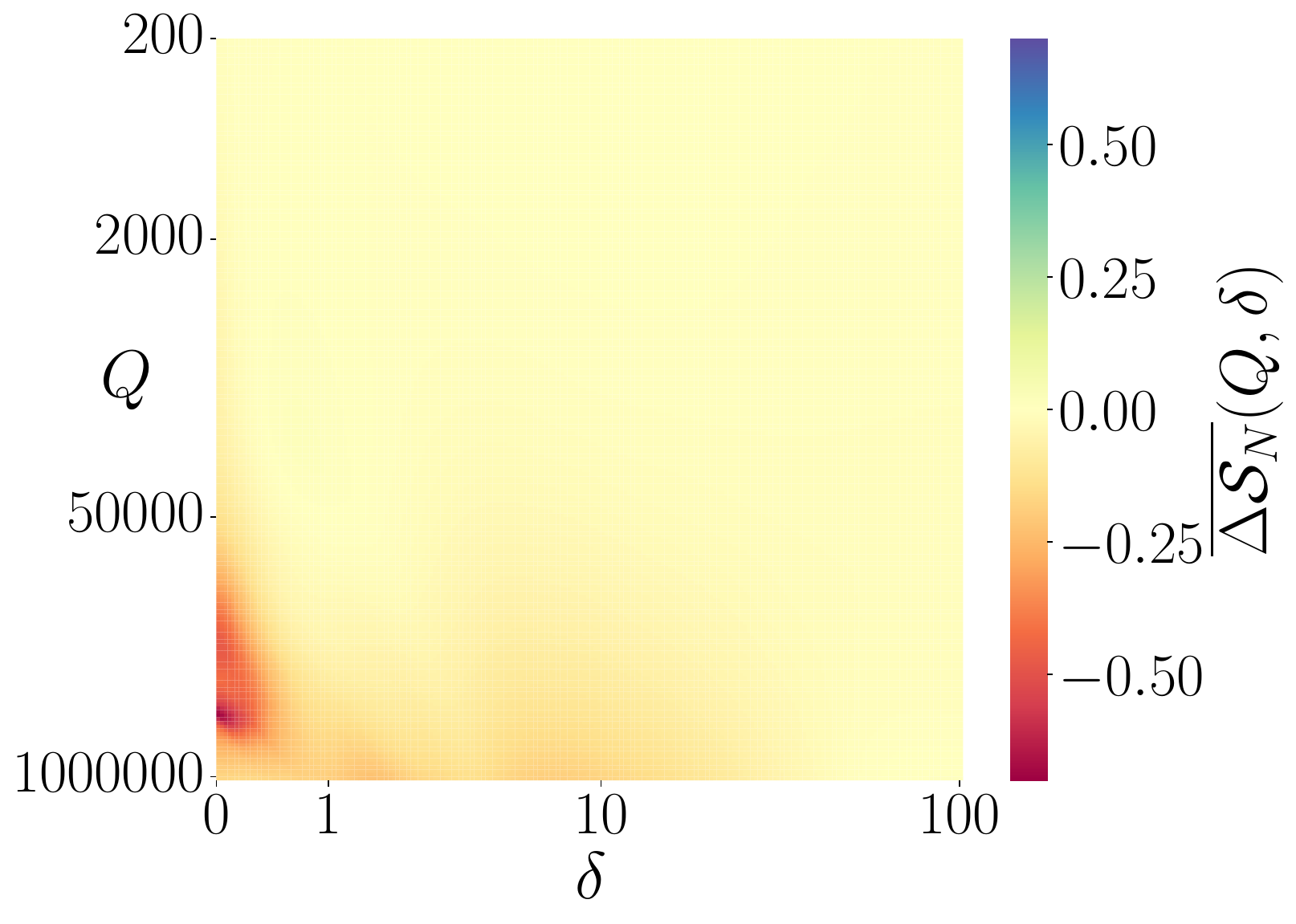}%
    }~~~
    \subfloat[$p$-value, Bid]{%
        \includegraphics[width=0.25\linewidth]{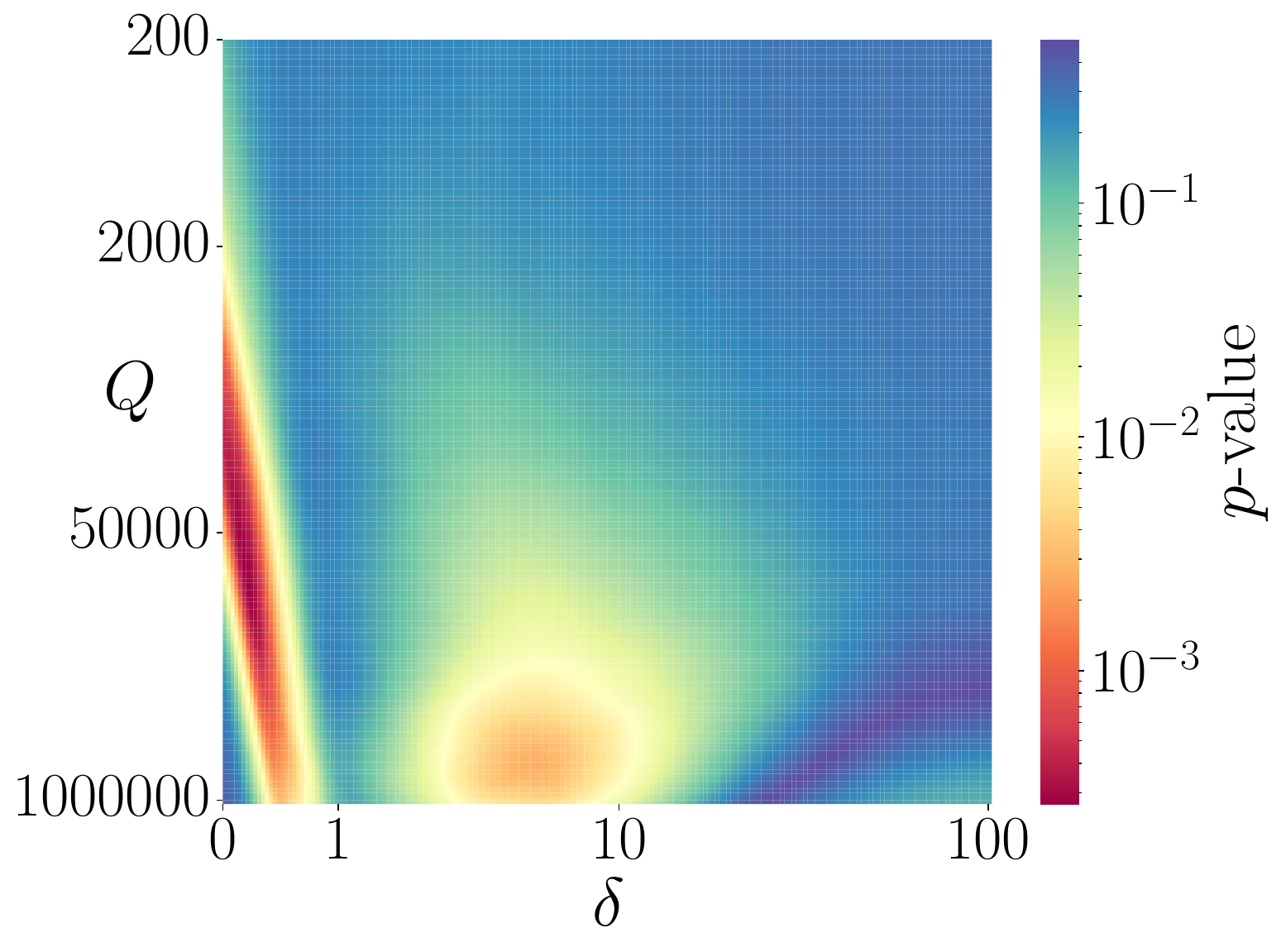}%
    }~~~
    \subfloat[$p$-value, Ask]{%
        \includegraphics[width=0.25\linewidth]{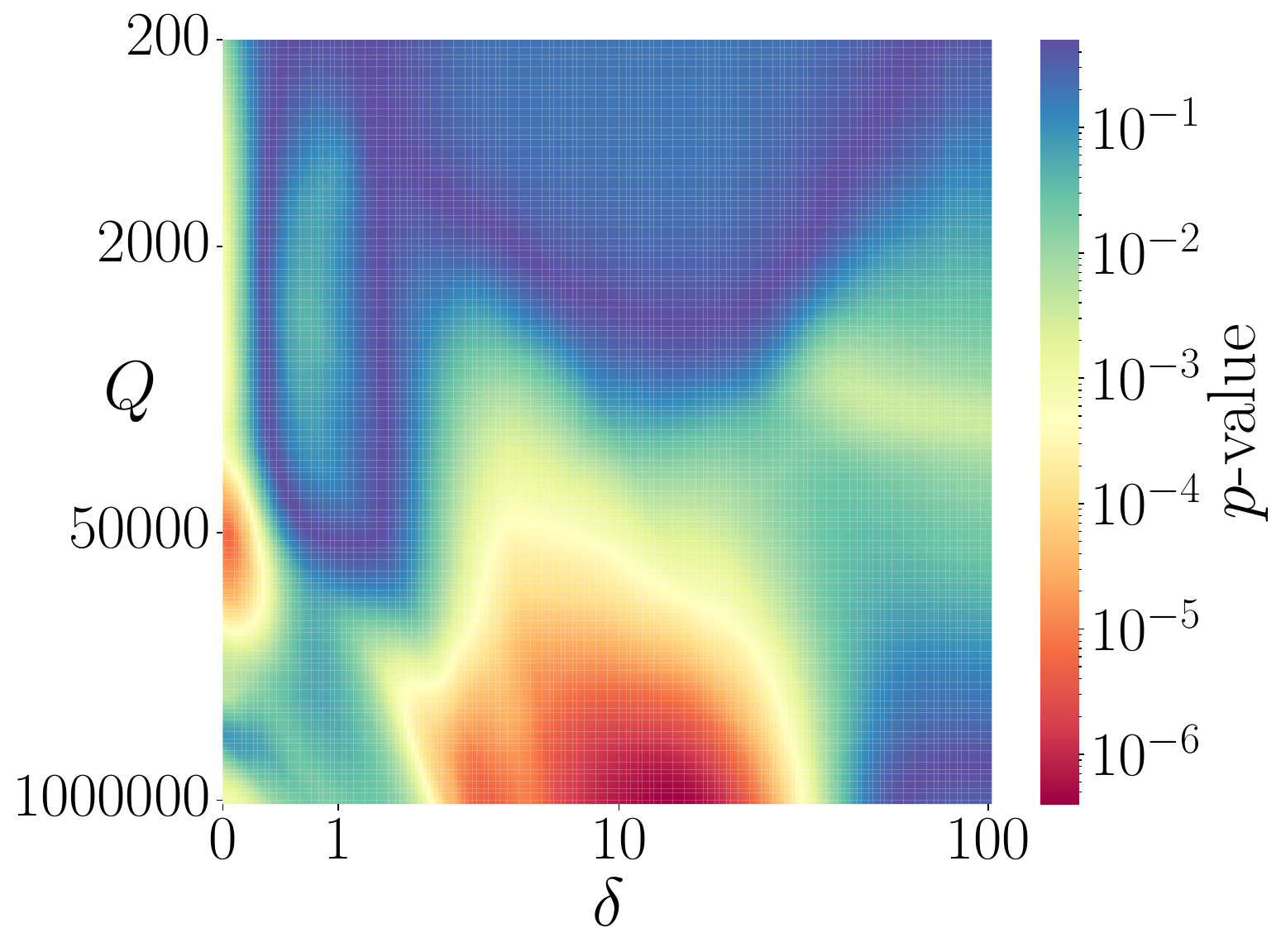}%
    }
    \caption{\textit{Cross-asset price response} --- Average variation of BTC-USD Sharpe ratio $\overline{\Delta\mathcal{S}_N}(Q,\delta)$ when an order of size $Q$ is posted in the ETH-USD LOB at a distance of $\delta$ basis points from the best queue, and $p$-value of the corresponding Student test.}
    \label{fig:ethusd_btcusd_price_response}
\end{figure}

We now proceed to the same experiment with the cross-asset case. We use the bivariate neural network model to analyse the impact of the insertion of an order in the order book of BTC (resp. ETH) over the Sharpe ratio of ETH (resp. BTC-USD). Results are reported in Figs \ref{fig:btcusd_ethusd_price_response} and \ref{fig:ethusd_btcusd_price_response}.

Strikingly, the impact of BTC orders over the ETH Sharpe ratio is much more statistically significant than the over way around. We observe the same property as the single asset case for BTC orders: the larger the size, the larger the impact on ETH and the larger the posting distance, the smaller the impact. While posting limit orders on ETH LOB seems to generate impact in some configurations, we do not observe a monotonic relationship of the impact with respect to the order size and the distance. Although this could be a data artefact, it is more likely a sign that spoofing ETH using BTC orders is easier than spoofing BTC with ETH orders.

\section{Spoofing detection}

\subsection{Single asset case}

At the beginning of the trading period, the agent observes the best bid price $p^b$, the best ask price $p^a$, and the bid-ask spread $\Psi$. Let us define the spoofing tactic for an agent willing to sell. The manipulative agent simultaneously places a \textit{bona fide} sell order of small size $q$ at price $p^a+\delta^a$ and a non-\textit{bona fide} buy order of large size $Q$, with $Q\gg q$, at price $p^b-\delta^b$.

We make the following assumptions.

\begin{assumption}
No partial fill occurs. At the trading horizon, the orders are either completely filled or not filled at all.
\end{assumption}

This assumption is not particularly restrictive for the \textit{bona fide} order as its size is supposed to be sufficiently small, but it does not hold for the large manipulative order for obvious reasons. We will discuss this assumption later in the paper and propose an approach to extend the model to partial fills.

\begin{assumption}
The execution of an order is triggered if the fair price crosses the price of this order. Mathematically, the execution of the bid order is observed if $p_T<p^b-\delta^b$, and the execution of the ask order is observed if $p_T>p^a+\delta^a$. This translates into $\{\Delta p < -(\delta^b+\frac{1}{2}\Psi)\}$ for the bid side and $\{\Delta p > \delta^a+\frac{1}{2}\Psi\}$ for the ask side.
\end{assumption}

Our characterization of the execution rule is equivalent to the first-passage time method using price time series \citep{eisler2009diffusive}. Although the execution of an order can be observed even when the mid price does not cross the price limit, we will see later in this section that this specification allows us to reduce the complexity of the cost function computation.

\begin{assumption}
The agent has to execute the target quantity $q$ within a preset trading time frame. Hence, if the \textit{bona fide} order is not filled at the horizon, a marketable order is sent on the opposite side, crossing the spread, and incurring extra transaction costs. The agent is not willing to keep the inventory that results from the undesired execution of the manipulative order. If this order is filled, the inventory will be liquidated at the trading horizon by sending a marketable order to the opposite side.
\end{assumption}

In this framework, the manipulative agent will penalize her expected cost with respect to the risk of execution of the non-\textit{bona fide} order, and will not spoof the LOB if the expected loss produced by the execution of the non-\textit{bona fide} order is too large compared to the expected gain resulting from the execution of the \textit{bona fide} order. Thus, a trade-off should be made in order to make spoofing profitable. On the one hand, lowering the distance of placement of the non-\textit{bona fide} order, \textit{i.e.} quoting closer to the best quote increases the expected price impact, but the risk of execution of the non-\textit{bona fide} order becomes significant. On the other hand, if the agent chooses a large distance of placement for the non-\textit{bona fide} order, \textit{i.e.} quotes deeper in the LOB, the risk of execution becomes negligible, but the expected price impact vanishes as demonstrated in our previous experiments.

We denote by $\varepsilon^+$ and $\varepsilon^-$ the maker fee and the taker fee, and by $C_{\text{spoof}}^a(\delta^b, \delta^a, Q, q)$ the cost incurred by the spoofer that behaves as described above. From the above assumptions, we write the expected cost of the spoofer as follows
\begin{align}\label{eq:cost_bid_spoofer}
    \mathbb{E}\left(C_{\text{spoof}}^a(\delta^b, \delta^a, Q, q)\right)&=\underbrace{-\mathbb{P}\left(\Delta p > \delta^a+\frac{1}{2}\Psi\right)(1-\varepsilon^+)q\left(p^a+\delta^a\right)}_{\text{Execution of the \textit{bona fide} order}}\notag\\
    &\underbrace{+\mathbb{P}\left(\Delta p < -(\delta^b+\frac{1}{2}\Psi)\right)(1+\varepsilon^+)Q\left(p^b-\delta^b\right)}_{\text{Execution of the non \textit{bona fide} order}}\notag\\
    &\underbrace{-\mathbb{P}\left(\Delta p \leq \delta^a+\frac{1}{2}\Psi\right)(1-\varepsilon^-)q\left(p^b+\mathbb{E}\left(\Delta p|\Delta p \leq \delta^a+\frac{1}{2}\Psi\right)\right)}_{\text{Terminal liquidation of the unfilled \textit{bona fide} order}}\notag\\
    &\underbrace{-\mathbb{P}\left(\Delta p < -(\delta^b+\frac{1}{2}\Psi)\right)(1-\varepsilon^-)Q\left(p^b+\mathbb{E}\left(\Delta p|\Delta p < -(\delta^b+\frac{1}{2}\Psi)\right)\right)}_{\text{Terminal liquidation of the filled non \textit{bona fide} order}}.
\end{align}

In the same manner, we denote by $C_{\text{spoof}}^b(\delta^b, \delta^a, q, Q)$ the cost incurred by the spoofer with buying intentions and write

\begin{align}\label{eq:cost_ask_spoofer}
    \mathbb{E}\left(C_{\text{spoof}}^b(\delta^b, \delta^a, q, Q)\right)&=\mathbb{P}\left(\Delta p < -(\delta^b+\frac{1}{2}\Psi)\right)(1+\varepsilon^+)q\left(p^b-\delta^b\right)\notag\\
    &-\mathbb{P}\left(\Delta p > \delta^a+\frac{1}{2}\Psi)\right)(1-\varepsilon^+)Q\left(p^a+\delta^a\right)\notag\\
    &+\mathbb{P}\left(\Delta p \geq -(\delta^b+\frac{1}{2}\Psi)\right)(1+\varepsilon^-)q\left(p^a+\mathbb{E}\left(\Delta p|\Delta p \geq -(\delta^b+\frac{1}{2}\Psi)\right)\right)\notag\\
    &+\mathbb{P}\left(\Delta p > \delta^a+\frac{1}{2}\Psi)\right)(1+\varepsilon^-)Q\left(p^a+\mathbb{E}\left(\Delta p|\Delta p > \delta^a+\frac{1}{2}\Psi\right)\right).
\end{align}

Using the pre-trained neural network model, the computation of these cost functions is easy thanks to the specification of the fill probability with a first passage time approach. Using the predicted parameters of the distribution of $\Delta p$, all the terms of the cost function are easily calculated for a wide range of parametric distributions. We provide in the appendix some formulas regarding the computation of the terms of the cost functions in our framework. Note that for most of the common parametric distributions, closed-form formulas or polynomial approximations exist, leading to a quick estimation of the expected cost, which makes it suitable to high-frequency live computation.

\subsection{Cross asset spoofing}

Let us briefly discuss the cross-asset case. In this framework, the agent spoofs asset 1 by posting a large non-\textit{bona fide} order in its LOB and posts a \textit{bona fide} order in the order book of asset 2. We previously demonstrated the cross-asset spoofability of BTC and ETH, especially the spoofability of the price of ETH by posting a large non-\textit{bona fide} order in BTC LOB. Denote by $\Psi_1$ (resp. $\Psi_2$) the initial bid-ask spread of asset 1 (resp. asset 2), $p_1^b$ (resp. $p_2^b$) and $p_1^a$ (resp. $p_2^a$) the initial best bid price and the best ask price of asset 1 (resp. asset 2), $\Delta p_1$ (resp. $\Delta p_2$) the price move of asset 1 (resp. asset 2). Using the same idea as the single asset case, we write the expected cost of the cross-asset spoofer as
\begin{align}\label{eq:cost_bid_cross_spoofer}
    \mathbb{E}\left(C_{\text{spoof}}^a(\delta^b, \delta^a, Q, q)\right)&=\underbrace{-\mathbb{P}\left(\Delta p_2 > \delta^a+\frac{1}{2}\Psi_2\right)(1-\varepsilon^+)q\left(p_2^a+\delta^a\right)}_{\text{Execution of the asset 2 \textit{bona fide} order}}\notag\\
    &\underbrace{+\mathbb{P}\left(\Delta p_1 < -(\delta^b+\frac{1}{2}\Psi_1)\right)(1+\varepsilon^+)Q\left(p_1^b-\delta^b\right)}_{\text{Execution of the asset 1 non-\textit{bona fide} order}}\notag\\
    &\underbrace{-\mathbb{P}\left(\Delta p_2 \leq \delta^a+\frac{1}{2}\Psi_2\right)(1-\varepsilon^-)q\left(p_2^b+\mathbb{E}\left(\Delta p_2|\Delta p_2 \leq \delta^a+\frac{1}{2}\Psi_2\right)\right)}_{\text{Terminal execution of the unfilled asset 2 \textit{bona fide} order}}\notag\\
    &\underbrace{-\mathbb{P}\left(\Delta p_1 < -(\delta^b+\frac{1}{2}\Psi_1)\right)(1-\varepsilon^-)Q\left(p_1^b+\mathbb{E}\left(\Delta p_1|\Delta p_1 < -(\delta^b+\frac{1}{2}\Psi_1)\right)\right)}_{\text{Terminal execution of the filled asset 1 non-\textit{bona fide} order}},
\end{align}
for the buyer, and analogously

\begin{align}\label{eq:cost_ask_cross_spoofer}
    \mathbb{E}\left(C_{\text{spoof}}^b(\delta^b, \delta^a, q, Q)\right)&=\mathbb{P}\left(\Delta p_2 < -(\delta^b+\frac{1}{2}\Psi_2)\right)(1+\varepsilon^+)q\left(p_2^b-\delta^b\right)\notag\\
    &-\mathbb{P}\left(\Delta p_1 > \delta^a+\frac{1}{2}\Psi_1)\right)(1-\varepsilon^+)Q\left(p_1^a+\delta^a\right)\notag\\
    &+\mathbb{P}\left(\Delta p_2 \geq -(\delta^b+\frac{1}{2}\Psi_2)\right)(1+\varepsilon^-)q\left(p_2^a+\mathbb{E}\left(\Delta p_2|\Delta p_2 \geq -(\delta^b+\frac{1}{2}\Psi_2)\right)\right)\notag\\
    &+\mathbb{P}\left(\Delta p_1 > \delta^a+\frac{1}{2}\Psi_1)\right)(1+\varepsilon^-)Q\left(p_1^a+\mathbb{E}\left(\Delta p_1|\Delta p_1 > \delta^a+\frac{1}{2}\Psi_1\right)\right),
\end{align}
for the seller.

\subsection{Identification of suspicious order flow}

A problem that arises in detecting spoofing behavior is that we do not directly observe $q$, since the spoofing attempt can be done without inserting the \textit{bona fide} order synchronously. For example, if a market maker has already posted orders on the ask side and spontaneously tries to manipulate the price after posting these orders, we cannot identify the corresponding \textit{bona fide} orders together with the suspicious order. Furthermore, the lack of anonymized ids of the traders inherently limits our ability to detect manipulative behavior with certainty. However, we will show that our detection rule can identify suspicious activity, which is a crucial step before delving deeper into further investigations. Note that it is even more difficult for cross-asset spoofing, whose detection is a truly hard task if no anonymized ids are available. Hence, we will discard the application of our methodology to cross-asset spoofing detection and focus on the single asset case. Nevertheless, one can apply our methodology to the cross-asset case --- and, in the same manner, to the cross-venue case --- with data of better quality using the cost functions defined in Equations \eqref{eq:cost_bid_cross_spoofer} and \eqref{eq:cost_ask_cross_spoofer}.

We set the size of the \textit{bona fide} order to $q=100$ USD. As regards the distance of placement, we set $\delta^a=0$ bp indicating that the \textit{bona fide} order is quoted at touch, that is, at the current best ask. The fees are set according to Coinbase fee policy and we select the highest level, \textit{i.e.} a taker fee of $\varepsilon^-=5$ bps, and no maker fee, \textit{i.e.} $\varepsilon^+=0$ bp.

\begin{figure}
    \centering
    \includegraphics[width=0.95\linewidth]{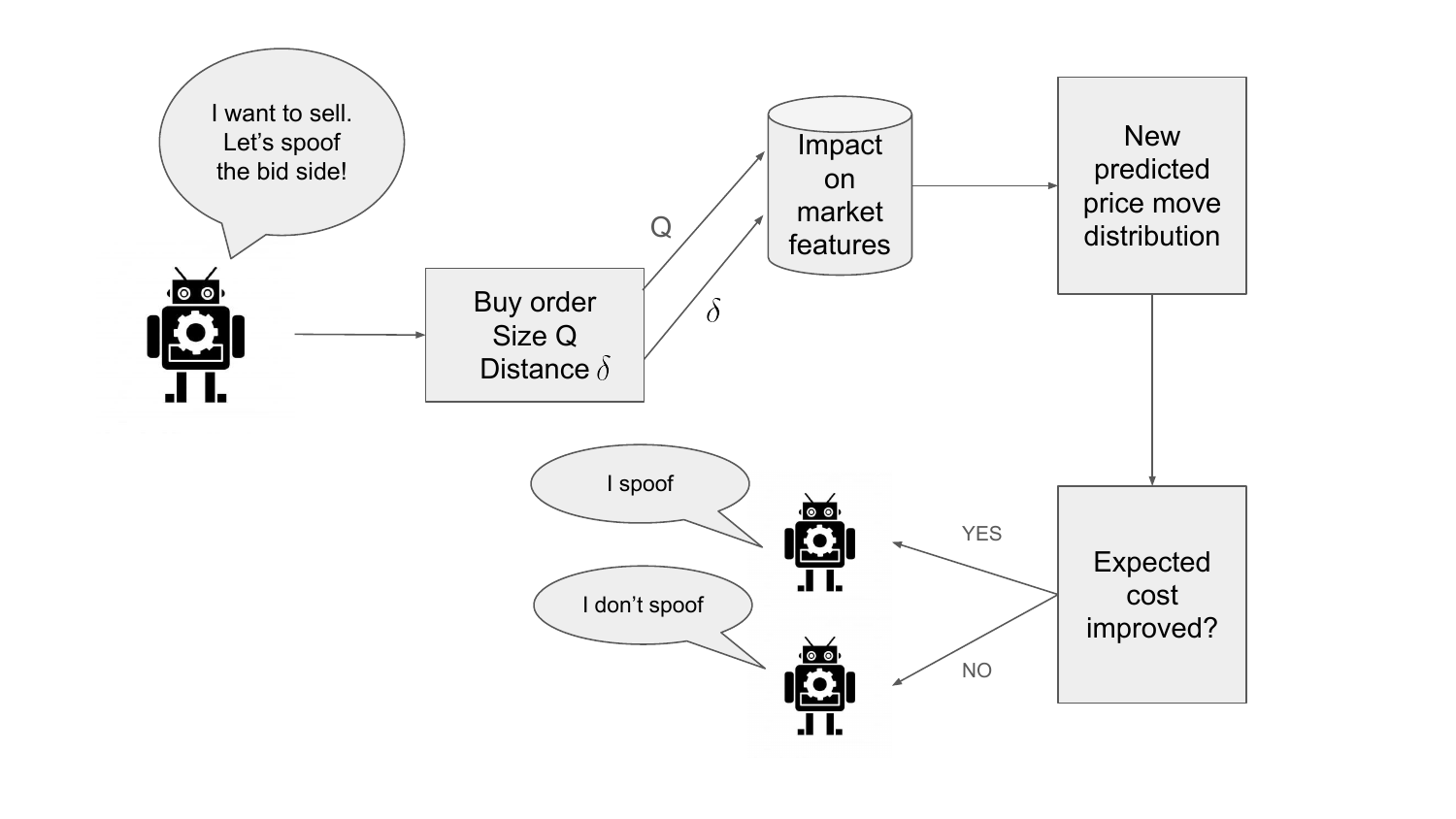}
    \caption{\textit{Spoofing detection diagram} --- An illustration of the detection method using the probabilistic neural network model and the spoofer's cost function.}
    \label{fig:spoofer_diagram}
\end{figure}

Suppose that a buy order is posted in the order book with size $Q$ and distance $\delta$. Let $x$ be the input vector right after the insertion of this order and denote by $x^0$ the corresponding corrected vector if the order was not inserted. Then, we evaluate the expected gain induced by the insertion of this order from the spoofer's point of view by computing the difference

\begin{equation}\label{eq:expected_gain_spoofer}
    \Delta \mathcal{C}(Q,\delta):=\mathbb{E}\left(C_{\text{spoof}}^a(0, \delta^a, 0, q)|x^0\right)-\mathbb{E}\left(C_{\text{spoof}}^a(\delta, \delta^a, Q, q)|x\right),
\end{equation}

which is the difference between the expected cost of a \textit{bona fide} agent and the expected cost of an agent who spoofs with size $Q$ and distance $\delta$. When $\Delta \mathcal{C}(Q,\delta)>0$ it indicates that it is, on average, profitable to spoof the bid side with an order of size $Q$ and distance $\delta$ when one seeks to fill an order of size $q$ and distance $\delta^a$ on the ask side. This condition forms the building block of our detection rule. By adopting the spoofer's point of view, we thus propose the live detection pipeline displayed in Figure \ref{fig:spoofer_diagram}.

We compute the expected gain of Equation \eqref{eq:expected_gain_spoofer} for all the 88,327,661 orders posted during the period  December 4, 2024 to December 7, 2024. We further call an order large if its size $Q\geq 4,500$\ USD (this roughly corresponds to twice the average event size as suggested in \cite{lee2013microstructure}) and which, as suggested by Fig.\ \ref{fig:btcusd_price_response} are more likely to cause a favorable mid price move; there were 8,601,227 large orders during this period. Orders that are labeled as suspicious are orders of size $Q\geq 4,500$ USD such that $\Delta \mathcal{C}(Q,\delta)>0$.

\begin{figure}
    \centering
    \subfloat[BBOs' dynamics and trade prices]{%
        \includegraphics[width=0.25\linewidth]{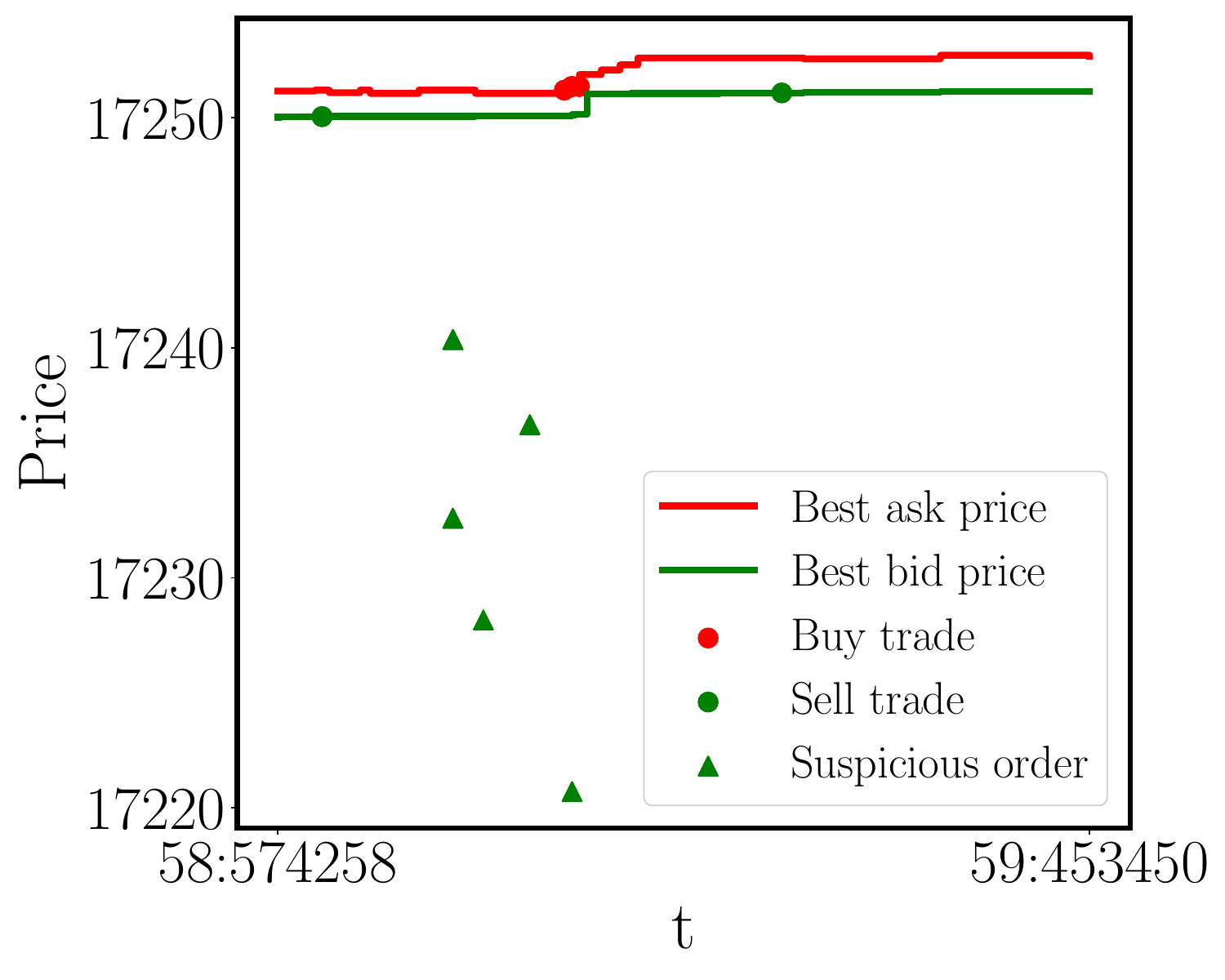}%
    }\hfill
    \subfloat[Order sizes (USD)]{%
        \includegraphics[width=0.25\linewidth]{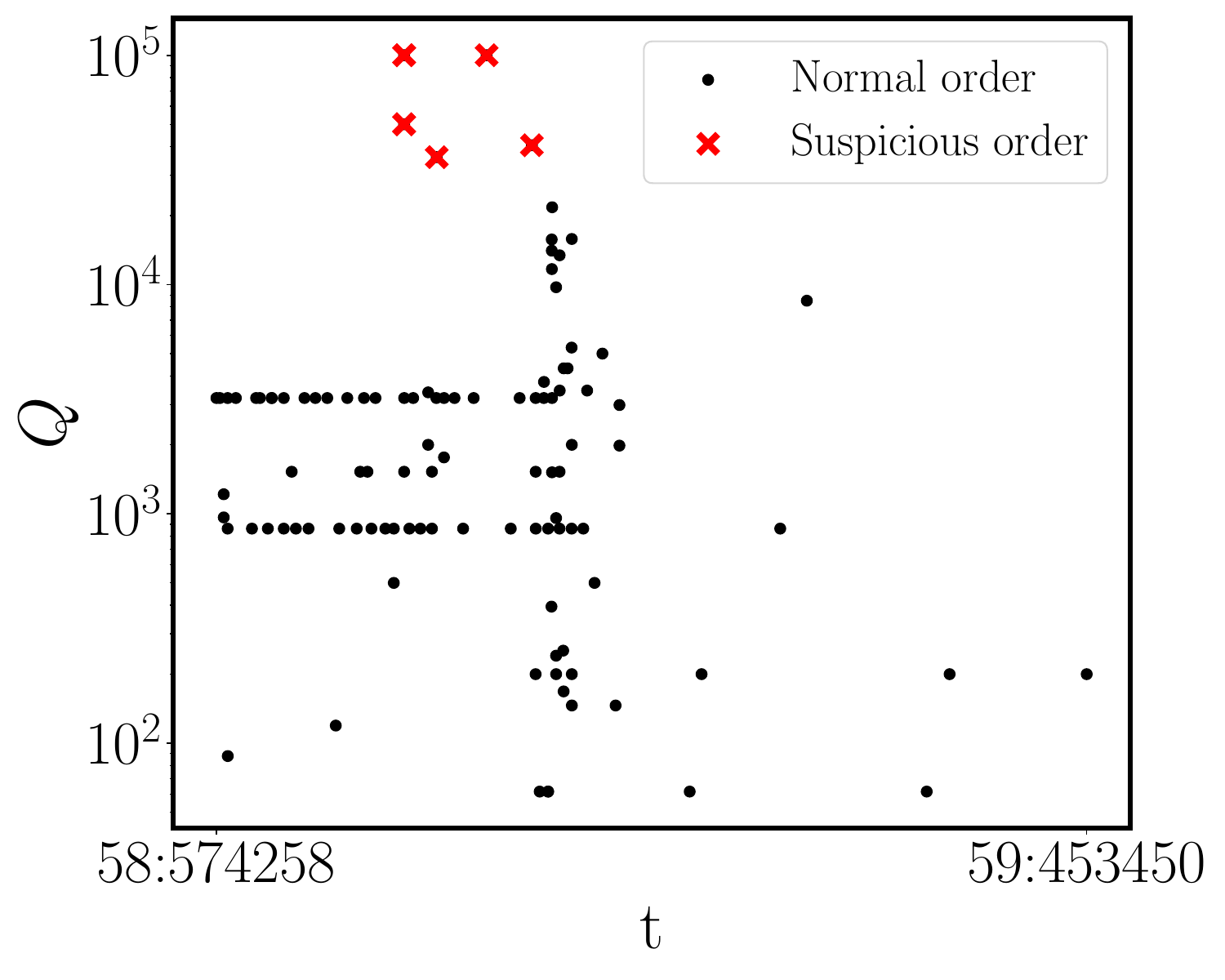}%
    }\hfill
    \subfloat[Order distances (bps)]{%
        \includegraphics[width=0.25\linewidth]{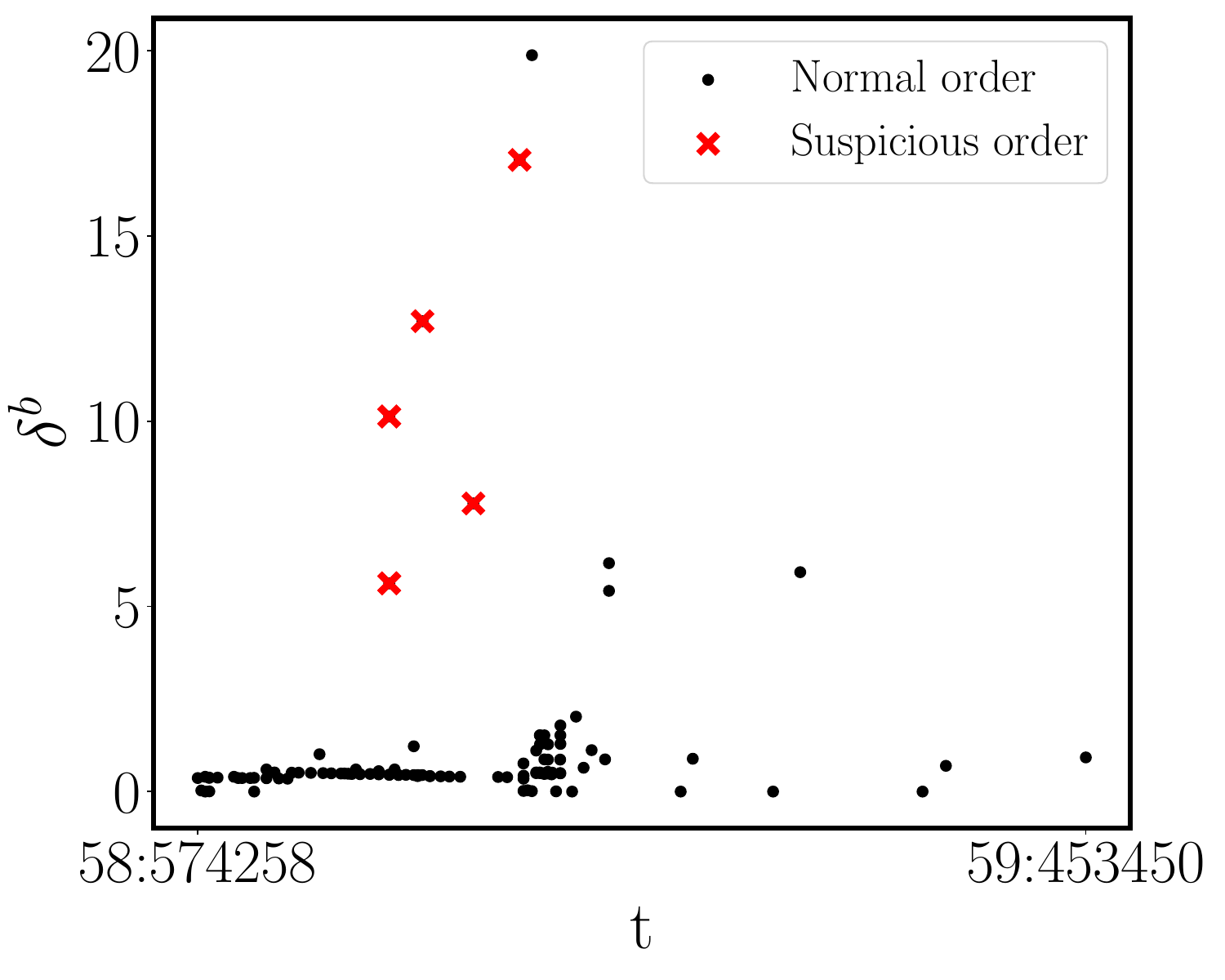}%
    }\hfill
    \subfloat[Spoofer's PnL (USD)]{%
        \includegraphics[width=0.25\linewidth]{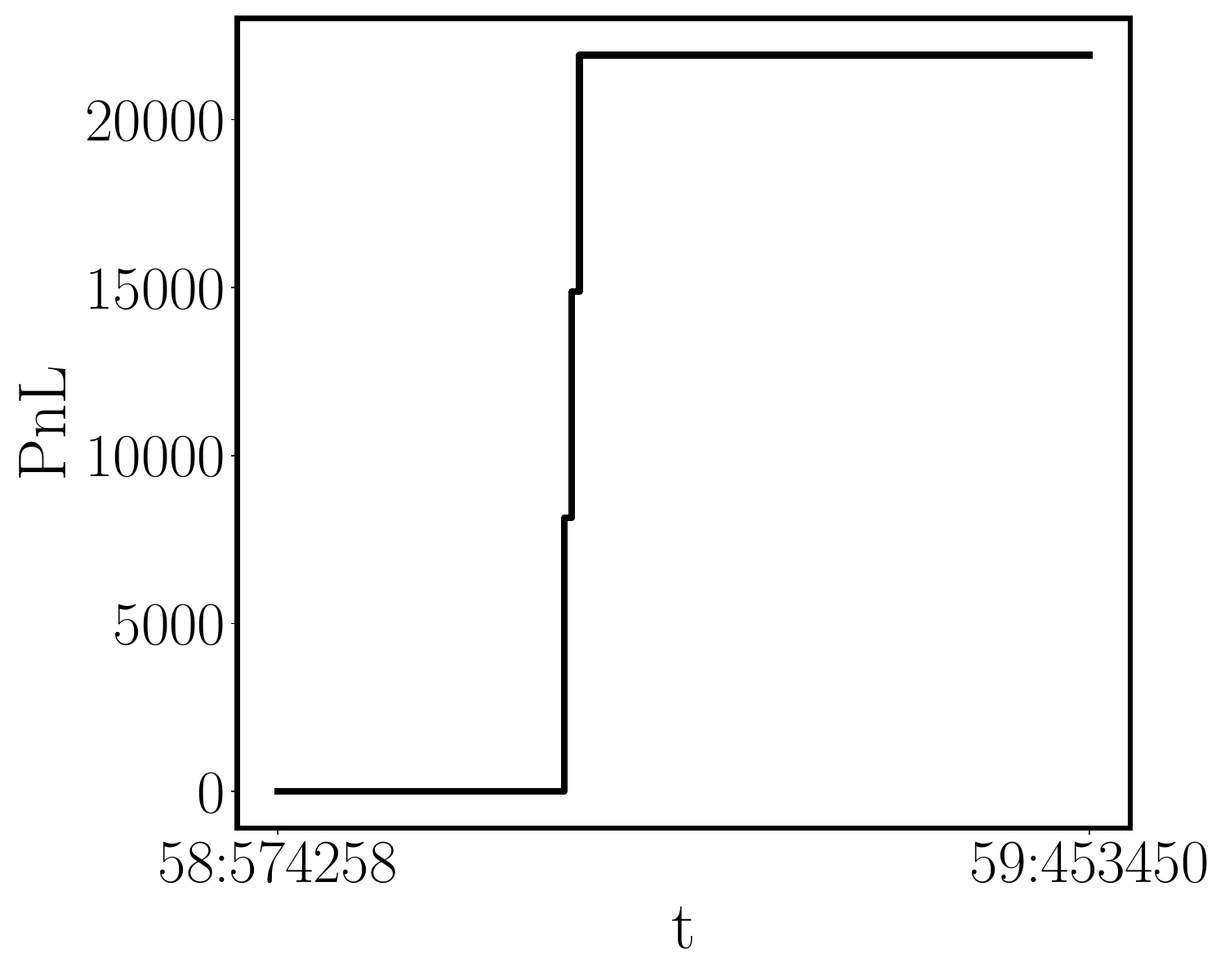}%
    }
    \caption{\textit{Suspicious orders} --- An example of ``suspicious'' behaviour identified by the model on the bid side of BTC-USD LOB. Note the logarithmic scale of the y-axis for the order sizes. The timestamps given in x-axis are given in second:microsecond format for information purpose. This anomaly was spotted on December 5$^{\text{th}}$ at 01:47:58 UTC.}
    \label{fig:bid_spoofing_btcusd_1}
\end{figure}

\begin{figure}
    \centering
    \subfloat[BBOs' dynamics and trade prices]{%
        \includegraphics[width=0.25\linewidth]{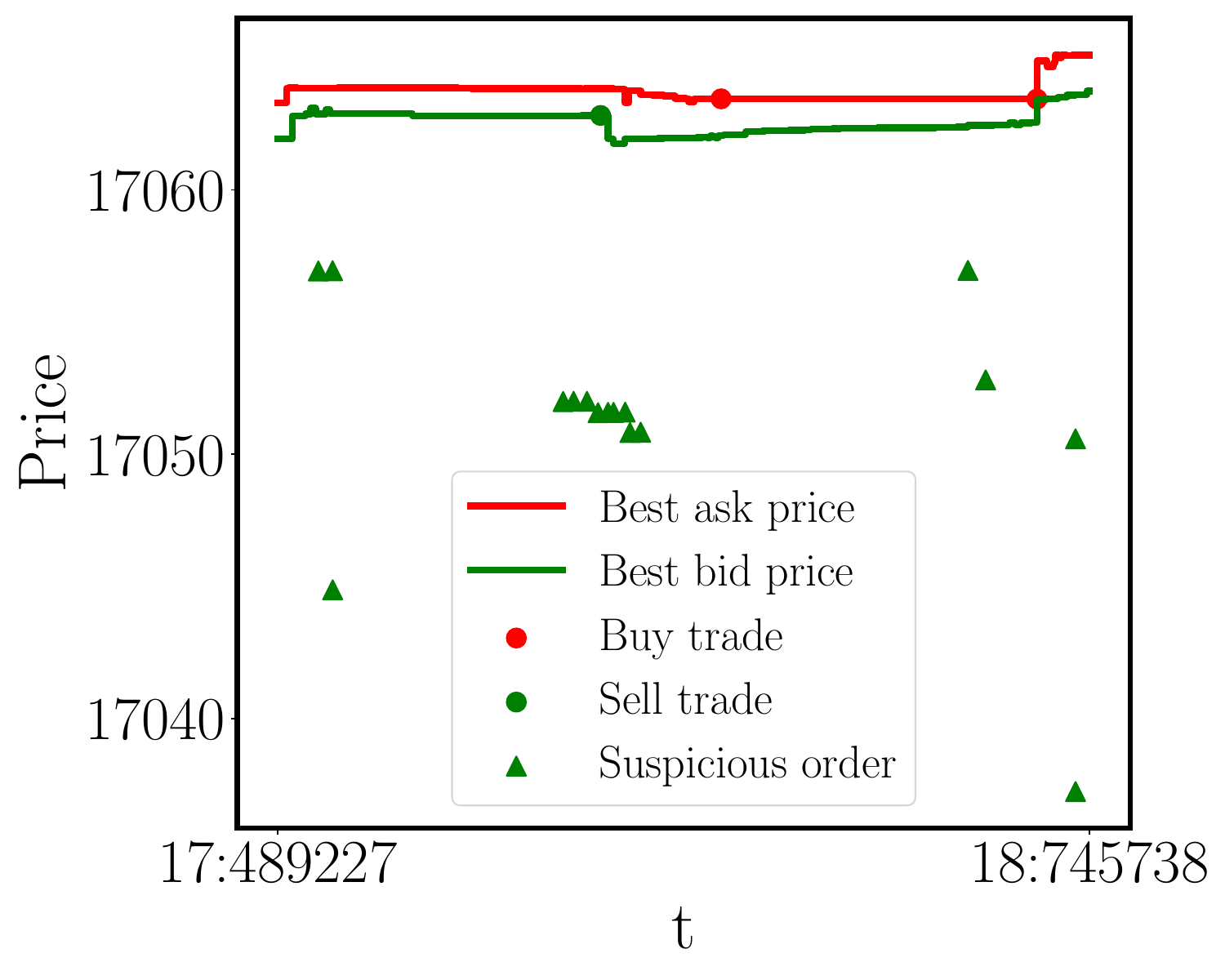}%
    }\hfill
    \subfloat[Order sizes (USD)]{%
        \includegraphics[width=0.25\linewidth]{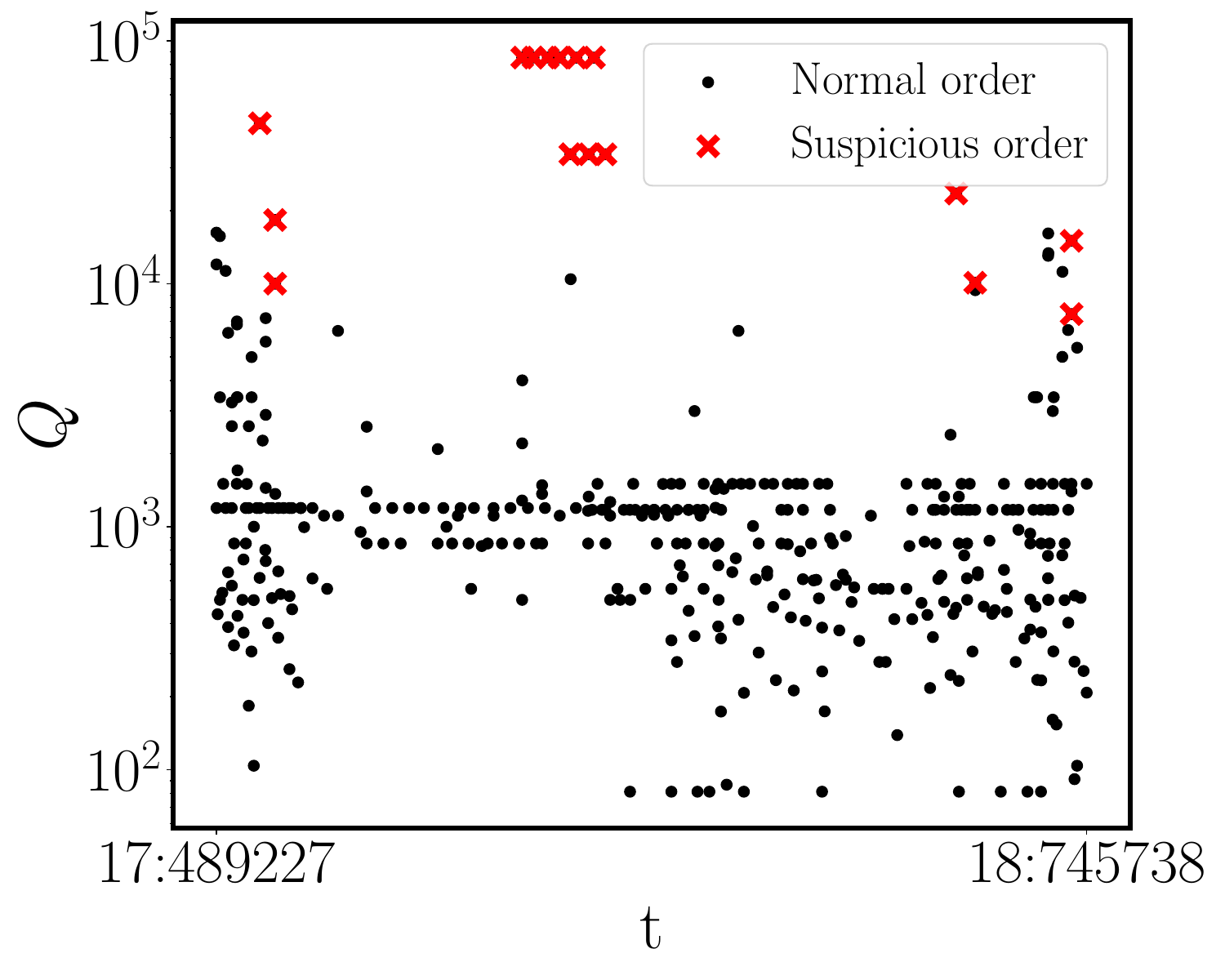}%
    }\hfill
    \subfloat[Order distances (bps)]{%
        \includegraphics[width=0.25\linewidth]{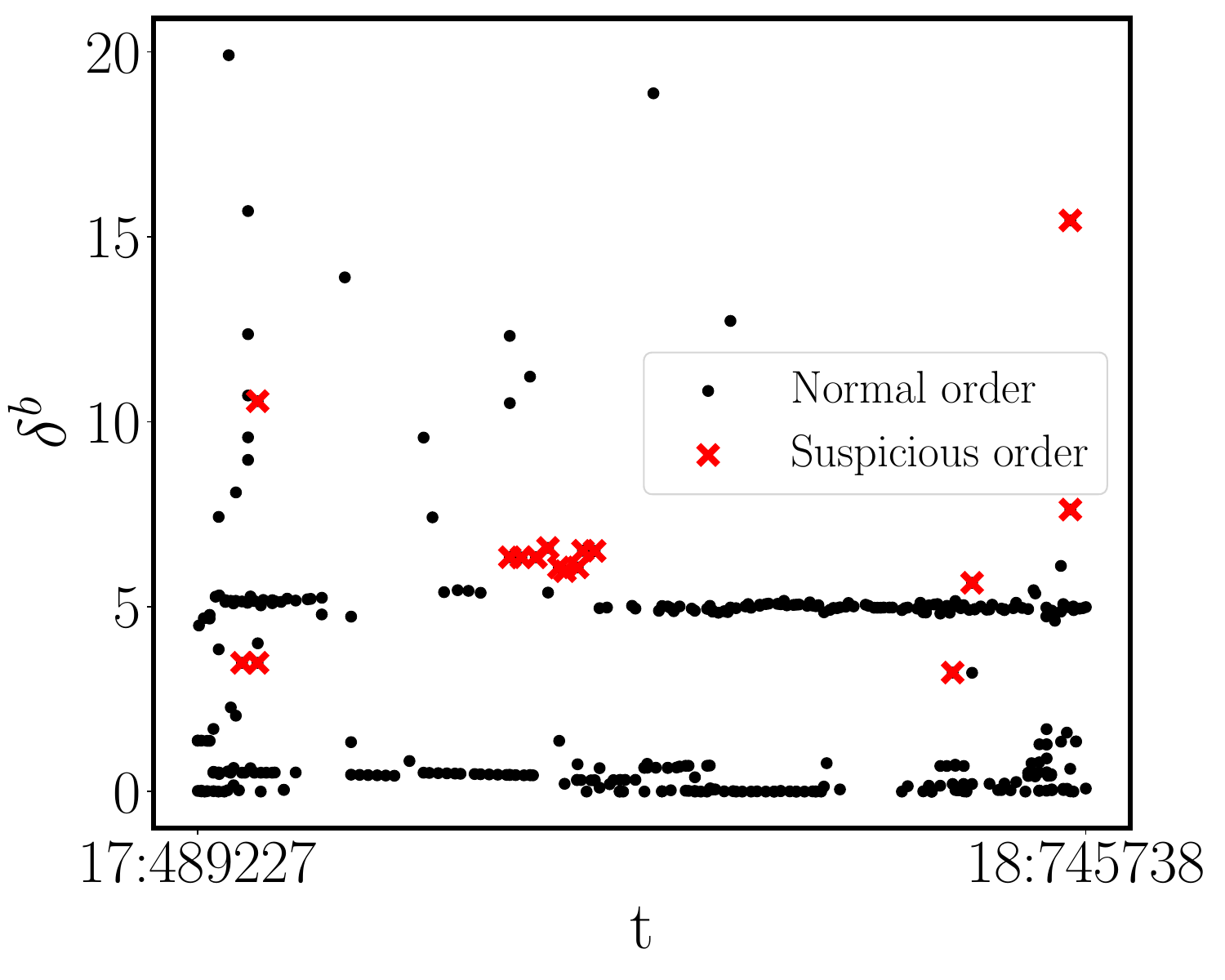}%
    }\hfill
    \subfloat[Spoofer's PnL (USD)]{%
        \includegraphics[width=0.25\linewidth]{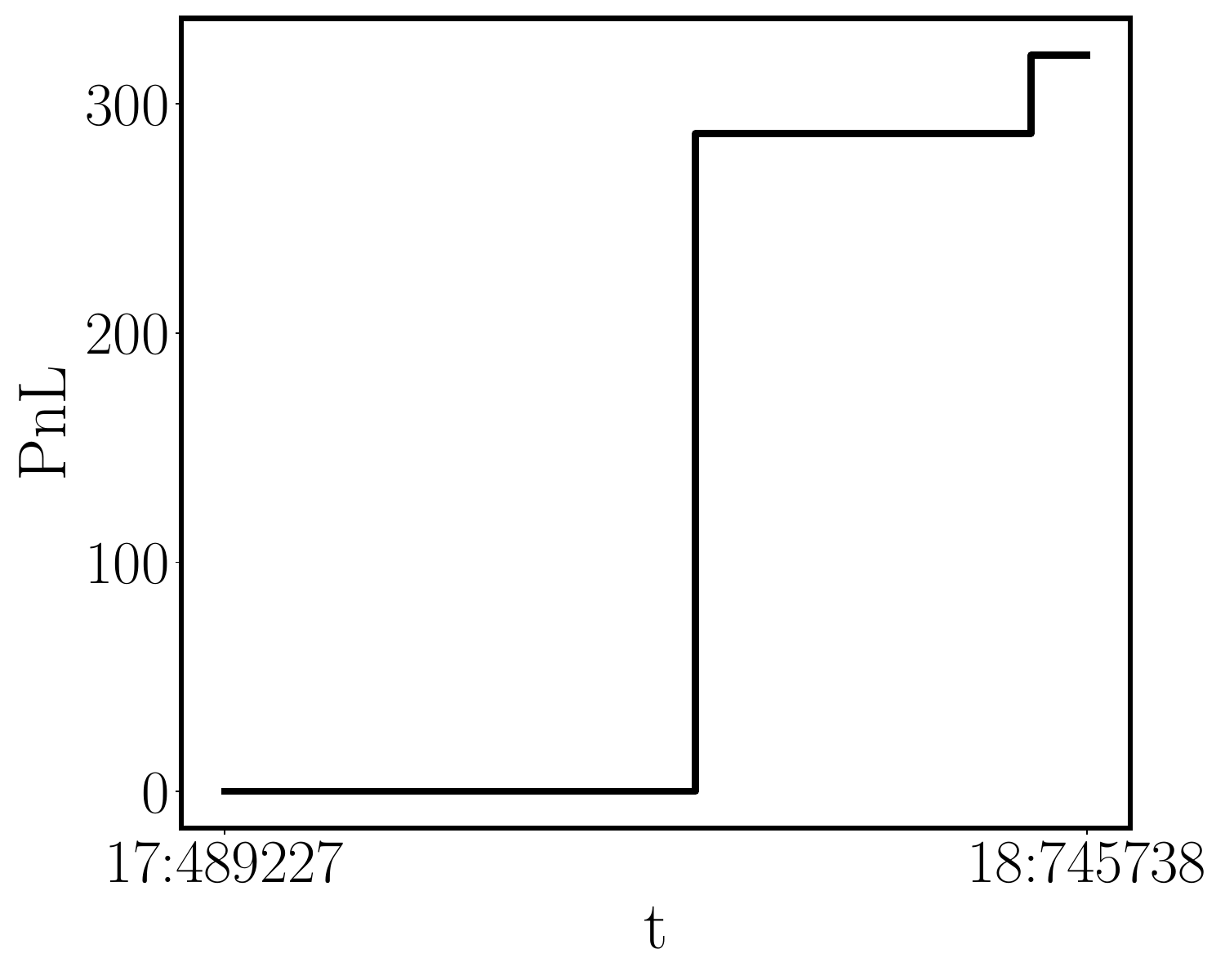}%
    }
    \caption{\textit{Suspicious orders} --- An example of ``suspicious'' behaviour identified by the model on the bid side of BTC-USD LOB. Note the logarithmic scale of the y-axis for the order sizes. The timestamps given in x-axis are given in second:microsecond format for information purposes. This anomaly was spotted on December 7$^{\text{th}}$ at 00:56:18 UTC.}
    \label{fig:bid_spoofing_btcusd_2}
\end{figure}

\begin{figure}
    \centering
    \subfloat[BBOs' dynamics and trade prices]{%
        \includegraphics[width=0.25\linewidth]{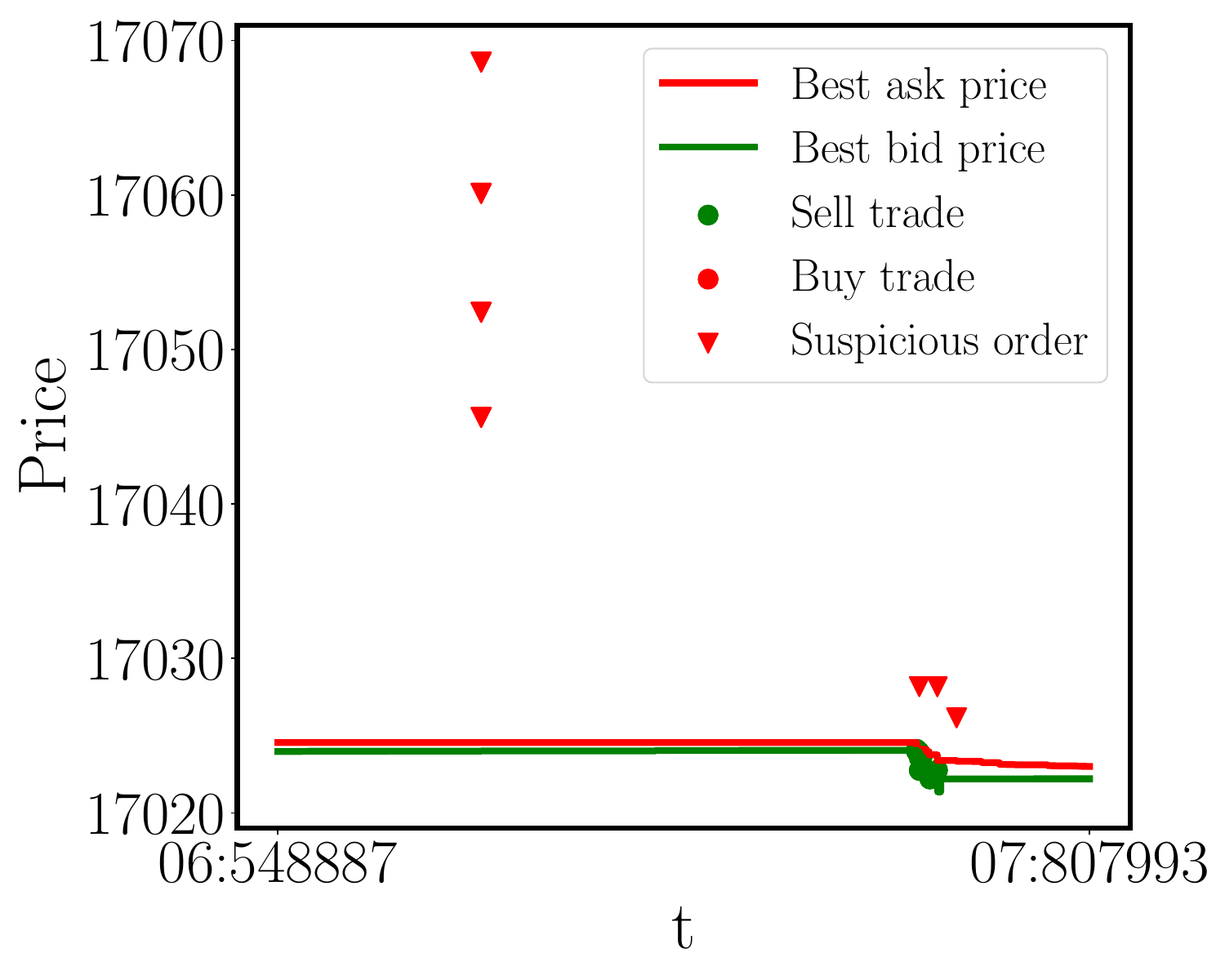}%
    }\hfill
    \subfloat[Order sizes (USD)]{%
        \includegraphics[width=0.25\linewidth]{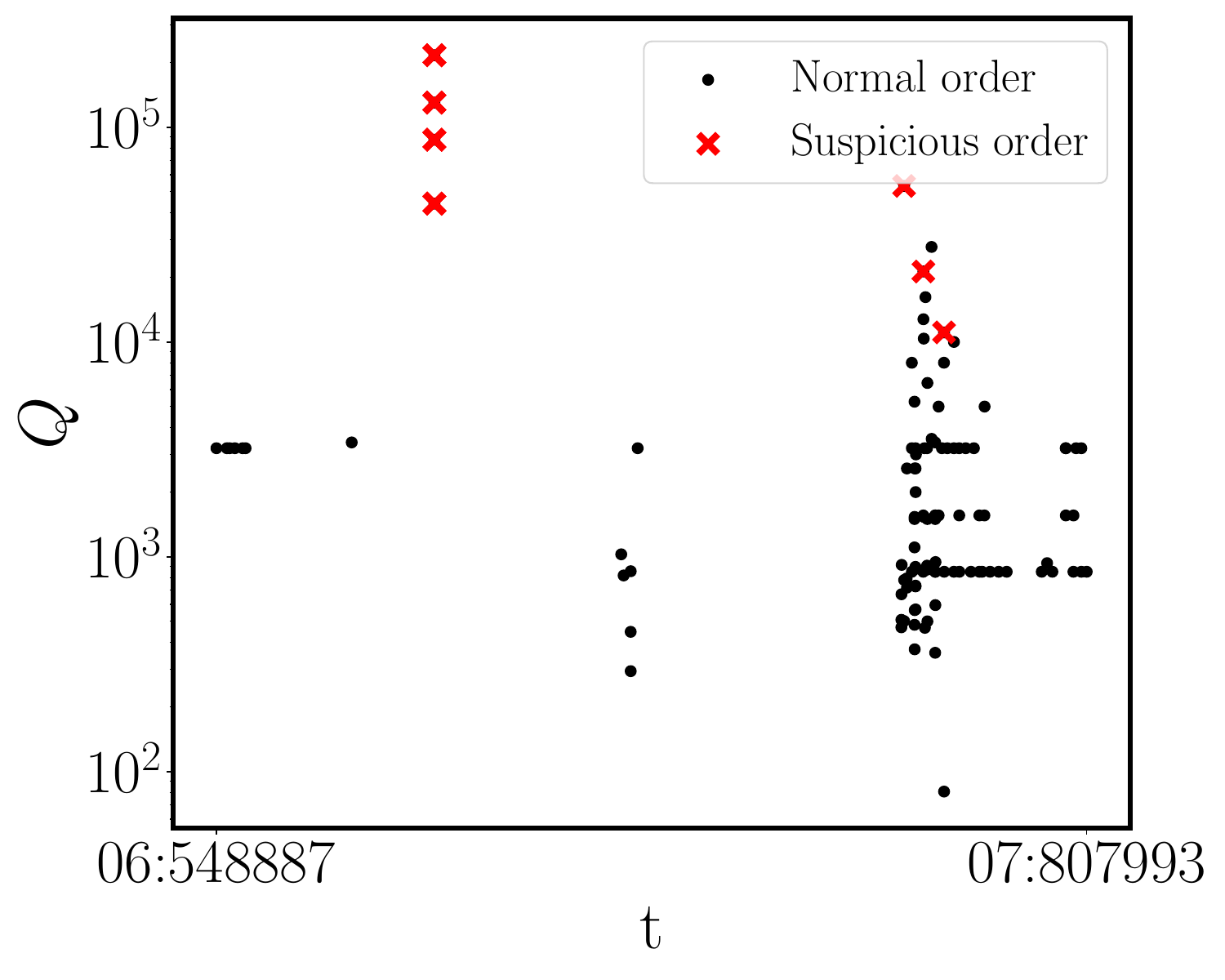}%
    }\hfill
    \subfloat[Order distances (bps)]{%
        \includegraphics[width=0.25\linewidth]{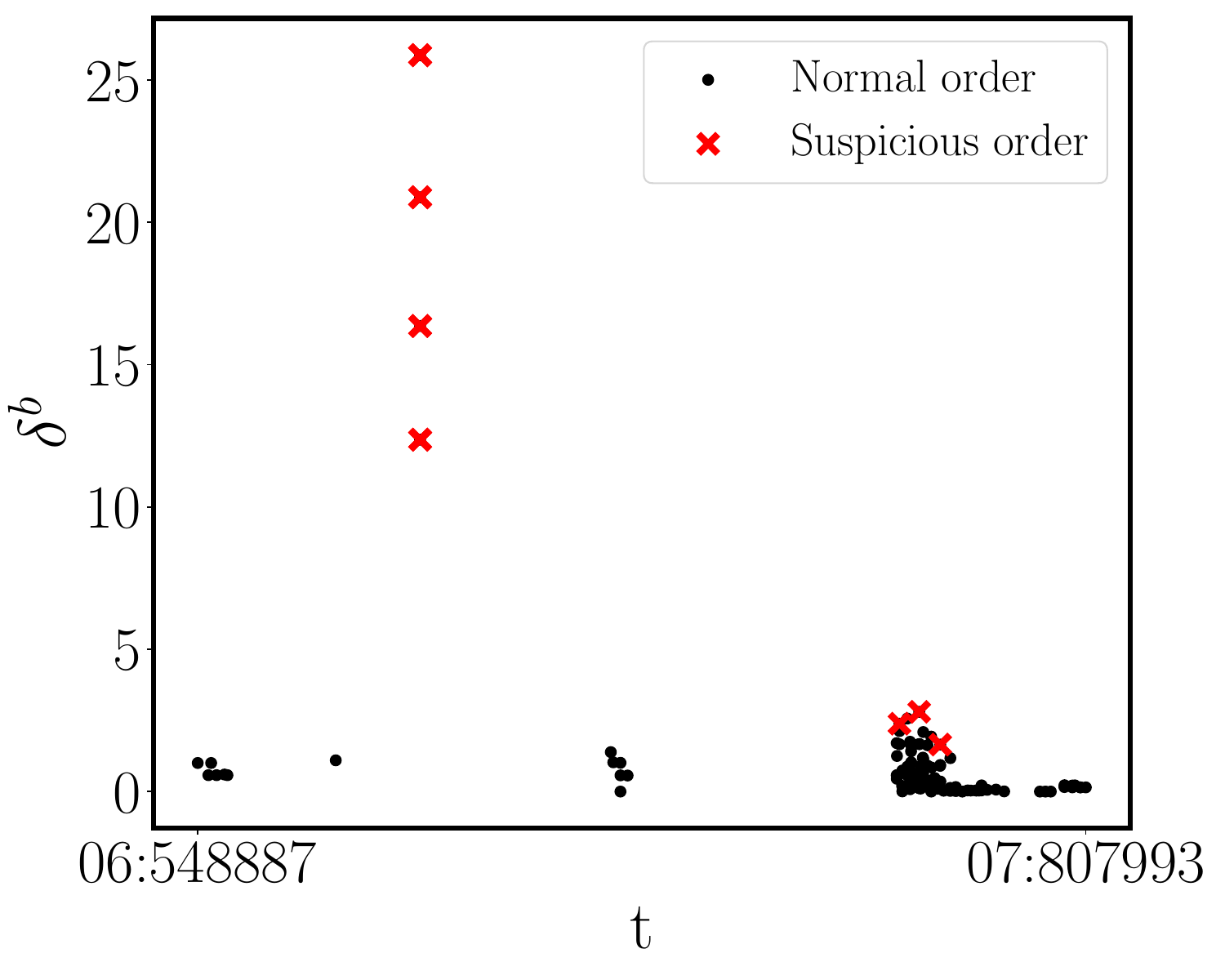}%
    }\hfill
    \subfloat[Spoofer's PnL (USD)]{%
        \includegraphics[width=0.25\linewidth]{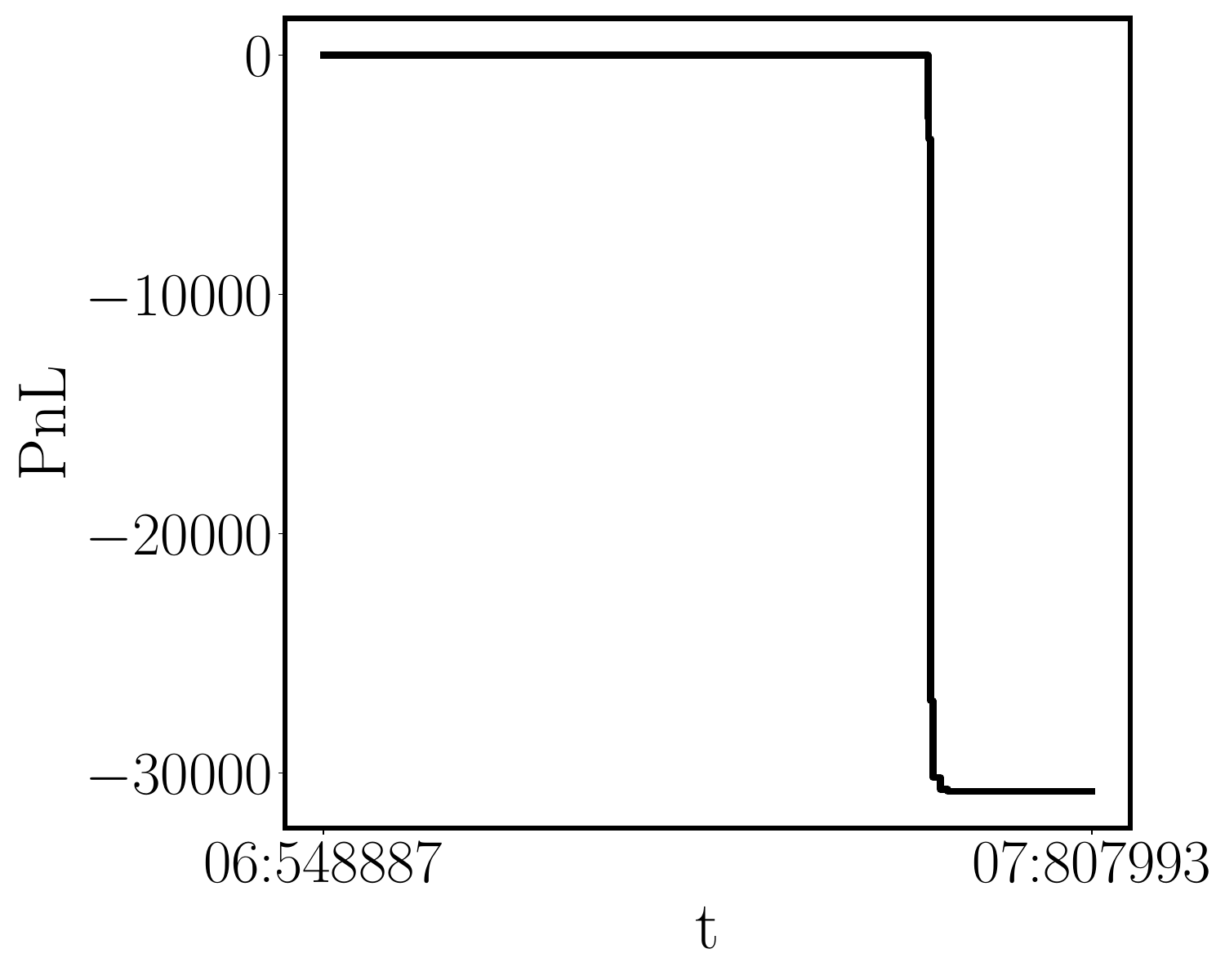}%
    }
    \caption{\textit{Suspicious orders} --- An example of ``suspicious'' behaviour identified by the model on the ask side of BTC-USD LOB. Note the logarithmic scale of the y-axis for the order sizes. The timestamps given in x-axis are given in second:microsecond format for information purposes. This anomaly was spotted on December 4$^{\text{th}}$ at 16:44:06 UTC.}
    \label{fig:ask_spoofing_btcusd_1}
\end{figure}

\begin{figure}
    \centering
    \subfloat[BBOs' dynamics and trade prices]{%
        \includegraphics[width=0.25\linewidth]{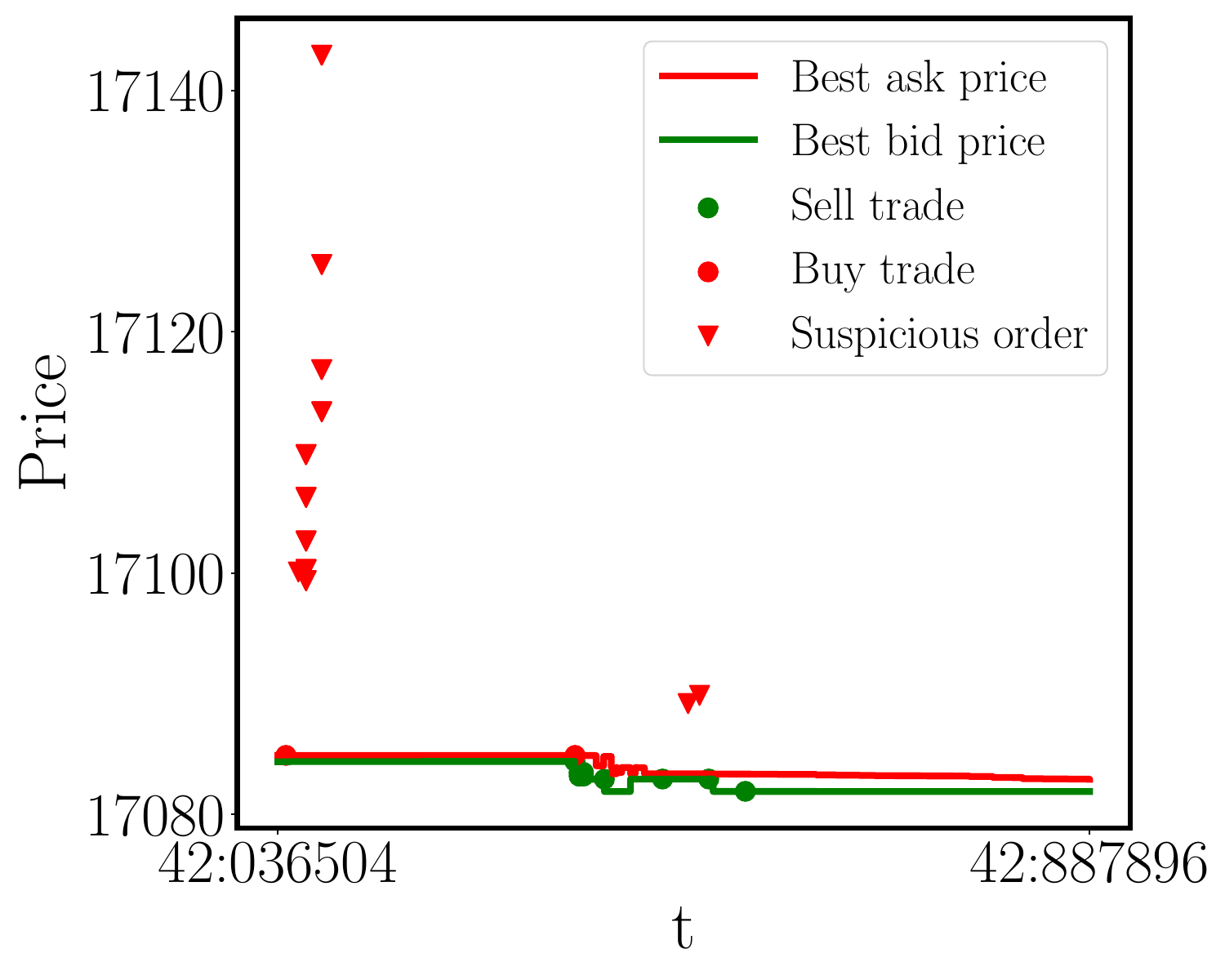}%
    }\hfill
    \subfloat[Order sizes (USD)]{%
        \includegraphics[width=0.25\linewidth]{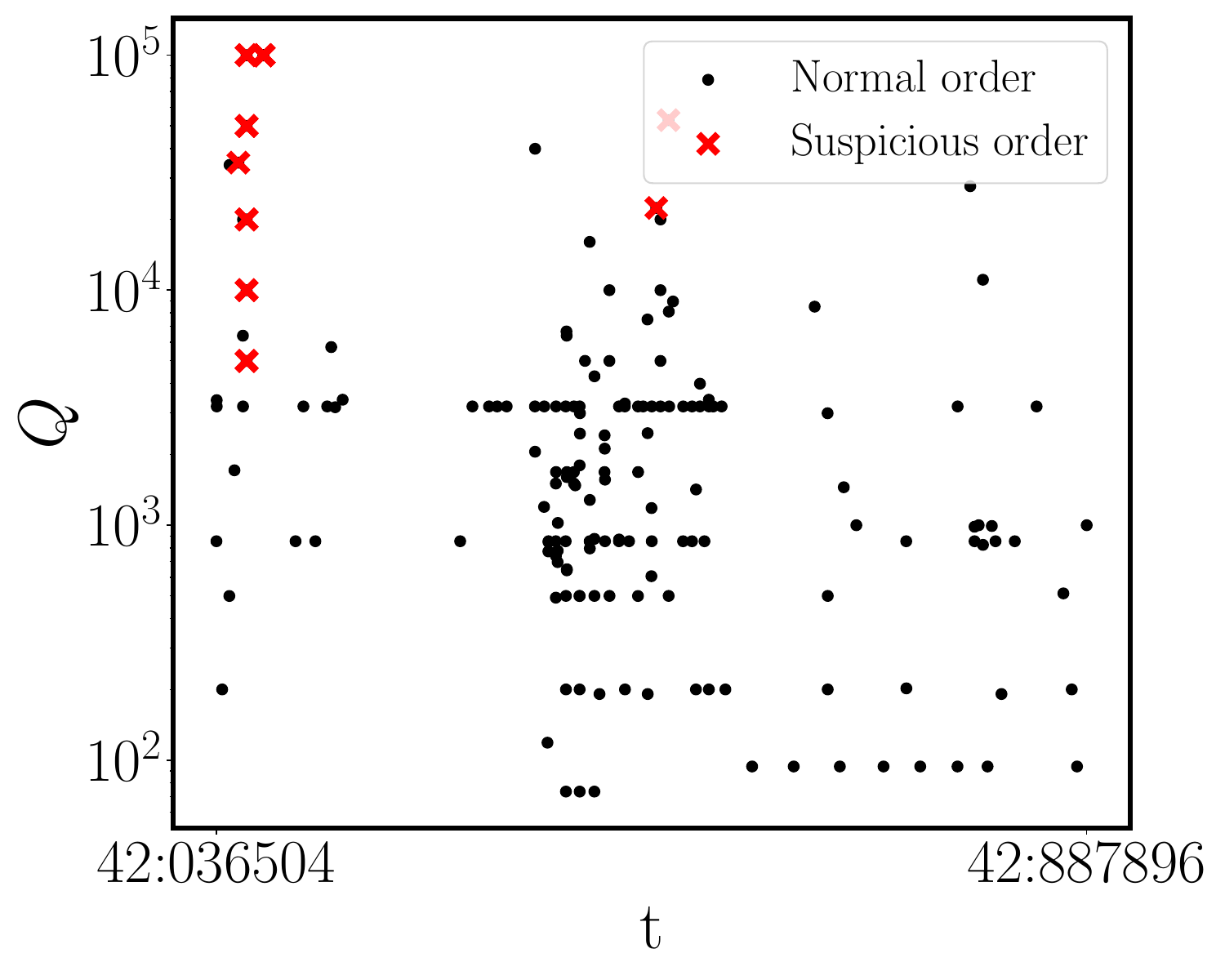}%
    }\hfill
    \subfloat[Order distances (bps)]{%
        \includegraphics[width=0.25\linewidth]{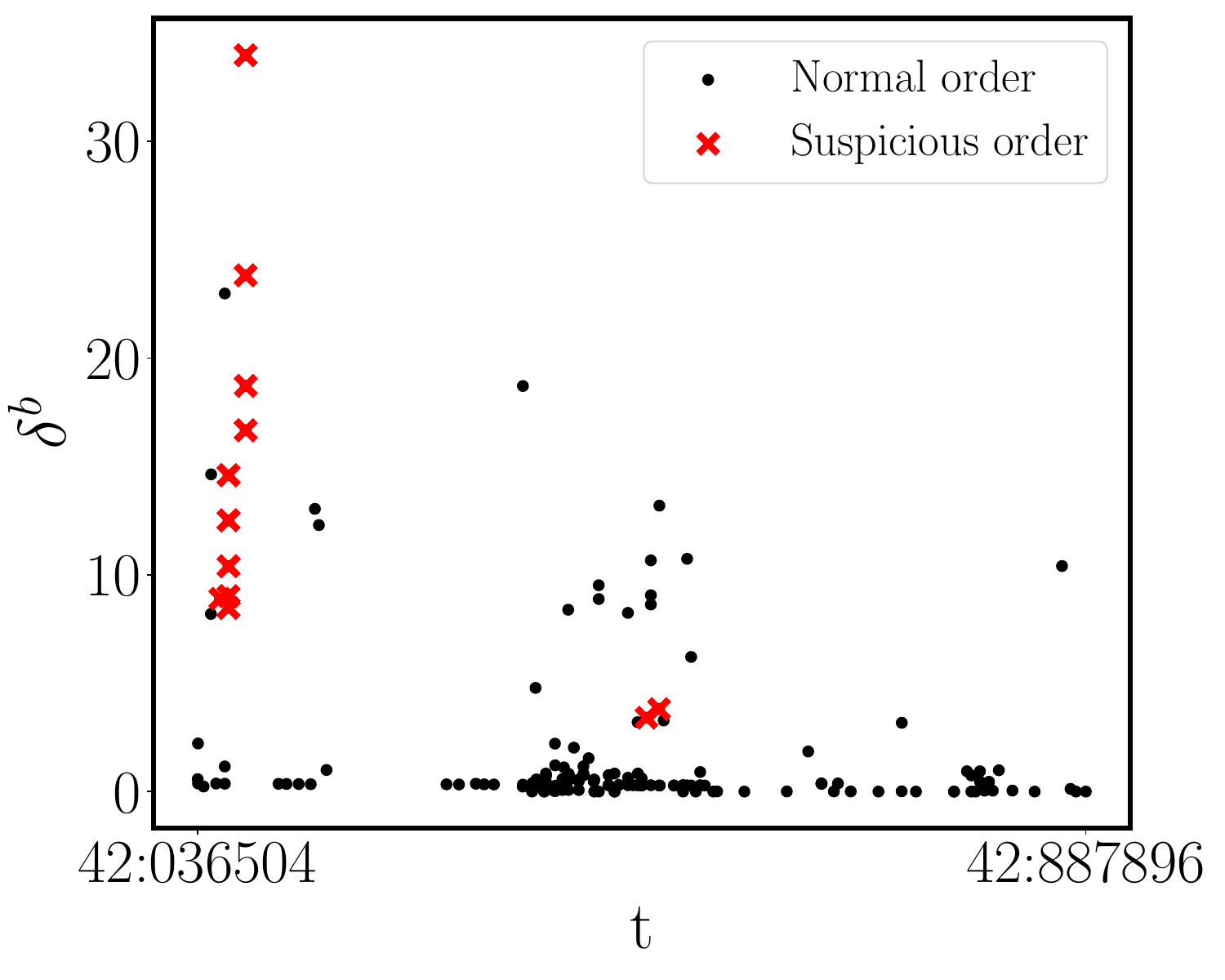}%
    }\hfill
    \subfloat[Spoofer's PnL (USD)]{%
        \includegraphics[width=0.25\linewidth]{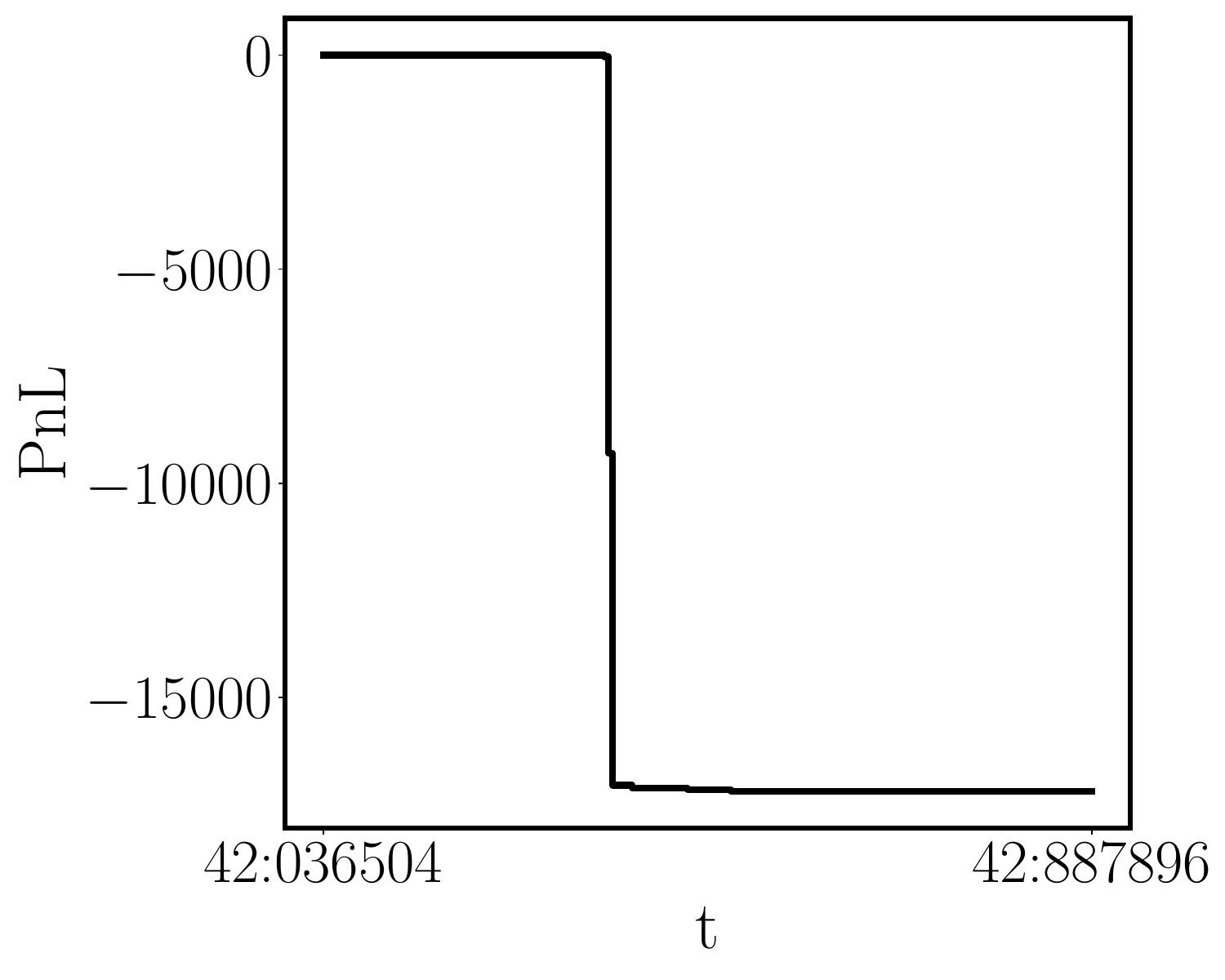}%
    }
    \caption{\textit{Suspicious orders} --- An example of ``suspicious'' behaviour identified by the model on the ask side of BTC-USD LOB. Note the logarithmic scale of the y-axis for the order sizes. The timestamps given in x-axis are given in second:microsecond format for information purposes. This anomaly was spotted on December 5$^{\text{th}}$ at 18:41:42 UTC.}
    \label{fig:ask_spoofing_btcusd_2}
\end{figure}

\begin{figure}
    \centering
    \subfloat[BBOs' dynamics and trade prices]{%
        \includegraphics[width=0.25\linewidth]{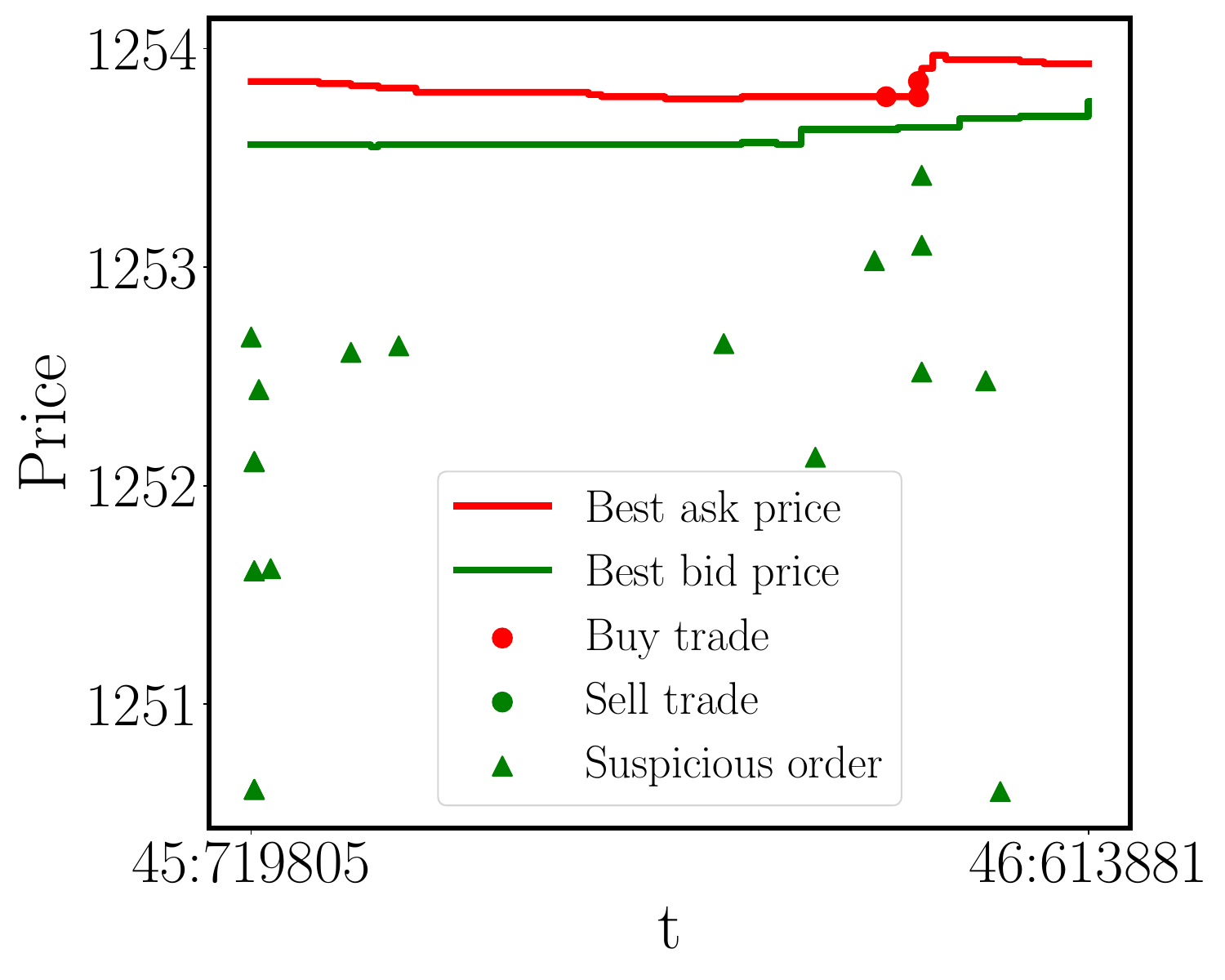}%
    }\hfill
    \subfloat[Order sizes (USD)]{%
        \includegraphics[width=0.25\linewidth]{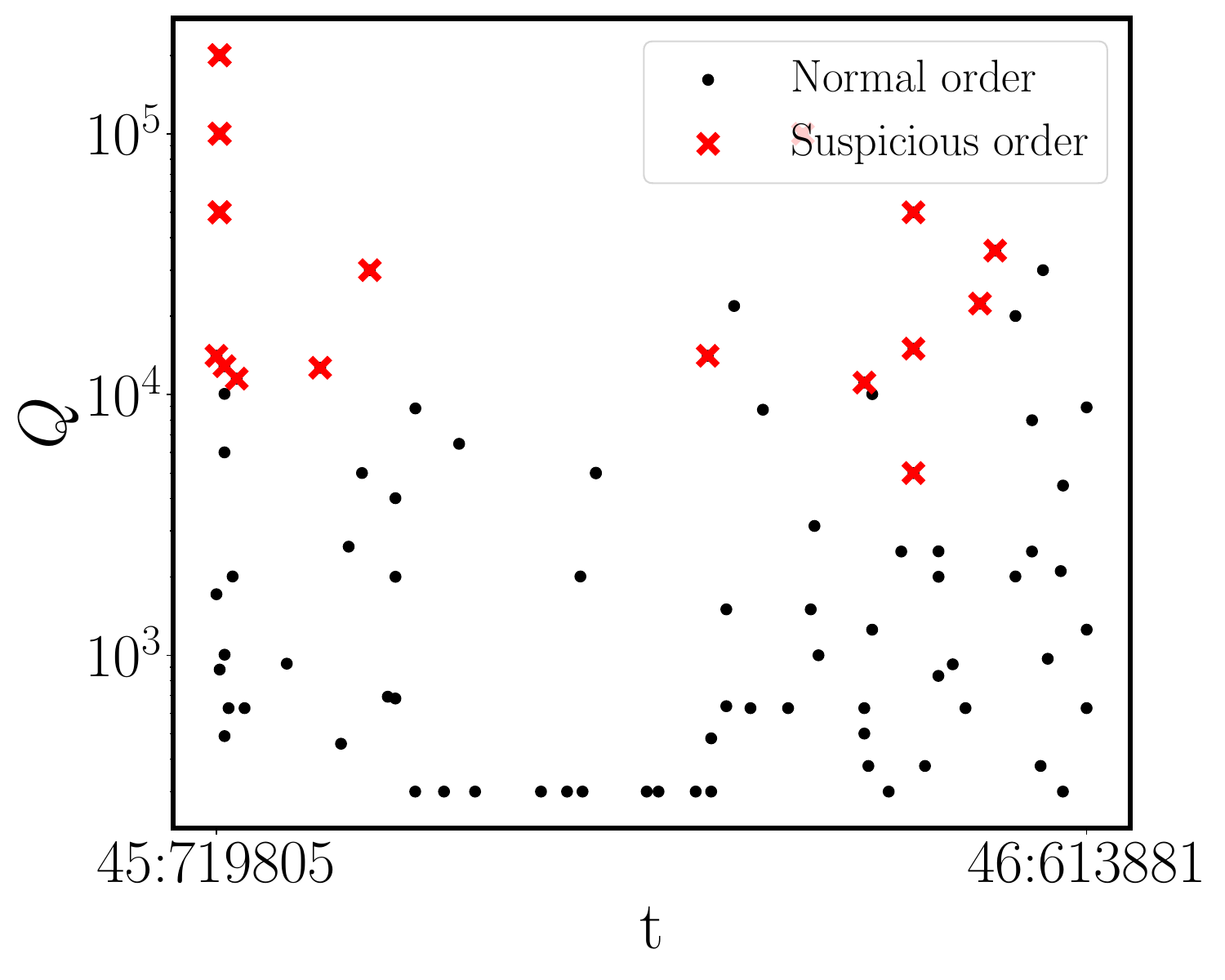}%
    }\hfill
    \subfloat[Order distances (bps)]{%
        \includegraphics[width=0.25\linewidth]{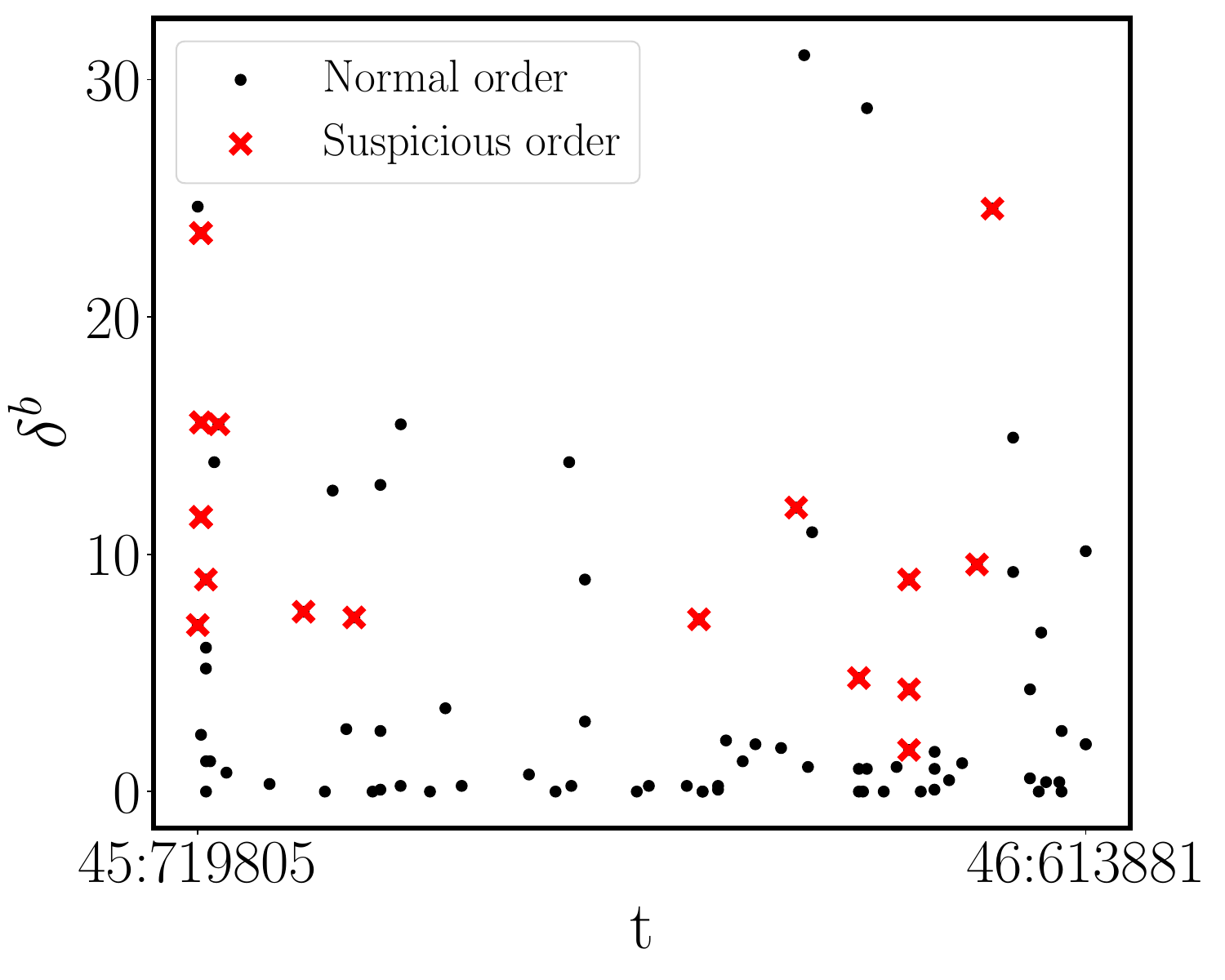}%
    }\hfill
    \subfloat[Spoofer's PnL (USD)]{%
        \includegraphics[width=0.25\linewidth]{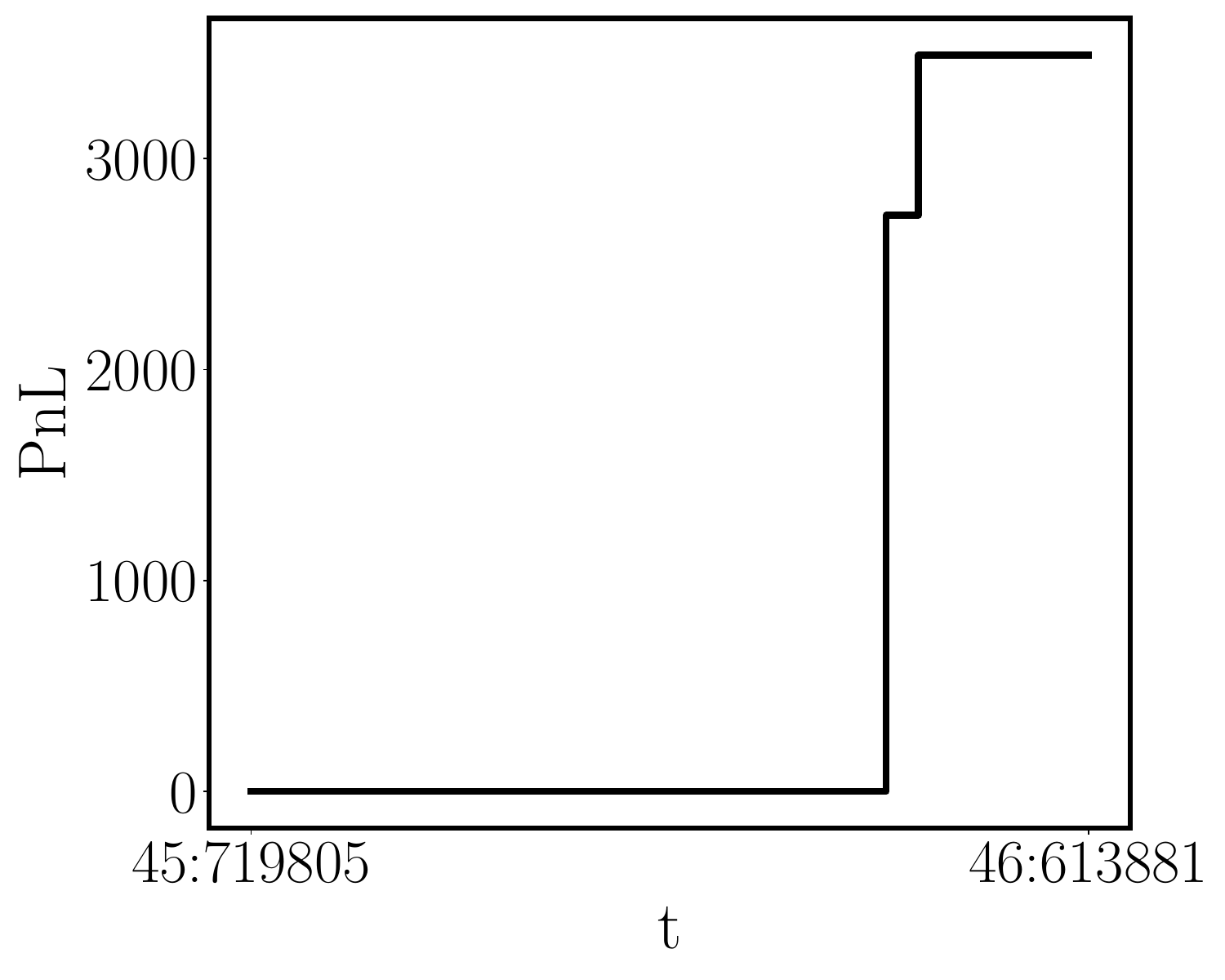}%
    }
    \caption{\textit{Suspicious orders} --- An example of ``suspicious'' behaviour identified by the model on the bid side of ETH-USD LOB. Note the logarithmic scale of the y-axis for the order sizes. The timestamps given in x-axis are given in second:microsecond format for information purpose. This anomaly was spotted on December 5$^{\text{th}}$ at 19:18:45 UTC.}
    \label{fig:bid_spoofing_ethusd_1}
\end{figure}

\begin{figure}
    \centering
    \subfloat[BBOs' dynamics and trade prices]{%
        \includegraphics[width=0.25\linewidth]{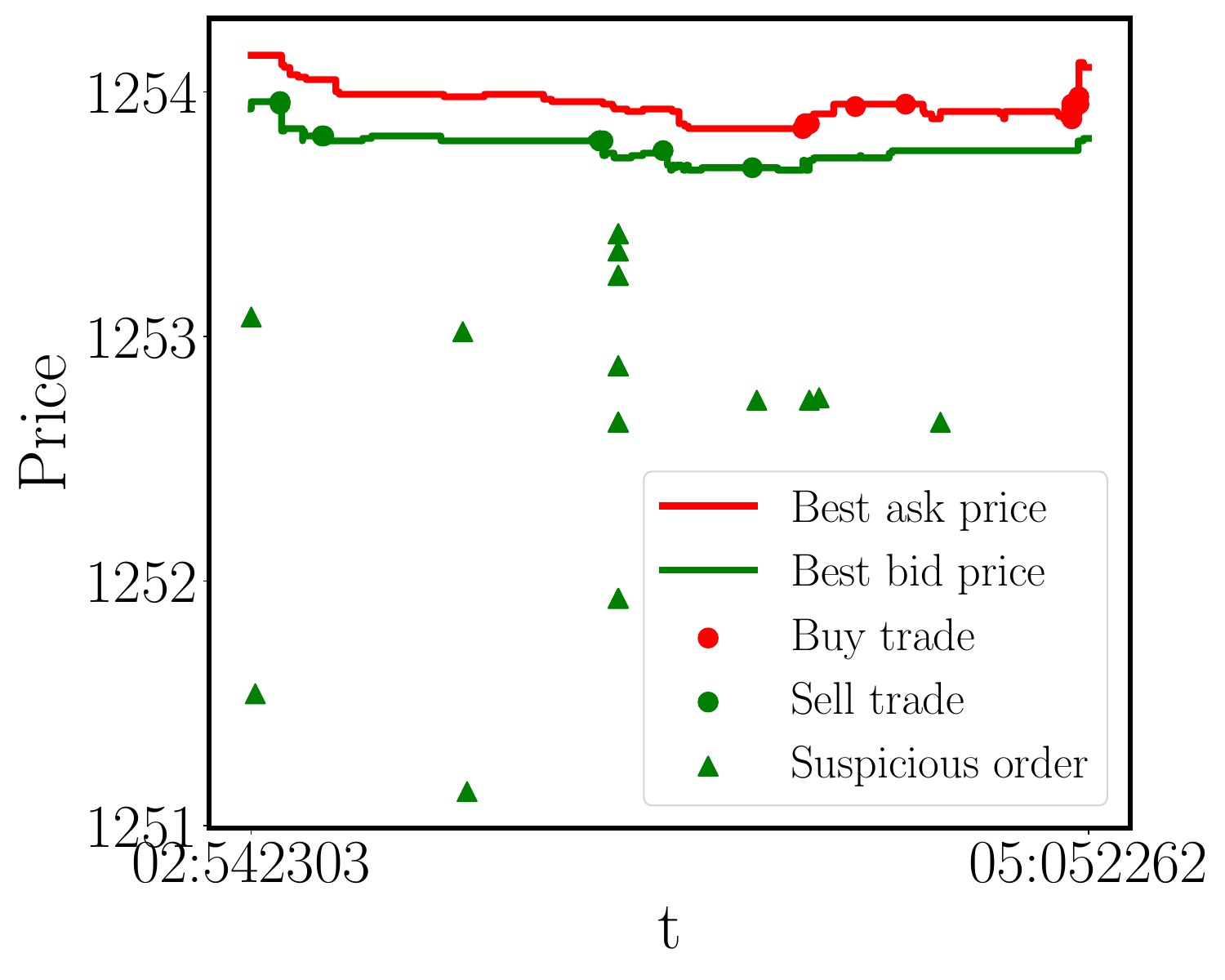}%
    }\hfill
    \subfloat[Order sizes (USD)]{%
        \includegraphics[width=0.25\linewidth]{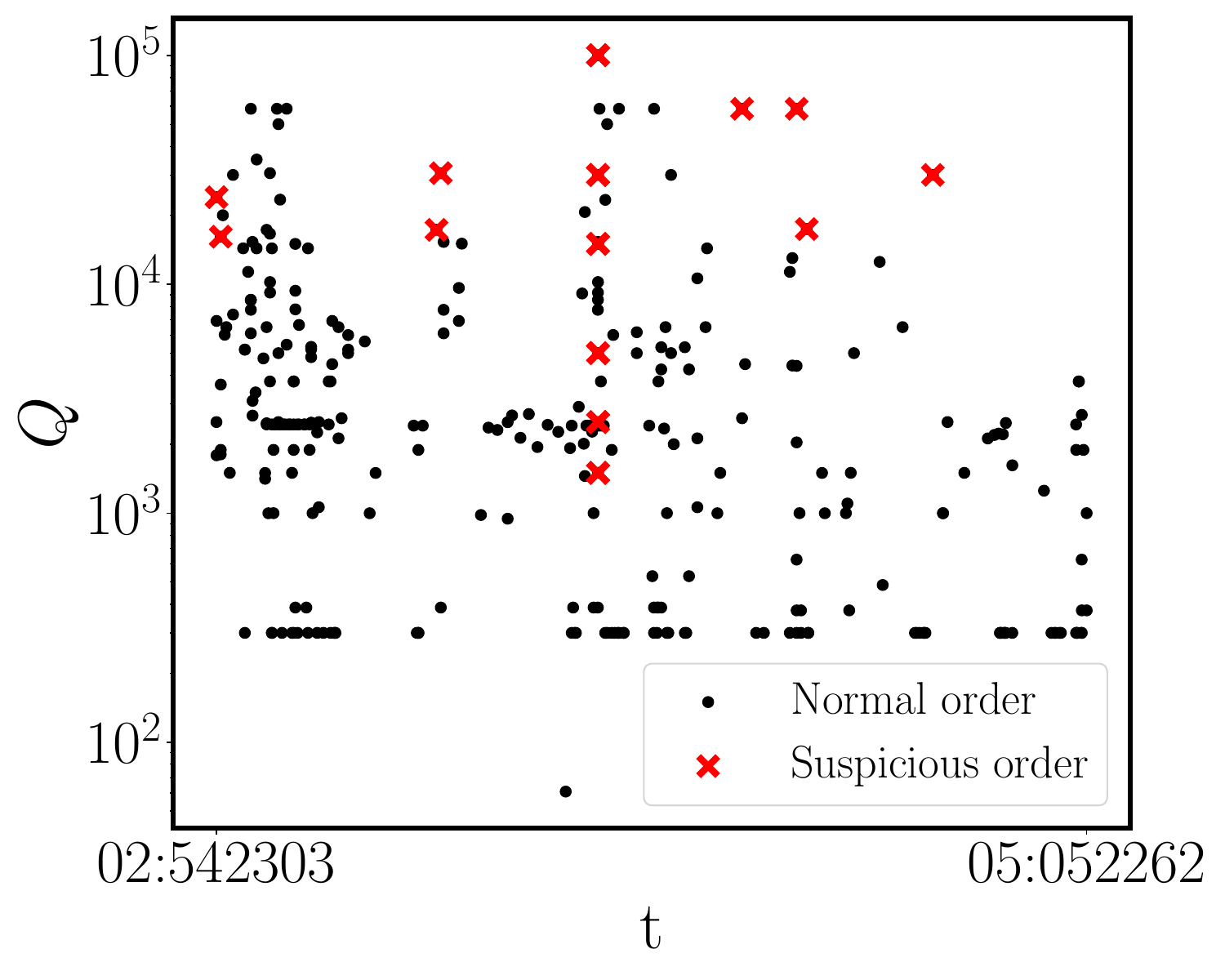}%
    }\hfill
    \subfloat[Order distances (bps)]{%
        \includegraphics[width=0.25\linewidth]{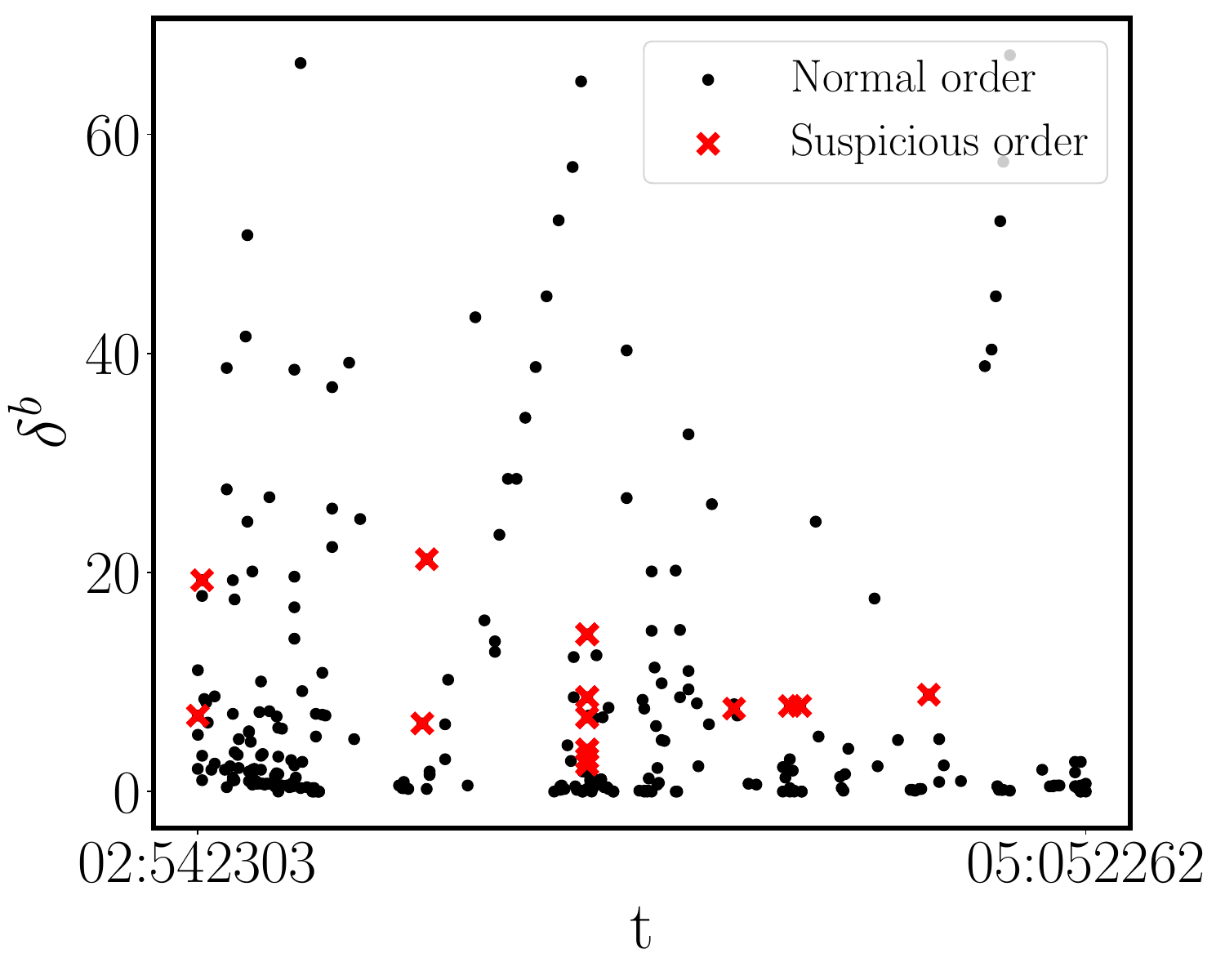}%
    }\hfill
    \subfloat[Spoofer's PnL (USD)]{%
        \includegraphics[width=0.25\linewidth]{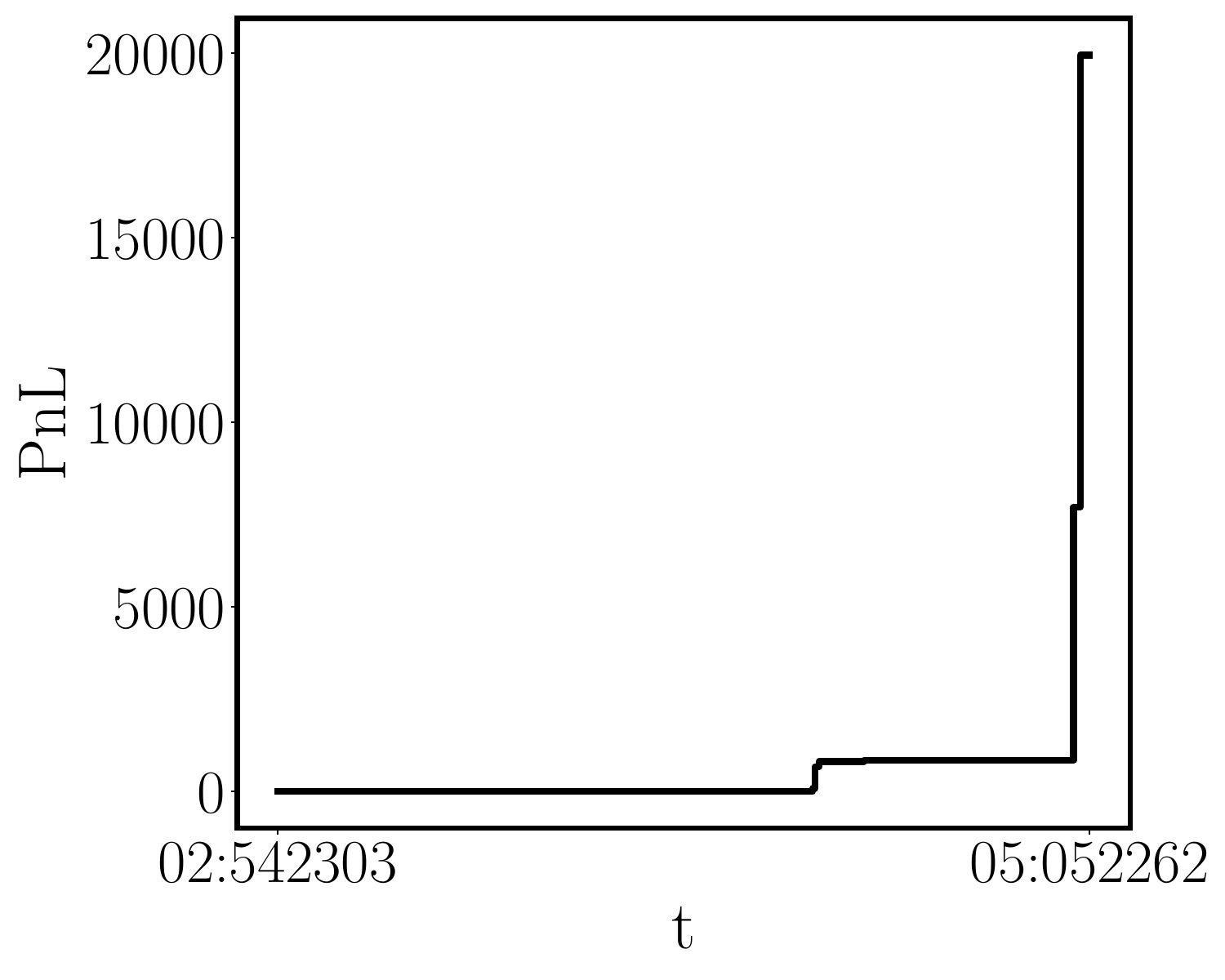}%
    }
    \caption{\textit{Suspicious orders} --- An example of ``suspicious'' behaviour identified by the model on the bid side of ETH-USD LOB. Note the logarithmic scale of the y-axis for the order sizes. The timestamps given in x-axis are given in second:microsecond format for information purposes. This anomaly was spotted on December 6$^{\text{th}}$ at 20:50:04 UTC.}
    \label{fig:bid_spoofing_ethusd_2}
\end{figure}

\begin{figure}
    \centering
    \subfloat[BBOs' dynamics and trade prices]{%
        \includegraphics[width=0.25\linewidth]{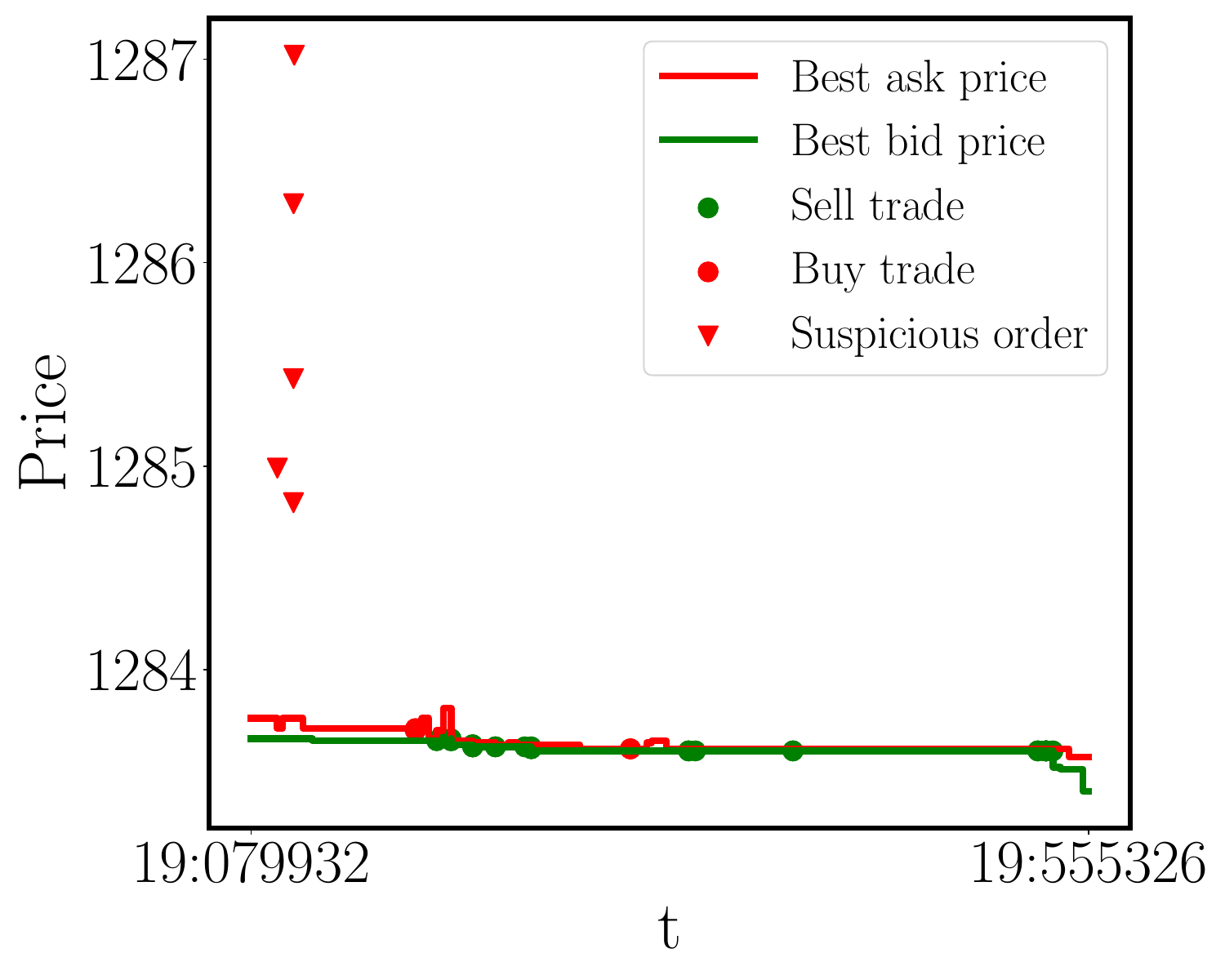}%
    }\hfill
    \subfloat[Order sizes (USD)]{%
        \includegraphics[width=0.25\linewidth]{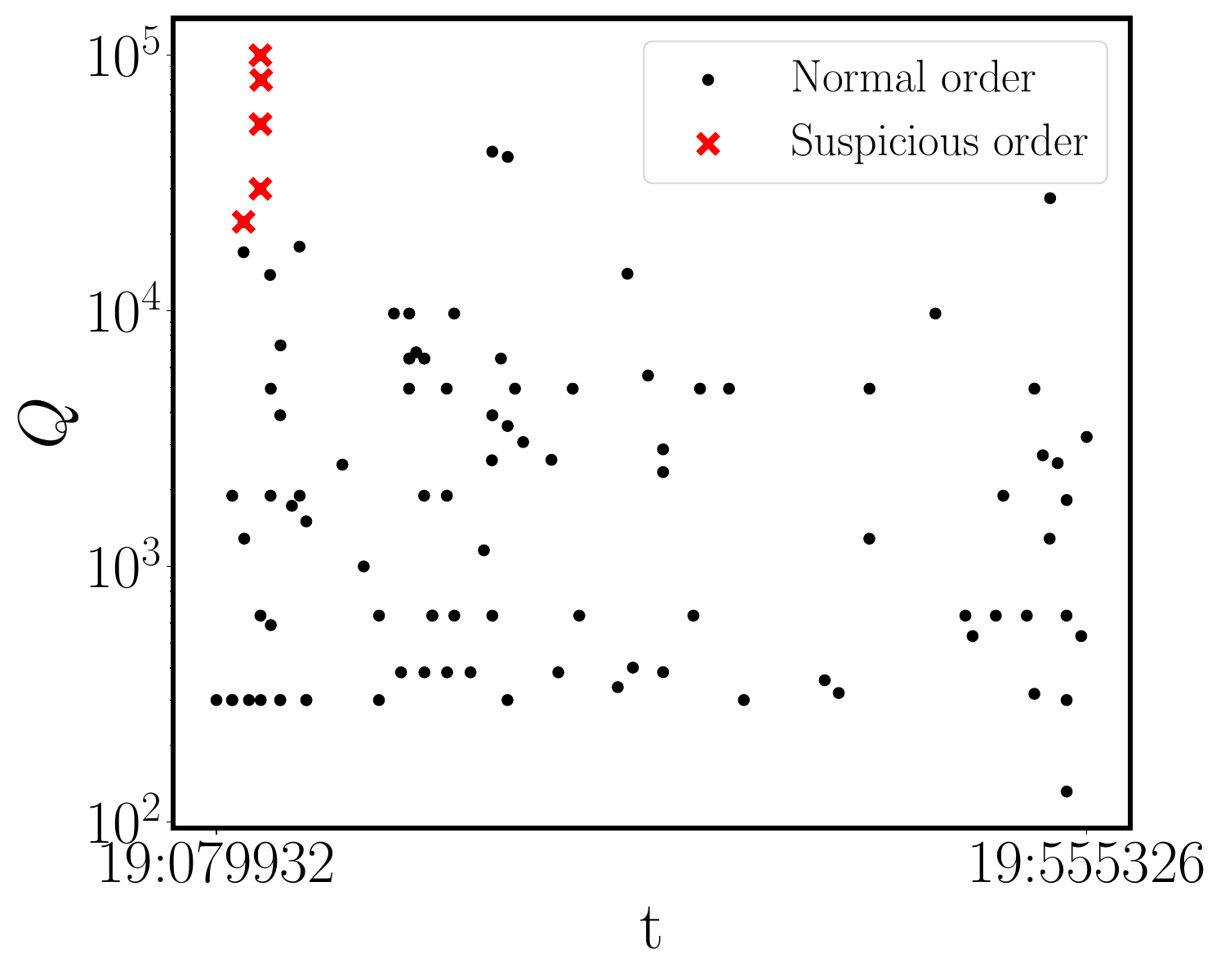}%
    }\hfill
    \subfloat[Order distances (bps)]{%
        \includegraphics[width=0.25\linewidth]{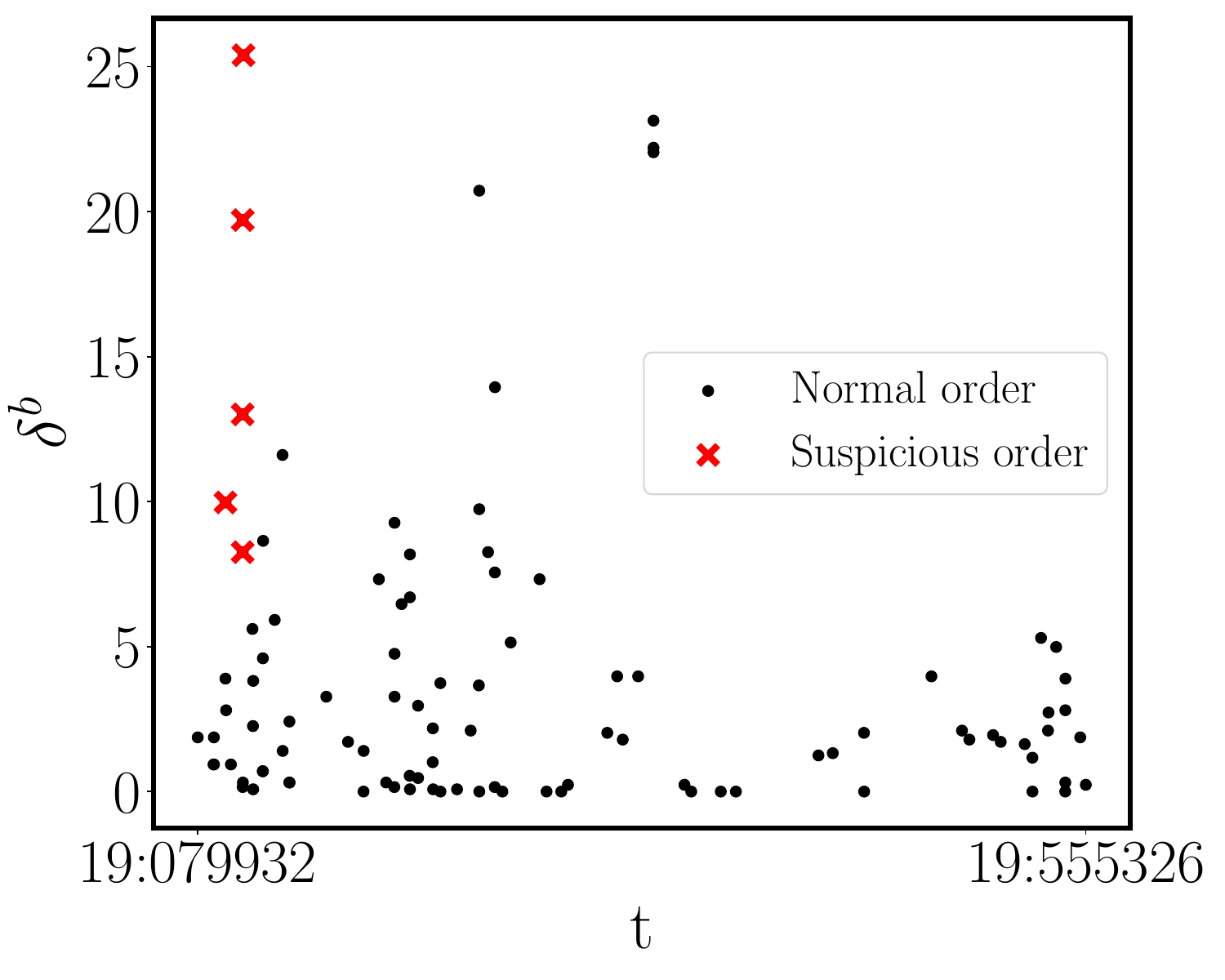}%
    }\hfill
    \subfloat[Spoofer's PnL (USD)]{%
        \includegraphics[width=0.25\linewidth]{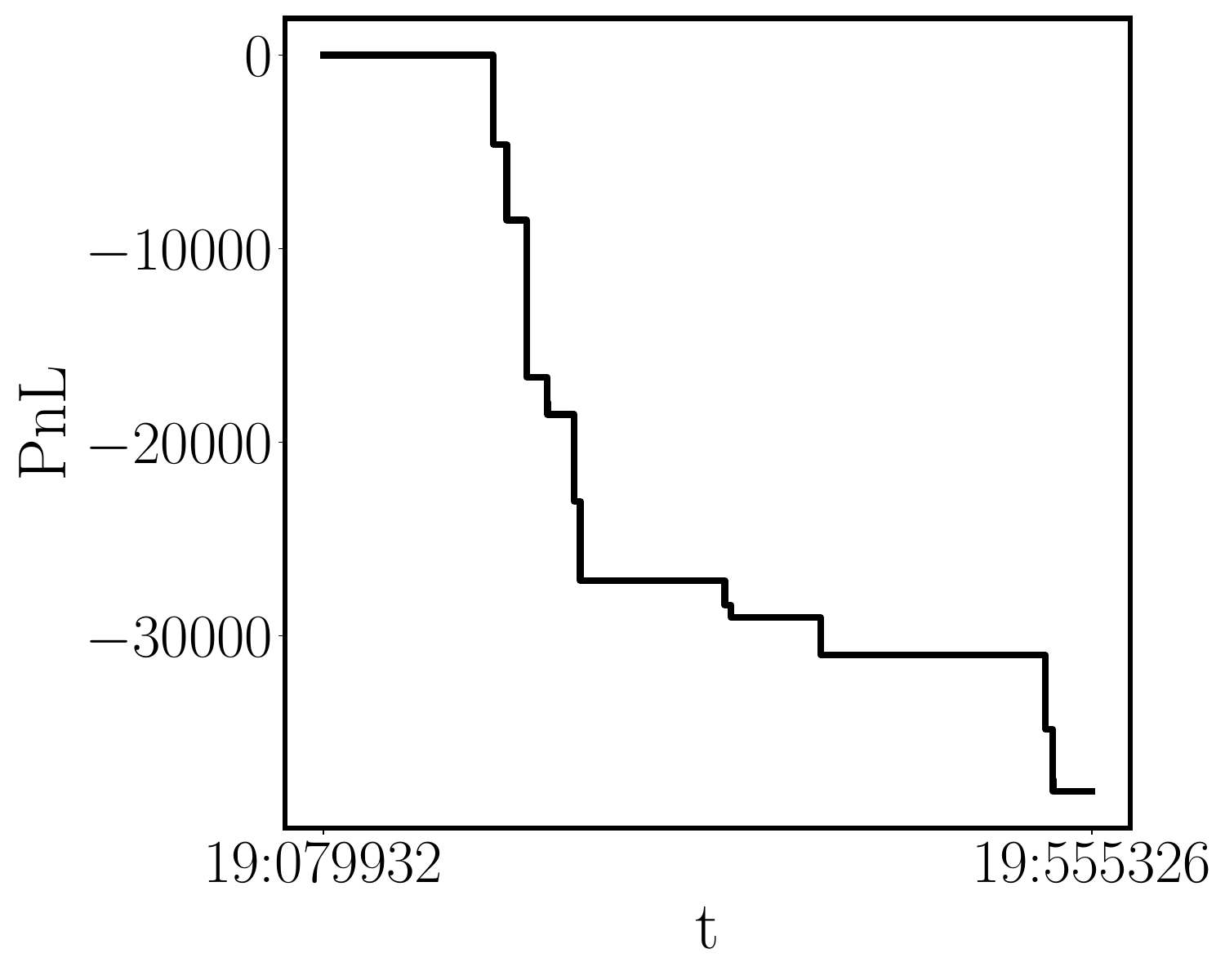}%
    }
    \caption{\textit{Suspicious orders} --- An example of ``suspicious'' behaviour identified by the model on the ask side of ETH-USD LOB. Note the logarithmic scale of the y-axis for the order sizes. The timestamps given in x-axis are given in second:microsecond format for information purposes. This anomaly was spotted on December 5$^{\text{th}}$ at 00:30:19 UTC.}
    \label{fig:ask_spoofing_ethusd_1}
\end{figure}

\begin{figure}
    \centering
    \subfloat[BBOs' dynamics and trade prices]{%
        \includegraphics[width=0.25\linewidth]{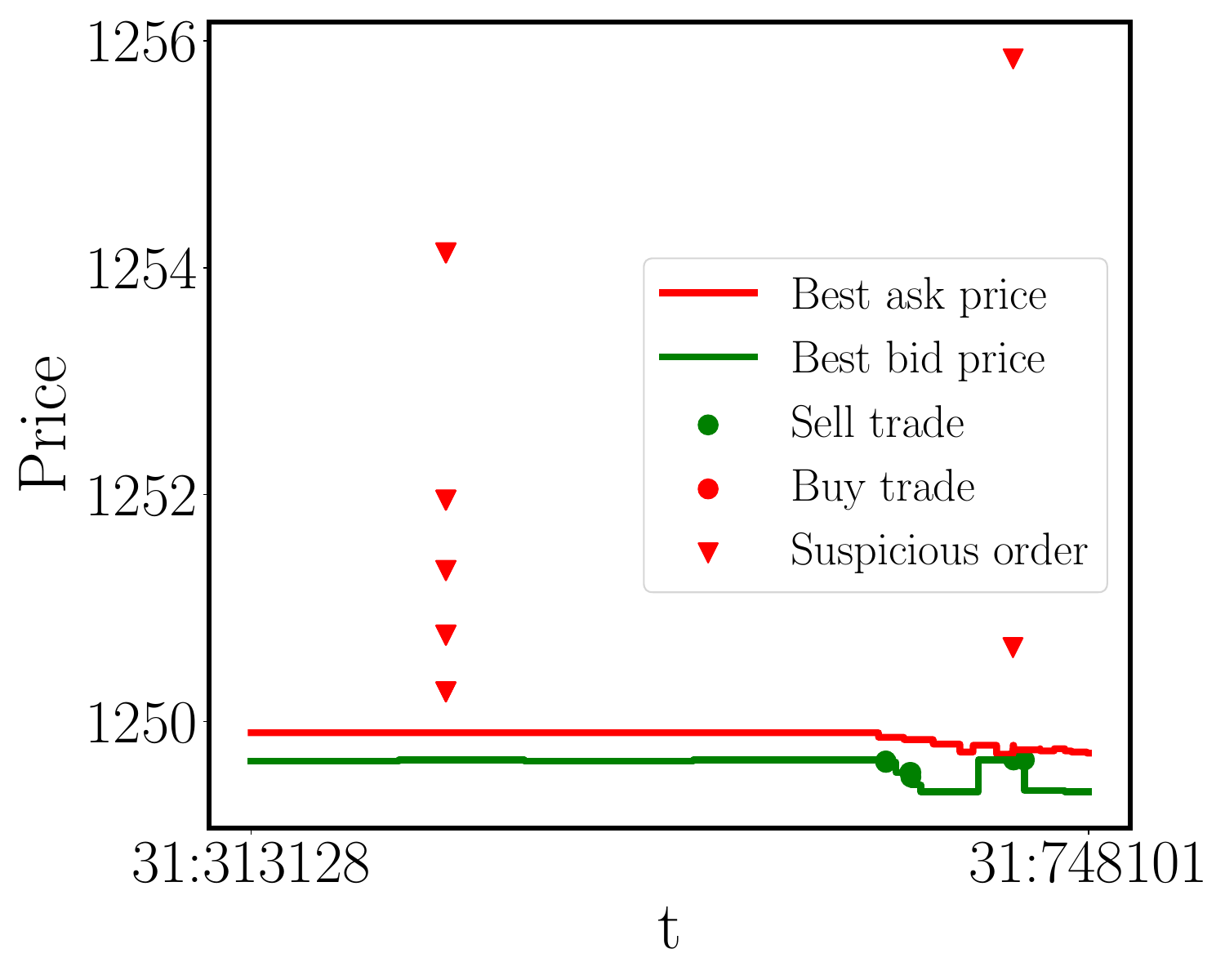}%
    }\hfill
    \subfloat[Order sizes (USD)]{%
        \includegraphics[width=0.25\linewidth]{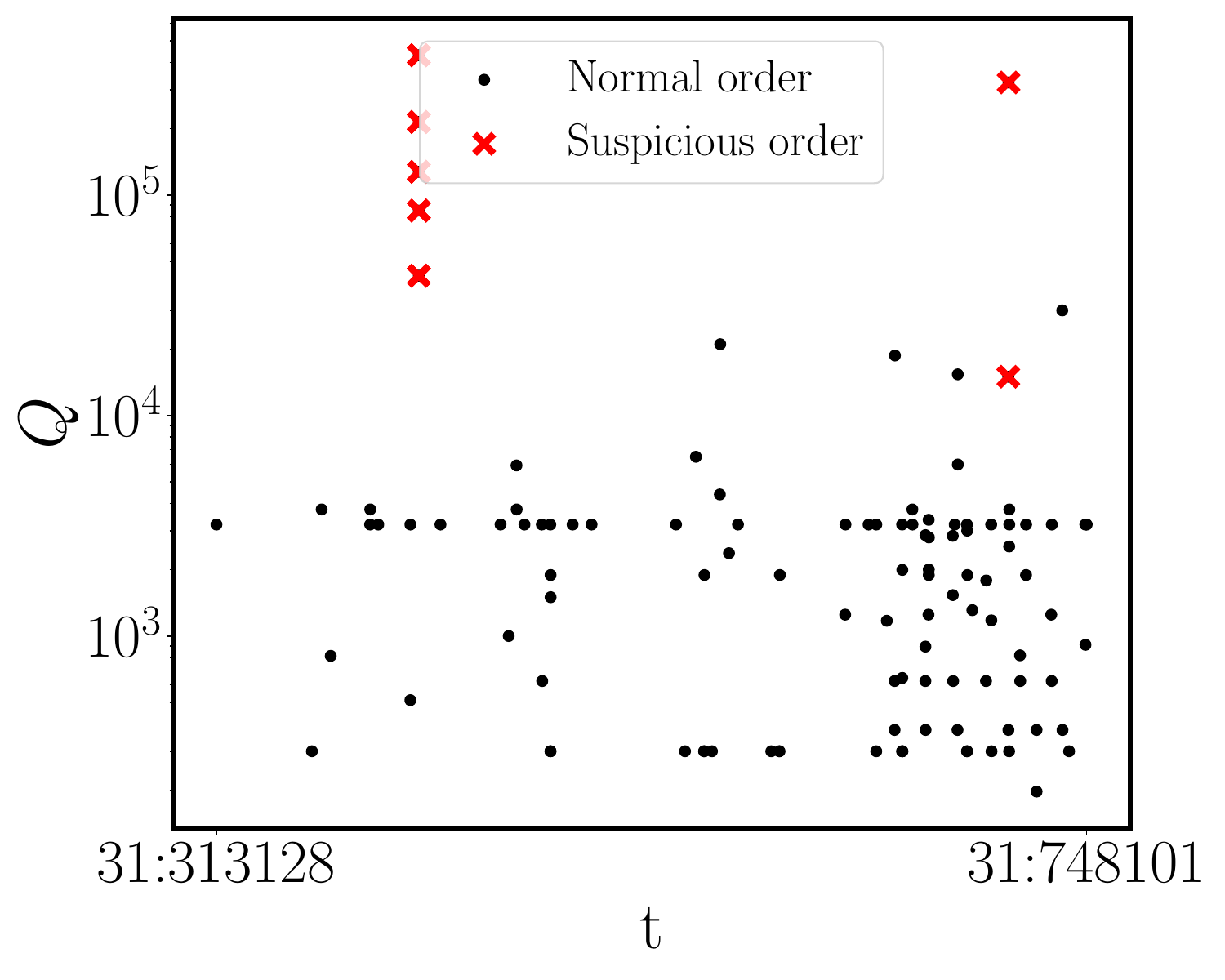}%
    }\hfill
    \subfloat[Order distances (bps)]{%
        \includegraphics[width=0.25\linewidth]{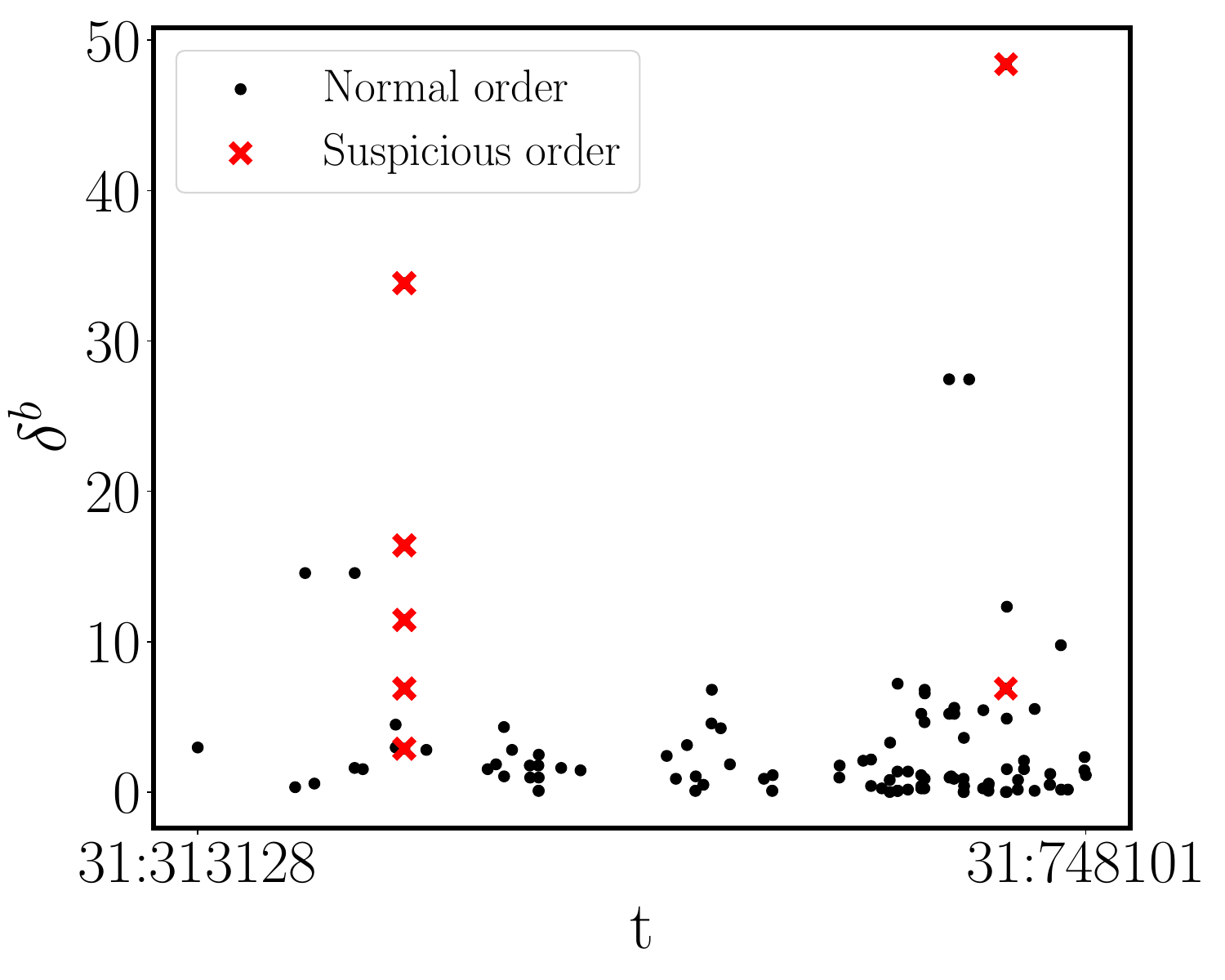}%
    }\hfill
    \subfloat[Spoofer's PnL (USD)]{%
        \includegraphics[width=0.25\linewidth]{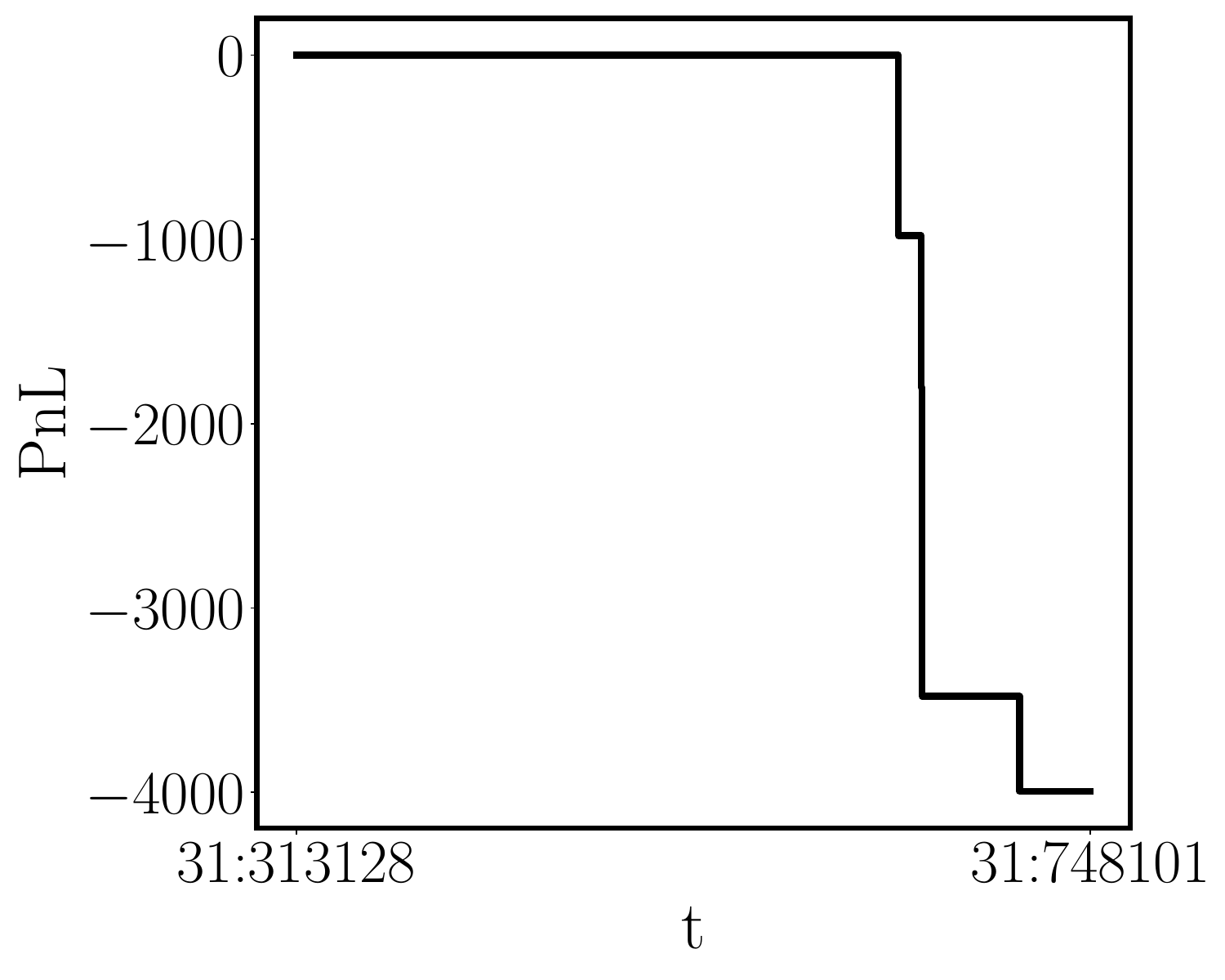}%
    }
    \caption{\textit{Suspicious orders} --- An example of ``suspicious'' behaviour identified by the model on the ask side of ETH-USD LOB. Note the logarithmic scale of the y-axis for the order sizes. The timestamps given in x-axis are given in second:microsecond format for information purposes. This anomaly was spotted on December 6$^{\text{th}}$ at 11:21:31 UTC.}
    \label{fig:ask_spoofing_ethusd_2}
\end{figure}

To provide first some illustrative examples, we display several time frames within which suspicious orders have been spotted. In Figures \ref{fig:bid_spoofing_btcusd_1}, \ref{fig:bid_spoofing_btcusd_2}, \ref{fig:ask_spoofing_btcusd_1}, \ref{fig:ask_spoofing_btcusd_2}, \ref{fig:bid_spoofing_ethusd_1} \ref{fig:bid_spoofing_ethusd_2}, \ref{fig:ask_spoofing_ethusd_1} and \ref{fig:ask_spoofing_ethusd_2}, we show the best prices (BBOs) dynamics with suspicious orders, order sizes, the posting distances of all orders and the profit and loss (PnL) curve of the spoofer. Note that the PnL curve is displayed for illustrative purposes and is the cumulative sum of executions on the \textit{bona fide} side over the time frame.

Remarkably, we observe that when many orders are labeled suspicious within a very short period of time, they are often posted with similar sizes that are round numbers; see, for example, Figures \ref{fig:bid_spoofing_btcusd_1}, \ref{fig:bid_spoofing_btcusd_2}, \ref{fig:ask_spoofing_btcusd_1}, \ref{fig:bid_spoofing_ethusd_1}, and \ref{fig:bid_spoofing_ethusd_2}. This observation indicates that these orders were probably sent by the same agent, which further reveals their suspicious nature. In all of the examples, we observe that suspicious orders are consistently followed by trades being executed on the opposite side, suggesting that their insertion triggers trading activity as intended. More interestingly, we frequently observe a layering behavior: the spoofer posts many spoof orders simultaneously at different price levels in the book, which is a well-documented price manipulation technique \citep{mark2019spoofing}. This pattern is particularly present in Figures \ref{fig:bid_spoofing_btcusd_1}, \ref{fig:ask_spoofing_btcusd_1}, \ref{fig:ask_spoofing_btcusd_2}, \ref{fig:bid_spoofing_ethusd_1}, \ref{fig:ask_spoofing_ethusd_1}, and \ref{fig:ask_spoofing_ethusd_2}. Note that it is not systematically abnormal to see market participants simultaneously post orders at different price levels on the same side. For example, market makers can skew their quote prices in order to be present at many price levels and mitigate adverse selection if an unexpected large price move occurs. In summary, our methodology is capable of detecting suspicious behavior and forms a complementary layer that helps lowering the false positive rates that may result from detection rules that are only based on the size and the distance of posted orders.

Table \ref{tab:descriptive_statistics_orders} reports some descriptive statistics about the size and distance of suspicious and normal orders, and the 1 second price move after they are posted. These statistics are computed for both the BTC-USD and the ETH-USD orders using the univariate skewed Gaussian model. We distinguish between large orders, that is, orders with a size larger than 4,500 USD, and the whole data set of orders. Suspicious orders represent approximately 31\% of all large orders, but 7\% of the entire data set. A negligible proportion of suspicious orders are placed at the top of the queue - 0\% among large orders and only 0.13\% overall. In contrast, 4.20\% of large normal orders and 10.14\% of all normal orders are quoted at the top-of-book level. These findings strongly suggest that spoofing activity is rarely conducted at the top of the book, probably due to the elevated risk associated with such action. Consequently, this challenges the assumptions underlying several mathematical frameworks for the study of spoofing behavior, particularly those relying on top-of-book imbalance signals (for example, \cite{cartea2023spoofing}). 

We observe that suspicious orders are, on average, substantially larger in size compared to normal orders. Similarly, the price distance from the best quote is also significantly greater for suspicious orders, indicating that spoofers post their non-\textit{bona fide} orders deep in the book as a way to minimize the risk of being filled. Furthermore, applying a size-based filter leads to notable differences in price distance for normal orders, but has minimal effect on the average price distance of suspicious orders. This suggests that suspicious orders consistently exhibit atypical placement strategies, regardless of their size. Suspicious orders, regardless of their size, are followed by significantly larger price movements compared to normal orders. This is true both in terms of average price response and skewness.

Together, these results validate the effectiveness of the proposed detection methodology. Orders flagged as suspicious exhibit distinctive behavioral patterns and have a demonstrably different impact on price formation. In addition, these characteristics are closely aligned with empirical studies of spoofing, see, \textit{e.g.}, \cite{lee2013microstructure}.

\begin{table}
    \caption{\textit{Descriptive statistics} --- Statistics regarding the spoof orders detected by the model using the cost function criterion. Distance of placement and realized price move are expressed in basis points of the mid price, the size in USD. The realized price move is multiplied by 1 for bid orders and -1 for ask orders. Bold fonts emphasize the largest value of a given column. BTC-USD, 2024-12-04 to 2024-12-07;  88,327,661 orders, of which  8,601,227 were classified as large (size $\ge 4500$\ USD.}
    \begin{center}
        \begin{tabular}{clcccc}
            \toprule
            \toprule
                     &    & \textbf{Avg. distance}  & \textbf{Avg. size} & \textbf{Avg. price move} & \textbf{Skew price move}\\
            \midrule
                     & Large orders  & \textbf{7.45}  & \textbf{25,291} & \textbf{0.15}  & 0.33  \\
            Suspicious  &  \\
                     & All orders    & 6.85  & 9,661  & 0.11  & \textbf{0.45}  \\
            \midrule
                     & Large orders  & 4.06  & 15,980 & 0.05 & -0.04  \\
            Normal  &  \\
                     & All orders    & 1.03  & 2,042  & 0.02  & 0.02  \\
            \bottomrule
        \end{tabular}
    \end{center}
    \label{tab:descriptive_statistics_orders}
\end{table}

% \begin{figure}[!ht]
%     \centering
%     \subfloat[BTC-USD, Bid side]{%
%         \includegraphics[width=0.25\linewidth]{figures/spoofing_cases/btcusd_bid_spoof_price_variation_density.pdf}%
%     }\hfill
%     \subfloat[BTC-USD, Ask side]{%
%         \includegraphics[width=0.25\linewidth]{figures/spoofing_cases/btcusd_ask_spoof_price_variation_density.pdf}%
%     }\hfill
%     \subfloat[ETH-USD, Bid side]{%
%         \includegraphics[width=0.25\linewidth]{figures/spoofing_cases/ethusd_bid_spoof_price_variation_density.pdf}%
%     }\hfill
%     \subfloat[ETH-USD, Ask side]{%
%         \includegraphics[width=0.25\linewidth]{figures/spoofing_cases/ethusd_ask_spoof_price_variation_density.pdf}%
%     }
%     \caption{\textit{Spoofing detection} --- Density histogram of the realized price move 1 second after the insertion of a normal large order or a suspicious order, computed over the test set.}
%     \label{fig:price_variation_distribution_spoofing}
% \end{figure}

\section{Conclusion}

In this work, we tackle the problem of real-time spoofing detection in a high-frequency trading environment. Our focus was on a simple spoofing strategy which consists of posting a large order on one side of the order book in order to inflate the latent liquidity in the book and ultimately mislead other market participants. We introduced novel order flow variables that are built on the intensity of self-exciting point processes. These variables were computed after limit orders' insertions and used as model features to be applied to the prediction of price move distribution. We explored the dependence of the predicted distribution with respect to the size and the distance of the limit order. We demonstrated that even though its impact on the price distribution vanishes quickly with the distance of placement, significant short-term price impact can still be generated by orders placed deep in the book. Based on this model, we considered a market manipulator willing to improve her expected cost by spoofing the order book. The decision of spoofing or not can thus be made by comparing the cost of trying to manipulate the price with the one of doing nothing. From this consideration a simple detection rule emerges.

Through the analysis of some suspicious orders that were detected by the model, we demonstrated that it can be successfully applied to the identification of abusive behavior, emphasizing the presence of layering patterns. We discussed the natural extension of our framework to cross-asset and cross-venue spoofing, and we provided the corresponding cost functions. Thereby, our model can act as a first layer of a large detection procedure of manipulative orders. We believe that this work brings new insights and contributions to the sparse literature on the detection of high-frequency market manipulation in cryptocurrency CEXs. 

Moreover, it is noteworthy that our framework allows for fast computations, thanks to the simple architecture of the neural network model and the semi-analytic formulas that come with the use of common parametric distributions. Using Python and some code optimization, we were able apply the detection rule for each order in less than 100 microseconds. This demonstrates the applicability of our framework to a live trading environment, where a low-level programming language would further reduce this computation time. Future research could focus on the time horizon of the spoofer and extend our model to multi-horizon spoofing or a horizon in trade time. %Note that in the latter case, the framework with a probabilistic neural network may not be directly applicable as some changes need to be made regarding the specification of the cost function. 

\section{Funding}

T. F. acknowledges funding from ANRT, under the CIFRE contract nr 2021/0905.

\bibliography{main}
\bibliographystyle{apalike}

\section{Appendix}
This appendix provides additional formulas for the calculation of the expected cost functions of the spoofer given by Equations \eqref{eq:cost_bid_spoofer} and \eqref{eq:cost_ask_spoofer}.

\paragraph{Gaussian:} If $\Delta p\sim\mathcal{N}(\mu, \sigma^2)$, then

\begin{equation}
    \mathbb{P}(\Delta p \leq x)=\Phi\left(\frac{x-\mu}{\sigma}\right),
\end{equation}

\begin{equation}
    \mathbb{E}(\Delta p | \Delta p \leq x)=\mu-\sigma\frac{\phi\left(\frac{x-\mu}{\sigma}\right)}{\Phi\left(\frac{x-\mu}{\sigma}\right)},
\end{equation}

\begin{equation}
    \mathbb{E}(\Delta p | \Delta p > x)=\mu+\sigma\frac{\phi\left(\frac{x-\mu}{\sigma}\right)}{1-\Phi\left(\frac{x-\mu}{\sigma}\right)},
\end{equation}

where $\phi(x):=\frac{1}{\sqrt{2\pi}}e^{-\frac{x^2}{2}}$ is the standard Gaussian density function and $\Phi(x):=\frac{1}{\sqrt{2\pi}}\int_{-\infty}^xe^{-\frac{t^2}{2}}\mathrm{d}t$ its cumulative distribution function.

\paragraph{Skewed Gaussian:} If $\Delta p\sim\mathcal{S}\mathcal{N}(\mu, \sigma^2, \alpha)$, then

\begin{equation}
    \mathbb{P}(\Delta p \leq x)=F_\alpha\left(\frac{x-\mu}{\sigma}\right),
\end{equation}

\begin{equation}
    \mathbb{E}(\Delta p | \Delta p \leq x)=\mu+\sigma\frac{2}{\pi}\frac{\beta\Phi\left(\sqrt{1+\alpha^2}\frac{x-\mu}{\sigma}\right)-e^{-\frac{1}{2}\left(\frac{x-\mu}{\sigma}\right)^2}\Phi\left(\alpha\frac{x-\mu}{\sigma}\right)}{F_\alpha(\frac{x-\mu}{\sigma})},
\end{equation}

\begin{equation}
    \mathbb{E}(\Delta p | \Delta p > x)=\mu+\sigma\frac{2}{\pi}\frac{\beta\left(1-\Phi\left(\sqrt{1+\alpha^2}\frac{x-\mu}{\sigma}\right)\right)+e^{-\frac{1}{2}\left(\frac{x-\mu}{\sigma}\right)^2}\Phi\left(\alpha\frac{x-\mu}{\sigma}\right)}{1-F_\alpha(\frac{x-\mu}{\sigma})},
\end{equation}

where $\beta:=\frac{\alpha}{\sqrt{1+\alpha^2}}$, and $F_\alpha$ is the cumulative distribution function of the standard skewed Gaussian distribution with parameter $\alpha$, \textit{i.e.} $\mathcal{S}\mathcal{N}(0, 1, \alpha)$.

\end{document}